\documentclass[a4paper,aps,rmp,reprint,superscriptaddress,notitlepage,floatfix,nofootinbib]{revtex4-2}
\usepackage[utf8]{inputenc}
\usepackage[T1]{fontenc}

\usepackage{greekctr}
\usepackage{xpatch}
\usepackage{amsmath}
\usepackage{amssymb}
\usepackage{stmaryrd}
\usepackage{fourier-orns}
\usepackage{commath}
\usepackage{mathrsfs}
\usepackage{braket}
\usepackage{xcolor}
\usepackage{bm}
\usepackage{csquotes}
\usepackage{graphicx}
\usepackage{booktabs}
\usepackage{enumitem}
\usepackage{siunitx}
\usepackage{microtype}
\usepackage[version=4]{mhchem}

\usepackage{tikz}

\usepackage{hyperref}
\usepackage{bookmark}

\definecolor{tab_blue}{HTML}{1F77B4}
\definecolor{tab_orange}{HTML}{FF7F0E}
\definecolor{tab_green}{HTML}{2CA02C}
\definecolor{tab_red}{HTML}{D62728}
\definecolor{tab_purple}{HTML}{9467BD}
\definecolor{tab_brown}{HTML}{8C564B}
\definecolor{tab_pink}{HTML}{E377C2}
\definecolor{tab_gray}{HTML}{7F7F7F}
\definecolor{tab_olive}{HTML}{BCBD22}
\definecolor{tab_cyan}{HTML}{17BECF}

\definecolor{text_red}{HTML}{AE5149}
\definecolor{text_blue}{HTML}{1F77B4}

\hypersetup{breaklinks=true,
  colorlinks=true,
  linkcolor=tab_blue,
  filecolor=tab_blue,
  urlcolor=tab_blue,
  citecolor=tab_blue,
}

\newcolumntype{L}[1]{>{\raggedright\let\newline\\\arraybackslash\hspace{0pt}}m{#1}}
\newcolumntype{C}[1]{>{\centering\let\newline\\\arraybackslash\hspace{0pt}}m{#1}}
\newcolumntype{R}[1]{>{\raggedleft\let\newline\\\arraybackslash\hspace{0pt}}m{#1}}

\def\ii{\text{i}}
\def\ee{\text{e}}

\def\Id{\text{Id}}

\def\ZZ{\mathbb{Z}}
\def\RR{\mathbb{R}}
\def\CC{\mathbb{C}}

\def\vec#1{\bm{#1}}

\let\strong\textbf

\definecolor{tab_blue}{HTML}{1F77B4}
\definecolor{tab_orange}{HTML}{FF7F0E}
\definecolor{tab_green}{HTML}{2CA02C}
\definecolor{tab_red}{HTML}{D62728}
\definecolor{tab_purple}{HTML}{9467BD}
\definecolor{tab_brown}{HTML}{8C564B}
\definecolor{tab_pink}{HTML}{E377C2}
\definecolor{tab_gray}{HTML}{7F7F7F}
\definecolor{tab_olive}{HTML}{BCBD22}
\definecolor{tab_cyan}{HTML}{17BECF}

\def\RR{\mathbb{R}}

\def\Id{\text{Id}}

\makeatletter\begin{document}
\title{Nonreciprocal many-body physics}
\author{Michel Fruchart}
\affiliation{\text{Gulliver, ESPCI Paris, Université PSL, CNRS, 75005 Paris, France}}
\author{Vincenzo Vitelli}
\affiliation{\text{James Franck Institute, University of Chicago, 60637 Chicago, Illinois, USA}}
\affiliation{\text{Leinweber Center for Theoretical Physics, University of Chicago, 60637 Chicago, Illinois, USA}}

\begin{abstract}
    Reciprocity is a fundamental symmetry present in many natural phenomena and engineered systems.
    Distinct situations where this symmetry is broken are typically grouped under the umbrella term  
    \enquote{nonreciprocity}, colloquially defined by: the action of A on B $\ne$ the action of B on A. In this review, we elucidate what nonreciprocity is by providing an introduction to its most salient classes: nonvariational dynamics, violations of Newton’s third law, broken detailed balance, nonreciprocal responses and nonreciprocity of arbitrary linear operators. Next, we point out where to find these manifestations of non-reciprocity, from ensembles of particles with field mediated interactions to synthetic neural networks and open quantum systems. Given this breadth of contexts and the lack of an all-encompassing definition, it makes it all the more intriguing that some general conclusions can be gathered, when distinct definitions of nonreciprocity overlap. We explore what these universal consequences are with a special emphasis on collective phenomena that arise in nonreciprocal many-body systems. The topics covered include non-reciprocal phase transitions and non-normal amplification of noise and perturbations. We conclude with some open questions.
\end{abstract}

\maketitle

\makeatletter

\def\subparagraph{\@startsection
    {subparagraph}{4}{\z@}{3.25ex \@plus1ex \@minus .2ex}{-1em}{\normalfont\normalsize\itshape\selectfont}}\def\thesubparagraph{\greek{subparagraph}}

\makeatother

\tableofcontents

\section{Introduction}

Reciprocity is a fundamental symmetry present in many natural phenomena and, in different incarnations, it permeates virtually all areas of physics and beyond. In a nutshell, a reciprocity relationship between two entities A and B refers to the property that the effect of A on B is equal to the effect of B on A. Conversely, distinct situations where this symmetry is broken are typically grouped under the umbrella term \textit{non-reciprocity} colloquially defined by 
\begin{equation}
    \text{the action of $A$ on $B$} \neq \text{the action of $B$ on $A$.}
    \label{nr_from_far_away}
\end{equation}   
It is difficult to pinpoint a single all-encompassing notion of non-reciprocity beyond Eq.~\eqref{nr_from_far_away}.
Instead, one can identify several precise definitions that apply in specific contexts.

In population dynamics, an increase in the number of predators reduces the number of preys whereas an increase in the number of preys increases the number of predators. When space is introduced in this ecological setting, a related but distinct facet of nonreciprocity emerges: predators chase preys whereas preys run away from predators. This intuitive example highlights a key consequence of non-reciprocity. When different components of a system have different goals, it may not be possible to fulfill them simultaneously. If the goals are incompatible, 
the two components will quite literally end up running in circles (more precisely limit cycles) because there isn't a stationary state that is optimal for both. Mathematically, models exhibiting this type of run and chase dynamics are examples of non-variational dynamical systems, irrespective of their biological or physical origin.

In the context of physics, these incompatible goals could be represented by two energies $E_A$ and $E_B$ (for each of the two components A and B) that cannot be simultaneously minimized. As a consequence, the non-variational dynamics of the composite system does not follow from minimizing the total energy $E_T=E_A + E_B$. This form of non-reciprocity is characteristic of physical systems composed of particles that interact through non-conservative forces, i.e. forces that cannot be expressed as a gradient of a potential, leading to a non-vanishing circulation. Since motion along a closed loop can generate work (of either sign), energy must be constantly exchanged with the surrounding to support such non-conservative forces at steady state. This is only possible in driven and active systems, in which energy is only conserved if one includes the environmental degrees of freedom from which it flows in or out. More broadly, non-variational dynamics generalizes the notion of steepest descent on an energy landascape also to problems where the landscape represents non-physical quantities, e.g. evolutionary dynamics that is not described as climbing up a fitness landscape or algorithms in computer science that are not governed by an optimization principle. 

Perhaps the first encounter with the notion of (non)reciprocity in a physics education is through 
Newton's third law, formally equivalent to conservation of linear momentum, which is a well-defined notion within mechanical contexts, but not necessarily applicable beyond, despite similarities at the level of Eq.~\eqref{nr_from_far_away}.
Contrary to the intuition gathered from our early exposures to Newtonian dynamics, violations of the third law are easy to find, even in relatively simple systems, if we take into account the inextricable link between particles and fields that underlie much of physics. While two charged particles \textit{at rest}
experience Coulomb forces that are equal and opposite, this reciprocity is violated if said charges \textit{move} along directions perpendicular to each other, as you can see from a glance at the schematic reproduced from \citet{Feynman1989} in section \ref{EM}.  
Here, a violation of Newton's third law is possible because linear momentum is exchanged with the electromagnetic field generated by the moving charges (that in turn mediates their interactions), very much like energy in the case of non-conservative forces analyzed in the previous paragraph. Note, however, that the net effect of these field-mediated forces vanishes at equilibrium where all directions of motion are equally weighted to determine the average behavior. This observation brings to the fore a more general conclusion: it is out of equilibrium that nonreciprocity typically manifests itself. 

The field-mediated interactions exemplified in the previous paragraph occur more broadly throughout physics and chemistry if different types of fields are considered. A very intuitive (but more complex to analyse mathematically) example is so-called hydrodynamic interactions between particles immersed in a fluid, see Sec. \ref{hydroint}. These forces seemingly violate Newton' third law because the particles, ranging from active colloids to swimming embryos, can exchange momentum (linear or angular) with the surrounding fluid. Such non-reciprocal forces can be mediated also by chemical fields triggered, for example, by reactions occurring on the particles themselves (e.g. oil droplets or enzymes) that affect how they interact with each other. 

Even in cases where the microscopic origin of non-reciprocal inter-particle forces has long been recognized, much remains unexplored about the collective properties of many such particles, and more broadly of extended media in which non-reciprocity survives at the macroscopic level in the form of peculiar responses, e.g. elastic or transport coefficients that appear in the corresponding hydrodynamic theories and in the lab, see Sec. \ref{hall_odd_effects}. The theoretical and experimental characterization of non-reciprocal many-body systems is the scope of this review. 

Examples of non-reciprocal many-body physics can be encountered even in everyday phenomena and industrial processes such as sedimentation, i.e. particles falling in a fluid, a time-honored but active subject, see Sec. \ref{hydroint}. Another cutting-edge area with deep roots in the history of neuroscience and machine learning, is how collective phenomena of learning and memorization occur in (typically disordered) systems composed of many neurons (synthetic or natural), see Sec. \ref{NN}. This is a problem with far reaching implications ranging from artificial and biological neural networks to neuromorphic computation. 

Aside from specific examples, it should not come as a surprise that, when the many-body problem is combined with the out of equilibrium character inherent to the notion of non-reciprocity, great complexity and largely unsolved mathematical challenges immediately arise. If anything, the cursory description of the experimental examples sketched above plays down their richness. As a case in point, revisit the fluid mediated interactions between particles. Aside from violating Newton's third law and being potentially non-conservative, these nonreciprocal interactions are often non-pairwise precisely because they are mediated by a field which introduces phenomena like non locality in space or time (through asymmetric time delays leading to non-Markovian dynamics). Even more stringent issues are at play when trying to model cognitive processes and even artificial neural networks using minimal models based on say non-reciprocal spin glasses. These difficulties notwithstanding, it is satisfying to discover how often qualitative conclusions about these complex systems can be traced directly to Eq.~\eqref{nr_from_far_away} in a system-independent manner. 

This universality may remind us of another fundamental context where (non)reciprocity enters the physics curriculum: the so-called Onsager reciprocity relations and their intimate connection with the thermodynamic notion of detailed balance, Sec.~\ref{broken_detailed_balance}. Onsager relations govern the response of transport coefficients such as diffusion or electrical and thermal conductivities
that arise by linearizing the response of physical systems close to equilibrium. Even without performing an explicit coarse-graining from particle-based models, one can deduce through general principles of statistical mechanics how violations (or generalizations) of Onsager reciprocity relations arise from operating near a non-equilibrium steady state. From a practical perspective, this results in generalized transport coefficients with tangible experimental consequences ranging from magnetohydrodynamics to active matter, Sec. \ref{hall_odd_effects}. 
Direct generalizations of Onsager's ideas to arbitrary linear operators have led to the formulation of more abstract notions of non-reciprocity, as explained in Sec. \ref{reciprocity_linear_system} with importat consequences for the engineering of optical and mechanical metamaterials, Sec. \ref{mat}.

The remainder of this review is organized in three main parts each addressing a broad question that guides our presentation. The first part, Sec. \ref{what} attempts to answer the question: What is nonreciprocity? Our answer is to provide a more mathematical exposition of the classes of non-reciprocity intuitively sketched in the introduction: non variational dynamics, violations of Newton's third law, broken detailed balance, nonreciprocal responses and nonreciprocity in arbitrary linear operators. Here, the mathematically inclined reader will find concise treatments of powerful techniques to systematically decompose a
generic dynamical system into “purely variational and non-variational” parts, handle symmetries and stochastic effects to name but a few.
Casual readers will find a summary of the underlying ideas with references to textbooks for further study.

Section \ref{where} answers the question: Where to find non-reciprocity? We mirror in the subheadings of the application section \ref{where}, the corresponding theoretical treatments of Sec. \ref{what}. Readers that opt for a non-sequential reading of this review 
could for example study the Onsager-Casimir theorem in Sec.~\ref{nonreciprocal_responses} and then turn to the identically named application section \ref{nrresponse} for concrete examples in say fluid mechanics, see also Fig.~\ref{figure_responses}. 
Similarly, Secs.~\ref{nonvariational_dynamics} and \ref{where_nonvariational_dynamics} (on nonvariational dynamics, Fig.~\ref{examples_nonvariational}) and Secs.~
\ref{violations_newton_third_law_general} and \ref{violations_newton_third_law} (on violations of Newton's third law, Fig.~\ref{examples_newton_third_law}) can be read in parallel.

The last part of the review, Sec. \ref{what_are_consequences} answers the question: What are the consequences of non-reciprocity? One of the primary topics we cover is the ongoing generalization to non-reciprocal many-body systems of the notions of phase transitions and universality classes familiar from equilibrium statistical physics. Simple models such as the nonreciprocal Ising model are used as an illustration. The concept of \enquote{time crystal} and its generalizations are introduced to explain the behavior of quantum and classical many-body systems with spontaneously broken time-translation symmetry. We also stress how the non-normal operators describing non-reciprocal systems generically lead to transient amplification of perturbations, not predicted by their eigenvalues, and hence to an increased sensitivity to noise. In the Conclusion, we sketch some promising directions for future research along with a brief summary.

\section{What is non-reciprocity?}
\label{what}

As pointed out in the Introduction, there is no unique definition of nonreciprocity. It makes it all the more intriguing that some general conclusions can be gathered, that apply when distinct definitions overlap. We start with a heuristic classification of the kinds of non-reciprocity that one can encounter and then focus on several classes of non-reciprocity in detail.

\subsection{Overview and classification}
\label{overview_classification}

\begin{figure*}
    \centering
    \includegraphics[width=172mm]{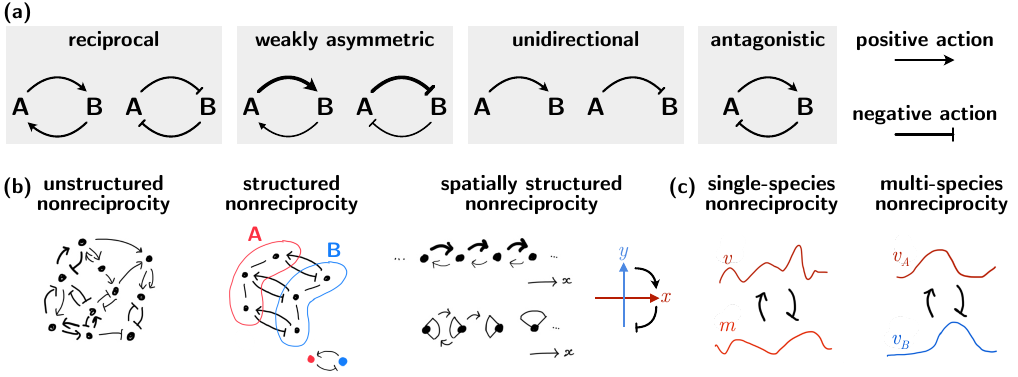}
    \caption{\textbf{Classes of nonreciprocity.}
    (a) Different classes of nonreciprocal behavior can be distinguished when the action of A on B has a strength that can be \enquote{positive} or \enquote{negative}.
    (See Fig.~\ref{nonreciprocal_scattering_elements} for a less binary notion of phase nonreciprocity.)
    We can then distinguish (i) reciprocal behavior where sign and amplitude are the same, (ii) weakly asymmetric behavior where amplitudes are different but signs identical, (iii) unidirectional nonreciprocity when A has an action on B and B has no action on A, and (iv) antagonistic nonreciprocity where the signs are different.
    (b) In addition, nonreciprocity may be completely unstructured within degrees of freedom, structured with two or more species, or structured in space, in several different ways.
    (c) Finally, nonreciprocity can occur within a species of agents, between different fields or modes, or between different species.
    Details are given in Sec.~\ref{overview_classification} of the main text.
\label{classes_of_nonreciprocity}
    }
\end{figure*}

Equation \eqref{nr_from_far_away} can be written as $\mathcal{A}_{BA} \neq \mathcal{A}_{AB}$ in which $\mathcal{A}_{BA}$ represents \enquote{the action of A on B}, to be better specified later in particular cases.
Interactions between two entities can often be classified as \enquote{positive} or \enquote{negative}, suggesting that we can think of $\mathcal{A}_{BA}$ as a number with a \textcolor{text_red}{positive} or \textcolor{text_blue}{negative} sign.
For example,
\begin{itemize}[nosep,left=0pt,label=--]
    \item a particle can mechanically \textcolor{text_red}{attract} or \textcolor{text_blue}{repel} another;
    \item a neuron can be \textcolor{text_red}{excitatory} or \textcolor{text_blue}{inhibitory};
    \item a spin may \textcolor{text_red}{align} or \textcolor{text_blue}{anti-align} with another;
    \item a gene may \textcolor{text_red}{activate} or \textcolor{text_blue}{repress} the expression of another;
    \item a species may \textcolor{text_red}{increase} or \textcolor{text_blue}{reduce} the population of another;
    \item someone may \textcolor{text_red}{love} or \textcolor{text_blue}{hate} somebody else.
\end{itemize}
Borrowing a graphical notation from systems biology \cite{Alon2019}, we represent positive interactions with a pointed head arrow (\textcolor{text_red}{$\to$}) and negative interactions with a blunt head arrow (\textcolor{text_blue}{$\raisebox{-.2ex}{\rotatebox{90}{\scalebox{1}[1.2]{$\bot$}}}\,$}), see Fig.~\ref{classes_of_nonreciprocity}a.
With this in mind, we can distinguish several cases:
\begin{itemize}[nosep,left=0pt,label=--]
    \item \textit{unidirectional non-reciprocity}: $\mathcal{A}_{ij} = 0$ while $\mathcal{A}_{j i} \neq 0$ 
    \item \textit{antagonistic non-reciprocity}: $\mathcal{A}_{ij}$ and $\mathcal{A}_{j i}$ have opposite signs
    \item \textit{weak non-reciprocity}: $\mathcal{A}_{ij}$ and $\mathcal{A}_{j i}$ have the same sign, but $\mathcal{A}_{ij} \neq \mathcal{A}_{j i}$
    \item \text{reciprocity}: $\mathcal{A}_{ij} = \mathcal{A}_{ji}$.
\end{itemize}
These are represented represented in Fig.~\ref{classes_of_nonreciprocity}a.
Intuitively, the case of unidirectional non-reciprocity is the strongest, because it cannot be modified by rescaling.
In the romance example, for instance, A loves B but B is indifferent to A. 
Even though the intuition provided by this classification is useful, one should bear in mind that it is often a simplification. 
For instance, $\mathcal{A}_{BA}$ may be a complex number or a vector, where positive and negative numbers only corresponds to limiting cases (see Sec.~\ref{nonreciprocal_devices} for examples of phase nonreciprocity).
In addition, the sign of an interaction may not always be the same (for instance, mechanical forces between two particles may be attractive or repulsive depending on their distance; or may depend on the presence or absence of another particle in nonpairwise interactions).

In many-body systems, there is a second layer of classification, depending on whether there is a structure in how all the entities interact.
As an example, random interactions (Fig.~\ref{classes_of_nonreciprocity}b, left) would correspond to an entirely unstructured class of nonreciprocity.
In contrast, an example of structured nonreciprocity is given by two populations A and B in which the action of A on B is always negative, while the action of B on A is always positive
(Fig.~\ref{classes_of_nonreciprocity}b, middle).
In the structured case, nonreciprocity is expected to persist at a coarse-grained level, either through the existence of macroscopic populations (inset of the panel, where there are only two) and/or of macroscopic fluxes.
This coarse-graining can be in principle be performed systematically by identifying network motifs in the graph of interactions \cite{Alon2019}.

A particular class of structuration involves physical space, which always plays a distinguished role in physics (right panel of Fig.~\ref{classes_of_nonreciprocity}b).
For instance, the spatial organization of physical entities may constrain interactions through locality and spatial symmetries, and select natural observables.
Interactions from the left to the right may be stronger, say, than the other way around (top panel). 
More complicated examples involve dynamic rearrangements of the graph, like in the case of entities with a cone of vision (bottom panel).
In addition, the interacting entities may move in physical space, thereby changing the interaction network. This can lead to a variety of phenomena studied in the field of self-propelled active matter \cite{Cates2015,Marchetti2013,Bechinger2016,Granek2024,OByrne2022}, which can also have non-reciprocal features.

Finally, there are several ways by which nonreciprocally coupled components can emerge in systems made of many units each carrying several degrees of freedom (Fig.~\ref{classes_of_nonreciprocity}c).
We can first distinguish single-species from multiple-species nonreciprocity.
In the former case, nonreciprocity can arise from the existence of multiple fields corresponding to different degrees of freedom (e.g., position and orientation) or from different parts of the \emph{same} field (e.g., different Fourier harmonics, or different components of the vector/tensor field).
In the latter case, the non-reciprocity arises from the existence of multiple species of units; one can further assess whether the nonreciprocal coupling happends between fields of the same nature (say, the magnetization of the different species), of different natures (say, the magnetization of species $i$ and the density of species $j$), or even between different parts of different fields.

\subsection{Nonvariational dynamics}
\label{nonvariational_dynamics}

\noindent\emph{\textbf{In a nutshell:} nonvariational first-order dynamics can be seen as non-reciprocal.}
\medskip

Consider a time-independent dynamical system
\begin{equation}
    \frac{d \vec{x}}{d t} = \vec{f}(\vec{x})
    \label{ds}
\end{equation}
defined by a collection of first-order ordinary differential equations (ODEs) collected in the vector field $\vec{f} : \mathbb{R}^N \to \mathbb{R}^N$ \footnote{It is sometimes useful to consider a more general notion of dynamical system in continuous phase space and time: a flow on a set $X$ is an action $\varphi : \mathbb{R} \times X$ of the group $(\mathbb{R}, +)$ on $X$. Elements $t \in \mathbb{R}$ of the group represent time, and $\varphi_t(x) \equiv \varphi(t,x)$ gives the state at $t$ of a system that was in state $x$ at time zero.\label{flow_footnote}}.
We say that this dynamical system is variational if there is a scalar function $V : \mathbb{R}^N \to \mathbb{R}$ such that
\begin{equation}
    \vec{f} = - \frac{d V}{d \vec{x}}
    \quad
    \text{(variational)}
    \label{gradient_ds}
\end{equation}
and the vector field $\vec{f}$ is called a gradient vector field. In this case, the dynamics is entirely defined by the tendency of the system to go straight down the gradient of potential $V$.
Conversely, the dynamical system is said to be nonvariational (or nongradient) if there is no potential $V$ such that Eq.~\eqref{gradient_ds} is satisfied. 

Nonvariational dynamical systems are nonreciprocal in the sense that the Jacobian
\begin{equation}
    J_{ij} = \frac{\partial f_i}{\partial x_j}
\end{equation}
of a non-variational dynamical system is typically not symmetric, namely
\begin{equation}
    \label{nonsymmetric_jacobian}
    J_{ij} \neq J_{ji}.
\end{equation}
In contrast, when $\vec{f} = - \nabla V$ for a smooth potential $V$, then we necessarily have $J_{ij} = - \partial_i \partial_j V = - \partial_j \partial_i V = J_{j i} $ in which $\partial_i = \partial/\partial x_i$ and where we have used Schwarz's theorem on the symmetry of second derivatives.
(The converse does not always hold: see Sec.~\ref{helmholtz_hodge}.)

\subsubsection{What makes nonvariational dynamics interesting?}
\label{why_nonvariational}

A dynamical system like Eq.~\eqref{ds} may exhibit structures called attractors, that are invariant under the dynamics, and towards which the system tends to evolve\footnote{All of these structures can also repel the dynamics or act as saddles. We refer to (some reference on Conley-Morse) for more details.}. 
These include stable fixed points, limit cycles, or more complicated sets like chaotic attractors (these can be seen as the dynamical system version of dissipative structures as defined by \citet{Nicolis1973}).
Crucially, non-trivial attractors that exhibit steady-state dynamics cannot exist in variational dynamical systems  (that can only have fixed points), nor in volume-conserving dynamics like Hamiltonian systems (that cannot have attractors). 

Indeed, the convergence of trajectories towards the attractor in its basin of attraction is manifested by the fact that the rate of change $\Lambda \equiv \nabla \cdot \vec{f}$ of phase space volume is negative (meaning that the dynamics contracts phase space). 
This excludes volume-preserving dynamics (for which $\Lambda = 0$ everywhere) right away.
In particular, this excludes Hamiltonian dynamics because of Liouville theorem \cite{Arnold1989}.
A variational dynamical system like Eq.~\eqref{gradient_ds} does not preserve phase-space volume, but its only attractors are minima of $V$, where $\bm{f}(\bm{x})$ vanishes by definition, so there is no dynamics.

\begin{figure}
    \centering
    \includegraphics[width=8cm]{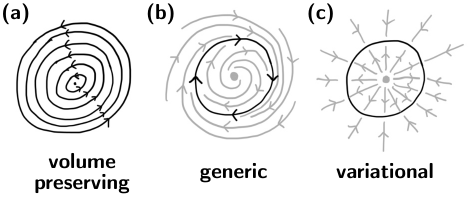}
    \caption{
    \label{different_dynamics}
    \textbf{Volume-preserving versus gradient-descent dynamics.}
    Sketch Eq.~\eqref{example_mix}
When $V = 0$, the phase space is made of concentric periodic orbits centered around a fixed point called a center.
The motion occurs on a given periodic orbit selected by the initial value of $H$, and any perturbation of the system will make it go from one periodic orbit to the other (panel a).
When $H = 0$, the variational dynamics drives the system towards the circle of stable fixed points ($r \equiv \lVert x \rVert = \sqrt{-a/b}$) at the bottom of the wine-bottle potential, where there is no motion (panel c).
    }
\end{figure}

Pictorially, this shows that one needs a mix of Hamiltonian-like and gradient-descent-like dynamics to produce non-trivial attractors.
This idea is illustrated in Fig.~\ref{different_dynamics} where we have sketched the phase portrait of the dynamical system
\begin{equation}
    \dot{x}_i
    =
    - \partial_i V
    + \epsilon_{ij} \partial_j H
    \label{example_mix}
\end{equation}
for $\bm{x}(t)=(x_1(t),x_2(t)) \in \RR^2$.
This dynamical system is the sum of a variational part controlled by the wine-bottle potential $V(\bm{x}) = a \lVert\bm{x}\lVert^2/2 + b \lVert\bm{x}\lVert^4/4$ (we take $a < 0$ and $b > 0$) and a volume-preserving part ruled by the Hamiltonian function $H(\bm{x}) = \omega x y$ ($\epsilon$ the fully antisymmetric matrix, which is the 2D standard symplectic matrix).
Only when both $V$ and $H$ are present does the system exhibits a limit cycle (panel b).
Else, the system either exhibits non-isolated periodic orbits (panel a, when $V=0$) or gradient-descent dynamics towards a circle of fixed point (panel c, when $H=0$)

\subsubsection{Decomposing dynamical systems}

We would like a systematic way of decomposing a generic dynamical system into a purely variational part and a \enquote{purely non-variational} part.
It turns out that doing so requires care. 
This is because we implicitly hope to use this decomposition for various purposes, which require the decomposition to satisfy constraints that have to be specified depending on the intended use.

For instance, as
\begin{equation}
    \bm{f}(\bm{x}) = - \nabla U(\bm{x}) + \bm{f}_{\text{nc}}(\bm{x})
    \label{decomposition}
\end{equation}
with a gradient part described by an effective potential $U$ plus a non-conservative part $\bm{f}_{\text{nc}}$, like in the example of Eq. \eqref{example_mix}.
As a consequence, several approaches have been proposed. 

First, it is desirable that the decomposition allows us to distinguish between transient and steady-state behavior.
In particular, we hope that the effective potential $U(\bm{x})$ is a Lyapunov function. That means that the potential satisfies $dU/dt < 0$ away from attractors and $dU/dt = 0$ on attractors.
Second, it is desirable that an effective potential gives information about the effect of noise on the system.
In particular, we hope that it gives access to the probability per unit time of jumping between attractors of the deterministic dynamical system in the small noise limit.
In equilibrium systems, this is given by the Arrhenius-Kramers-Eyring law \cite{Hanggi1990,Berglund2011,Bovier2004,Bovier2005}, which is generalized to irreversible systems through the Freidlin–Wentzell large deviation theory~\cite{Freidlin2012,Bradde2012,Bouchet2016}, see also \citet{Santolin2025b} for a stochastic thermodynamics perspective.
Finally, it should be as easy as possible to obtain the decomposition from the expression of $\bm{f}$.

\paragraph{Conley decomposition.} 
\label{conley_decomposition}

The dynamics of an arbitrary dynamical system can be fairly complicated, and this makes it difficult to assess whether some behavior is transient or corresponds to a non-trivial steady-state, such as a limit cycle or a chaotic attractor, whose behavior is recurrent rather than transient.
One of our hopes in decomposing the dynamics is to give a precise meaning to the difference between recurrent and transient behaviors. 

\citet{Conley1978} introduced a decomposition that precisely splits the dynamics into a \enquote{strongly gradient-like} part that describes transient behavior and a \enquote{non-gradient like part} that corresponds to recurrent behavior~\cite{Lewis2025}.
The latter is formalized through the \emph{chain-recurrent set} of the dynamical system. 
Colloquially, a point $x$ is said to be chain-recurrent if we would mistake it for a periodic point at any finite resolution $\epsilon$. 
Formally, $x$ is chain recurrent for the flow $\varphi$ (see footnote \ref{flow_footnote}) if for all $\epsilon > 0$ and $T >0$ there is a sequence of points $x_0, x_1, \dots, x_n$ and times $t_1, \dots, t_n$ with all $t_i > T$ such that the distance $\lVert x_i -  \varphi_{T_i}(x_i) \rVert < \epsilon$ for all $i=1,\dots,n$.
The chain-recurrent set, the set of all chain-recurrent points, contains all the interesting parts of the dynamical system, including all fixed points, limit cycles, tori, chaotic attractors and repellers, and so on.
Crucially, \citet{Conley1978} showed that, for flows on a compact metric space, the dynamics is gradient-like outside of the chain-recurrent set, meaning that if we collapse each connected component of the chain-recurrent set into a single point, then there is a global Lyapunov function for the flow that is strictly decreasing outside of the collapsed chain-recurrent set. 
Details and extensions to more general settings are provided in \citet{Lewis2025,Norton1995,Alongi2007} and references therein, and algorithmic approaches are discussed in \citet{Ban2006,Hafstein2015,Argaez2019}.

The Conley decomposition theorem provides a sound conceptual framework to separate transient and recurrent behavior in deterministic dynamical systems.  
However, it provides a decomposition of \emph{phase space} rather than a decomposition of the flow or the vector field $\bm{f}$. 
This makes it very powerful, but not always practical.

\paragraph{Helmholtz-Hodge decomposition.} 
\label{helmholtz_hodge}
Perhaps the most straightforward way of decomposing the vector field $\bm{f} : \Omega \subset \mathbb{R}^n \to \mathbb{R}^n$ defining the dynamical system consists in using a Helmholtz-Hodge decomposition to write
\begin{equation}
    \bm{f}(\bm{x}) = - \nabla U^{\text{HH}}(\bm{x}) + \bm{f}^{\text{HH}}_{\text{nc}}(\bm{x})
    \label{hh_decomposition}
\end{equation}
with $\nabla \cdot \bm{f}^{\text{HH}}_{\text{nc}} = 0$.
Such a decomposition exists when there is a solution $U^{\text{HH}}$ to the Poisson equation $\Delta U^{\text{HH}} = - \nabla \cdot \bm{f}$ on $\Omega$.
Under certain conditions, one can show that the solution of the Poisson equation and hence the Helmholtz-Hodge decomposition are unique, but this is not necessarily the case.
\citet{Bhatia2013} reviews the history and applications of the Helmholtz-Hodge decomposition; mathematical details can be found in \citet[\S~3.5]{Schwarz1995}.
Note that the Helmholtz-Hodge decomposition emphasizes variational dynamics (rather than relaxational dynamics, see Sec.~\ref{variational_vs_relaxational}). One of the shortcomings of the Helmholtz-Hodge decomposition is that the potential $U^{\text{HH}}$ cannot always be chosen to be a Lyapunov function \cite{Suda2019,Suda2020}.

As alluded in the introduction of Sec.~\ref{nonvariational_dynamics}, a necessary condition for a vector field $\bm{f}(\bm{x})$ to be variational is to satisfy
\begin{equation}
    \partial_j f_i = \partial_i f_j.
    \label{symmetry_derivatives}
\end{equation}
In certain conditions (for instance, on an open star-convex domain), the Poincaré lemma provides a converse and the condition Eq.~\eqref{symmetry_derivatives} is enough to guarantee that $\bm{f}$ is a gradient vector field \cite{Bott1982}.
Indeed, Eq.~\eqref{symmetry_derivatives} means that a generalized curl $\bm{\omega} = d\bm{f}^\flat = \frac{1}{2} (\partial_j f_i - \partial_i f_j) dx^i \wedge dx^j$ vanishes, which is the exterior derivative of a 1-form naturally associated to $\bm{f}$ ($\bm{\omega}$ is similar to the curl $\nabla \times \bm{f}$ in 3D).
However, the converse does not always hold: when phase space has a non-trivial topology, it is possible to have non-variational dynamical systems without having Eq.~\eqref{nonsymmetric_jacobian}. 
For instance, the dynamics $\dot{\theta} = \omega_0$
of an angle variable $\theta \in S^1$ is not variational, because this dynamics cannot be generated by any continuous periodic potential $U(\theta)$. 
This is true even though we don't have \eqref{nonsymmetric_jacobian} because the topology of phase space allows us to evade the Poincaré lemma.
The asymmetry in \eqref{nonsymmetric_jacobian} would be restored if this dynamics were embedded in a larger phase space on which Poincaré lemma applies (for instance by going back to the full dynamics if the equation was obtained by phase reduction). 

\citet{OByrne2020,OByrne2023,OByrne2024,OByrne2025} developed a functional version generalizing these ideas to spatially extended dynamical systems described by PDEs or field theories.

\paragraph{Transverse decomposition and quasipotential.}
\label{transverse_decomposition}

In this paragraph, we discuss a particular choice of decomposition called \enquote{transverse} and how it allows one to relate the deterministic dynamics to the weak-noise limit of a stochastic dynamics through a non-equilibrium equivalent of the free energy called the quasipotential \cite{Freidlin2012,Graham1985,Graham1986,Graham1989,Bouchet2016b}.
Other approaches based on stochastic dynamics are reviewed by \citet{Wang2015,Yuan2017,Fang2019}, with a particular focus on biological systems.

We can demand that $\nabla U$ and $\bm{f}_{\text{nc}}$ are orthogonal to each other at every point, namely that $\nabla U \cdot \bm{f}_{\text{nc}} = 0$.
This constraint takes the form of a (zero-energy) Hamilton-Jacobi equation
\begin{equation}
    \nabla U \cdot (\bm{f} + \nabla U) = 0.
    \label{hamilton_jacobi}
\end{equation}
When Eq.~\eqref{hamilton_jacobi} has a smooth solution $U^{\text{HJ}}$, it is guaranteed to be a Lyapunov function, for it implies that 
\begin{equation}
    \frac{dU}{dt} = (\nabla U) \cdot \bm{f} = - \lVert \nabla U\rVert^2 \leq 0.
\end{equation}
It turns out that the resulting transverse decomposition $\vec{f} = - \nabla U^{\text{HJ}} + \vec{f}_{\text{nc}}^{\text{HJ}}$, when it exists, also 
arises from looking at the deterministic dynamical system \eqref{ds} as the zero-noise limit of a stochastic dynamics
\begin{equation}
    \frac{d \vec{x}}{d t} = \vec{f}(\vec{x}) + \sqrt{2\epsilon} \, \bm{\eta}(t)
\label{random_perturbation}
\end{equation}
in which $\bm{\eta}(t)$ is a normalized Gaussian white noise and $\epsilon$ controls the amplitude of the noise.
Indeed, in the weak-noise limit $\epsilon \to 0$, the stationary probability distribution $p_{\text{ss}}(\bm{x})$ associated to Eq.~\eqref{random_perturbation} takes the form
\begin{equation}
    p_{\text{ss}}(\bm{x}) 
    \underset{\epsilon\to 0}{\propto}
    \ee^{U^{\text{QP}}(\bm{x})/\epsilon}
    \label{ldfw}
\end{equation}
in which $U^{\text{QP}}$ is called the quasipotential associated to Eq.~\eqref{random_perturbation}.
More precisely, $U^{\text{QP}}$ is a large deviation function \cite{Touchette2009} that arises in the theory of \citet{Freidlin2012}, which relates the behavior of the stochastic dynamics \eqref{random_perturbation} in the limit $\epsilon \to 0$ with the behavior of the deterministic dynamics \eqref{ds} (under some conditions, see \citet{Freidlin2012} for the theorems and \citet{Bouchet2012} for a situations where the Freidlin-Wentzell theory does not apply directly).

Crucially, the Freidlin–Wentzell quasipotential contains information about all the attractors of the deterministic systems and their relative \enquote{heights} in an effective potential landscape \cite{Graham1986}, which allows for a generalization the Arrhenius-Kramers-Eyring law to irreversible systems~\cite{Freidlin2012,Bradde2012,Bouchet2016}, giving the coarse-grained probability rate $p(a \to b)$ of jumping from an attractor $a$ of the deterministic dynamical system to another attractor $b$ in the $\epsilon \to 0$ limit.
Such a reduced Markov chain description is constructed explicitly in \citet{Lee2022b} for a particular class of diffusive processes where $\bm{f}_{\text{nc}}$ in the transverse decomposition is also incompressible (see also Sec.~\ref{accelerated_convergence}).

When it is defined, $U^{\text{HJ}}$ coincides with $U^{\text{QP}}$ (up to a constant).
This is always the case in equilibrium systems.
In systems with broken detailed balance (Sec.~\ref{broken_detailed_balance}), however, the quasipotential can exhibits singularities that prevent the existence of a smooth solution to Eq.~\eqref{hamilton_jacobi}.
These singularities of the large deviation function $U^{\text{QP}}$, reviewed by \citet{Baek2015}, are related to the behavior of rare events in the noisy system.
In short, the most probable trajectory of the stochastic system conditioned on the occurrence of a given fluctuation (namely, conditioned to start at $(t_1, \bm{x}_1)$ and to end at $(t_2, \bm{x}_2)$), called the optimal path or instanton, may abruptly change because it switches between two locally optimal paths that exchange global optimality.
The singularities of the quasipotential can be seen as phase transitions in
the fluctuations of physical observables, that have been termed \enquote{dynamical phase transitions} in the literature \cite{Falasco2025,Tsobgni2016}.

More recently, a generalization of the Freidlin–Wentzell to infinite-dimensional dynamical systems known as macroscopic fluctuation theory has been developed \cite{Bertini2015}.
It describes the weak noise limit of fluctuating diffusive hydrodynamics.
In the same spirit, procedures to obtain exact hydrodynamic descriptions of nonequilibrium lattice gases and the corresponding fluctuating hydrodynamics have been developed \cite{KourbaneHoussene2018,Agranov2023,Agranov2021}.
\citet{Bernard2021} discusses possible extensions to quantum systems.
As we shall see, this approach can be fruitful in understanding the possible behaviors of phase transitions in non-reciprocal systems (Sec.~\ref{nonreciprocal_flocking}).

Several numerical algorithms have been developed to compute the quasipotential, and we refer the reader to \citet{Maier1996,Weinan2002,Heymann2008,Kikuchi2020,Zakine2023,Simonnet2023} for details.

\paragraph{Relaxational dynamics and multiplicative noise.}
\label{variational_vs_relaxational}

The class of variational dynamics defined by Eq.~\eqref{gradient_ds} is not invariant under a change of coordinates in parameter space. This definition is useful when the coordinate system is fixed by something else (like observables or a noise term), but it is not a good basis for an intrinsic definition.
Instead, we can consider \emph{relaxational} dynamical system in which the vector field in Eq.~\eqref{ds} is of the form
\begin{equation}
    f_i(\bm{x}) = - \Gamma_{ij}(\bm{x}) \frac{\partial V}{\partial x_j}
    \label{relaxational_ds}
\end{equation}
in which $\bm{\Gamma}(\bm{x})$ is a positive-definite matrix.
Equation \eqref{relaxational_ds} can be obtained from Eq.~\eqref{gradient_ds} by a change of variable\footnote{
Starting with $\dot{\bm{y}} = - \nabla_{\bm{y}} V$ and making a change of variable $\bm{x} = \bm{x}(\bm{y})$, we get
\begin{equation*}
    \frac{dx_i}{dt} 
    = \frac{\partial x_i}{\partial y_j} \frac{d y_j}{dt} 
    = - \frac{\partial x_i}{\partial y_j} \frac{\partial V}{\partial y_j} 
    = - \frac{\partial x_i}{\partial y_j} \frac{\partial x_k}{\partial y_j} \frac{\partial V}{\partial x_k} 
\end{equation*}
and so we can define $\Gamma = J J^T$ in which $J_{ij} = \partial x_i/\partial y_j$ is the Jacobian of the change of variable to obtain Eq.~\eqref{relaxational_ds}.
}.
It can be seen as a gradient dynamics $\dot{\bm{x}} = - \nabla_{\bm{M}} V$ on a manifold with metric tensor $\bm{M} = \bm{\Gamma}^{-1}$, also known as natural gradient in the context of learning \cite{Amari1998}. 
Both variational and relaxational dynamics tend to contract the dynamics towards stable fixed points.
In the variational case, for instance, the potential is a Lyapunov function as $\dot{V} = - \lVert \nabla V(\bm{x}_t) \rVert \leq 0$; when $V$ is analytic around a non-degenerate critical point $x^*$, the Łojasiewicz inequality guarantees that $\lVert \nabla V(\bm{x}_t) \rVert \geq C \lVert V(\bm{x}_t) - V(\bm{x}^*) \rVert$ for some constant $C$, which implies exponential convergence towards the fixed point \cite{LojasiewiczEoM}.
Similar statements can be made for relaxational dynamics, see \citet{Wensing2020} for a review.

Now, in the presence of multiplicative noise, Eq.~\eqref{random_perturbation_multiplicative} becomes
\begin{equation}
    \frac{d \vec{x}}{d t} = \vec{f}(\vec{x}) + \sqrt{2\epsilon} \, \bm{g}(\bm{x}) \bm{\eta}(t)
    \label{random_perturbation_multiplicative}
\end{equation}
in which $\bm{g}(\bm{x})$ is a matrix\footnote{This equation can be interpreted as the Itô SDE
\begin{equation*}
    d x_i = f_i(\vec{x}) dt + g_{i j}(\vec{X}) dW_j
\end{equation*}
where $dW_i$ are independent Wiener processes.
}.
The discussion of Sec.~\ref{transverse_decomposition} still holds in this case, but the Hamilton-Jacobi Eq.~\eqref{hamilton_jacobi} arising from the weak-noise limit instead reads
\begin{equation}
    \nabla U \cdot [\bm{Q} \nabla U + \bm{f}] = 0
    \label{hamilton_jacobi_multi}
\end{equation}
in which the matrix $\bm{Q} \equiv \bm{g} \bm{g}^T$\footnote{Note that additional care has to be taken when $Q$ is positive-semidefinite rather than positive-definite \citet{Graham1985,Graham1986}.}, so the transverse decomposition in general reads
\begin{equation}
    \bm{f}(\bm{x}) = - \bm{Q}(\bm{x}) \nabla U^{\text{HJ}}(\bm{x}) + \bm{f}^{\text{HJ}}_{\text{nc}}(\bm{x})
    \label{transverse_decomposition_multiplicative}
\end{equation}
where $\bm{f}^{\text{HJ}}_{\text{nc}} = \bm{f} + \bm{Q} \nabla U^{\text{HJ}}$, and Eq.~\eqref{hamilton_jacobi_multi} again guarantees that $U^{\text{HJ}}$ is a Lyapunov function.
Here, the \enquote{gradient-like} part of the transverse decomposition \eqref{transverse_decomposition_multiplicative} is relaxational rather than variational.
This illustrates that there is some arbitrariness in the decomposition of the deterministic dynamics (lifted by the presence of noise), as $\bm{g}$ and $\bm{Q}$ are absent from the deterministic part.

\paragraph{Bracket-based approaches.}

Geometric approaches based on the combination of Poisson-like and dissipative brakets have been developed, in particular in the context of thermodynamic and kinetic theory, but they also apply to finite-dimensional systems.
These approaches, called metriplectic dynamics \cite{Morrison1986,Morrison1984,Morrison1998} or GENERIC \cite{Ottinger1997,Grmela1997}, can be seen as a combination of metric and symplectic flows.
The dynamics is expressed in the form
\begin{equation}
    \frac{d \bm{x}}{dt} = \{\bm{x},F\} + (x, F) = \{\bm{x},H\} + (x, S)
\end{equation}
in which $\{\bm{x},H\}$ is the Poisson bracket on $\bm{x}$ with a Hamiltonian function $H$ while $(\bm{x}, S)$ is the dissipative bracket of $\bm{x}$ with a generalized entropy function $S$, and $F = H + S$.
In coordinates, 
\begin{equation}
    \frac{d x_i}{dt} = J_{ij} \frac{\partial F}{\partial x_j} + g_{ij} \frac{\partial F}{\partial x_j}
\end{equation}
where $J_{ij}$ is a skew-symmetric Poisson bivector while $g_{ij}$ is a symmetric metric.
This structure guarantees that $dH/dt = 0$ and that $dS/dt \geq 0$.
Note however that not all dynamics can be cast in this form \cite{Pavelka2014,Kraaij2017}.

\subsubsection{Symmetries and nonvariational dynamics}
\label{symmetry_dynamics}

Symmetries play a key role in shaping the behavior of physical systems, and nonreciprocal systems are no exception.
In classical mechanics or equilibrium phase transitions, symmetries are usually phrased in terms of potentials -- think, for instance, of the way one constructs a Landau-Ginzburg free energy.
This formulation is not directly applicable to nonvariational dynamics.

Symmetries of a dynamical system such as Eq.~\eqref{ds} are encoded in the vector field $\bm{f}$. This vector field is said to be equivariant with respect to a symmetry group $G$ (acting linearly on $\RR^N$) or $G$-equivariant when
\begin{equation}
    \bm{f}(g \bm{x}) = g \bm{f}(\bm{x})
\end{equation}
for every $g \in G$ and $\bm{x} \in \RR^N$ (this can be generalized to other spaces), implying that $\gamma^{-1}J(\gamma \bm{x}) \gamma = J(\bm{x})$ where $J = D_x f$ is the Jacobian.

A theory of equivariant dynamical systems focused on bifurcations has been developed in the context of hydrodynamic instabilities and is reviewed by \citet{Chossat2000}, see also  \citet{Golubitsky2002,Golubitsky1985b,Golubitsky1988,Crawford1991} for more details and \cite{Marsden2013} for Hamiltonian systems. 
It formalizes the concepts of spontaneous symmetry breaking and forced (explicit) symmetry breaking and provides technical tools including 
(i) results known as equivariant branching lemmas that classify the possible patterns of spontaneous symmetry breaking, e.g. dynamic states like limit cycles or tori (ii) techniques to systematically construct the most general $G$-equivariant vector field, generalizing Landau-like constructions of symmetric potentials or (iii) model reduction approaches like normal forms that take the symmetry into account.

Another approach to symmetries in ODEs and PDEs is reviewed by \cite{Bluman2002,Bluman2010,Cantwell2002,Olver1993}, which originate from Lie theory of differential equations.
In particular, generalization of Noether’s theorem to nonvariational ODEs and PDEs provide a relation between certain classes of generalized symmetries and conservation laws \cite{KosmannSchwarzbach2011,Anco2017}.

\subsection{Violations of Newton's third law}
\label{violations_newton_third_law_general}

\noindent\emph{\textbf{In a nutshell:} mechanical systems where linear momentum is not conserved can be seen as non-reciprocal.}
\medskip

In classical mechanics, the evolution of the positions $\bm{r}_n$ of a collection of particles interacting pairwise is given by
\begin{equation}
    \frac{d \bm{p}_i}{d t} = \bm{F}^{\text{ext}}_i + \sum_{j} \bm{F}_{j \to i}
\end{equation}
in which $\bm{p}_i \equiv m_i \bm{\dot{r}}_i$ is the linear momentum of particle $i$,  $\bm{F}^{\text{ext}}_i$ is an external force imposed on the particle (that we set to zero in the following), and $\bm{F}_{j \to i}$ is the force that particle $j$ exerts on $i$. 
In this context, Newton's third law states that
\begin{equation}
    \bm{F}_{j \to i} = - \bm{F}_{i \to j}.
    \label{newton_third}
\end{equation}
This relation between action and reaction is also known as the law of reciprocity and is a manifestation of the conservation of linear momentum. 
Indeed, the total momentum $\vec{p}$ of the system evolves according to
\begin{equation}
    \vec{\dot{p}} = \sum_{i,j} \bm{F}_{i \to j}
\end{equation}
which is only guaranteed to vanish when Eq.~\eqref{newton_third} is satisfied.

When considering interactions mediated by an environment, it turns out that \emph{effective} interactions that violate Newton's third law can appear; in fact, they are the rule rather than the exception. 
In this case, one typically has
\begin{equation}
    \bm{F}_{j \to i} \neq - \bm{F}_{i \to j}.
    \label{newton_third_violated}
\end{equation}
The effective lack of linear momentum conservation (that is, the fact that the momentum of the particles is not conserved) is possible because some momentum has been exchanged with the field mediating the interaction.
The overall conservation of momentum always holds when the momentum of the field itself is taken into account; in practice, doing so requires some care and we refer to \citet{Pfeifer2007,Bliokh2025,Stone2002,Maugin2015,Maugin1993} for details and references.
Note that the mere existence of a fluctuating field is not enough: the field has to be driven out of equilibrium to lead to average effective forces that violate Newton's third law \cite{Dzubiella2003,Buenzli2008}.

\subsubsection{Nonreciprocal interactions versus external forces}

Interactions that violate Newton's third law are different from an external force because they depend on the internal state of the system. 
This can be seen on the example of two particles interacting through a nonreciprocal spring, following the equation of motion
\begin{subequations}
\label{example_nr_newton}
\begin{align}
    m \ddot{x}_{1} + \gamma \dot{x}_1 &= k_{12}(x_2 - x_1) \\
    m \ddot{x}_{2} + \gamma \dot{x}_2 &= k_{21}(x_1 - x_2)
\end{align}
\end{subequations}
where the spring constants can be decomposed into $k_{\pm} = (k_{12} \pm k_{21})/2$, where $k_{+}$ represents the usual reciprocal spring constant, while $k_{-}$ is a Newton-third-law-violating spring constant.
We find that the position $X = (x_1 + x_2)/2$ of the center of mass and the distance $\Delta x = x_2 - x_1$ between the particles evolve as
\begin{subequations}
\label{example_nr_newton_com}
\begin{align}
    m \ddot{X} + \gamma \dot{X} &= - k_{-} \Delta x \\
    m \Delta \ddot{x} + \gamma \Delta\dot{x} &= 2 k_{+} \Delta x
\end{align}
\end{subequations}
where, crucially, the motion of the center of mass depends on the distance between the particles. 
This coupling between the internal structure and the overall motion of the system is at the core of many of the noteworthy features of nonreciprocal mechanics.

As an aside, we note that having different masses $m_1$ and $m_2$ for the two particles described by Eq.~\eqref{example_nr_newton} would not violate Newton's third law (linear momentum is indeed conserved in this case).

\subsubsection{Beyond pairwise forces}
\label{beyond_pairwise}

Interactions are not all pairwise additive.
This happens in systems ranging from plasma and colloids~\cite{Ivlev2012} and optical matte \cite{Parker2025} to acoustically levitated grains \cite{Lim2022} and interatomic potentials in materials~\cite{Tadmor2011}. 
Then, Eq.~\eqref{newton_third} does not apply as is, and it is easier to directly check whether the interactions conserve linear momentum.
In this case, we have
\begin{equation}
    \frac{d \bm{p}_i}{d t} = \bm{F}^{\text{ext}}_i +\bm{F}_{*\to i}
\end{equation}
in which $\bm{F}_{*\to i}$ is the total force acting on the particle $i$ from all other particles, and linear momentum conservation can be stated as
\begin{equation}
    \label{conservation_momentum_mb}
    \sum_i \bm{F}_{*\to i} = 0
\end{equation}

The link with nonreciprocity can be understood when forces derive from a many-body potential $U(r_1, r_2, \dots, r_N)$. 
In systems with translation invariance, $U$ only depends on the relative positions $r_{ij} = r_i - r_j$ \footnote{If this is not the case (e.g. when an external force is present), the system is not translation invariant and we do not expect linear momentum to be conserved.}.
The force on particle $i$ is then
\begin{equation}
    F_{*\to i}
    = - \frac{\partial U}{\partial r_i}
    = - \sum_{k < \ell} \frac{\partial U}{\partial r_{k \ell}} \frac{\partial r_{k \ell}}{\partial r_i}
= - \sum_{j \neq i} F^{[ij]}
\end{equation}
in which
\begin{equation}
    F^{[ij]}
    \equiv
    \begin{cases}
    \displaystyle\frac{\partial U}{\partial r_{i j}} & \text{if $i < j$}
    \\[1em]
    -\displaystyle\frac{\partial U}{\partial r_{j i}}& \text{if $i > j$}
    \end{cases}
\end{equation}
satisfies $F^{[ij]} = - F^{[ji]}$ \footnote{In general, the quantities $F^{[ij]}$ do not correspond to pairwise interactions because they depend on the positions of all particles.}.
This guarantees that $\sum_i f_i = 0$ and hence that the total linear momentum is conserved.
We indeed recover Eq.~\eqref{newton_third} in the case of a pairwise additive potential.
When the interaction does not derive from a potential, then \eqref{conservation_momentum_mb} can be violated: this is the generalization of the violation of Newton's third law to nonpairwise potentials.

The calculation above shows that a translation invariant conservative system must conserve momentum. 
Hence a translation invariant system that does not conserve momentum is not conservative, and must break detailed balance (Sec.~\ref{broken_detailed_balance}).
We make this connection more precise in Sec.~\ref{broken_detailed_balance_newton_third_law}.

\subsection{Broken detailed balance}
\label{broken_detailed_balance}

\noindent\emph{\textbf{In a nutshell:} stochastic systems with broken detailed balance can be seen as non-reciprocal.}
\medskip

\def\WW{\mathbb{W}}

Detailed balance is a manifestation of time-reversal invariance on stochastic processes.
In the steady-state of a stochastic process satisfying detailed balance, the chance of observing a trajectory must, by definition, be equal to the chance of observing the time-reversed trajectory.

Mathematically, the continuous-time evolution of the probability distribution $p(t)$ of a Markovian (aka memoryless) system can be described by a master equation of the form
\begin{equation}
    \frac{d}{dt} p(t) = \hat{W} p(t)
    \label{master_equation}
\end{equation}
in which $\hat{W}$ is a linear operator acting on the space of probability vectors $p$.
Equation \eqref{master_equation} encompasses the Fokker-Planck equation for continuous state spaces, Markov chains for discrete state spaces, and quantum master equation if $p(t)$ is understood as a density matrix.
The action of time-reversal is represented by an anti-linear operator $\hat{\Theta}$ on the space of probability vectors\footnote{In the case of classical systems, both $p(t)$ and $\hat{W}$ are real-valued, so there is no practical difference between linear and anti-linear, provided that we ensure all quantities are real.}.

In all of these cases, detailed balance manifests itself as (i) the requirement that Eq.~\eqref{master_equation} admits a time-reversal invariant steady-state $p_{\text{ss}}$ [so $\hat{W} p_{\text{ss}} = 0$ and $\Theta p_{\text{ss}} = p_{\text{ss}}$] and (ii) a constraint \citet{Kurchan1998}
\begin{equation}
    \label{db_op_intro}
    \hat{{W}}^\dagger
    = 
    \mathcal{Q}^{-1}
    \hat{{W}} \mathcal{Q}
    \qquad
    \text{with}
    \qquad
    \mathcal{Q} \equiv \Theta \hat{p}_{\text{ss}}
\end{equation}
on the infinitesimal generator of the dynamics $\hat{W}$, where 
($\hat{p}_{\text{ss}}$ is the operator multiplication by ${p}_{\text{ss}}$)\footnote{In Eq.~\eqref{db_op_intro}, the adjoint is defined with respect to the standard inner product.}.
Note that condition (i) implies that $\mathcal{Q} \equiv \Theta \hat{p}_{\text{ss}} = \hat{p}_{\text{ss}} \Theta$.
Hence, detailed balance implies that the infinitesimal generator of the dynamics $\hat{W}$ is reciprocal in the sense of linear operators (compare with Eq.~\eqref{reciprocal_linear_operator} in Sec.~\ref{reciprocity_linear_system}).

In certain cases, we can make a connection between non-variational dynamics (Sec.~\ref{nonvariational_dynamics}) and broken detailed balance.
To see that, let us add a random noise to Eq.~\eqref{ds} to obtain the set of coupled Langevin equations
\begin{equation}
    \frac{d x_i}{d t} = f_i(\vec{x}) + \sqrt{2 D_i} \, \eta_i(t)
    \label{coupled_langevin}
\end{equation}
in which $\eta_i(t)$ are uncorrelated standard Gaussian white noises with unit standard deviation (see Sec.~\ref{sde_fp} for details), and assume that all degrees of freedom are even under time-reversal.
Then detailed balance implies that for all $i$ and $j$, one has
\begin{equation}
  D_j \partial_j f_i = D_i \partial_i f_j
\end{equation}
which reduces to Eq.~\eqref{nonsymmetric_jacobian} when all $D_i$'s are equal.

\subsubsection{Thermodynamics is a harsh mistress}
\label{harsh}

In the steady-state of a stochastic process satisfying detailed balance, the chance of observing a trajectory must, by definition, be equal to the chance of observing the time-reversed trajectory.
Formally, we may ask that
\begin{equation}
\begin{split}
    \mathbb{P}[ 
    X(t_1) = x_1
    \text{ and $\cdots$ and }
    X(t_N) = x_N
    ]\\
    =
    \mathbb{P}[ 
    X(-t_N) = \Theta x_N
    \text{ and $\cdots$ and }
    X(-t_1) = \Theta x_1
    ]
\end{split}
\end{equation}
in which the $t_i$ are consecutive times, and where we have introduced the time-reversal operator $\Theta$ on phase phase, which must satisfy $\Theta^2 = 1$.
As $\Theta^2 = 1$, it is possible to choose a coordinate system in phase space so that variables are either odd or even under time-reversal, namely $\Theta x_{i} = \varepsilon_i x_i$ with $\varepsilon_i = \pm 1$.
For time-translation invariant Markovian processes, it would be enough to require that
\begin{subequations}
\begin{equation}
    \mathbb{P}[ X(t) = x ] = \mathbb{P}[ X(t) = \Theta x ]
\end{equation}
which means that the steady-state distribution is symmetric under time-reversal, and that
\begin{equation}
\begin{split}
    \mathbb{P}[ X(0) = x \text{ and } X(t) = y ] \\
    =
    \mathbb{P}[ X(0) = \Theta y \text{ and } X(t) = \Theta x ]
\end{split}
\end{equation}
\label{db_proba}
\end{subequations}
which means that the probability of observing $x$ at time zero and then $y$ at time $t$ is the same as the probability of observing $\Theta y$ at time zero and then $\Theta x$.

While intuitive, this statement is not so clear in practice, in particular because it requires specifying what \enquote{time-reversal} actually means.
Doing so requires physical modeling going beyond mathematical model itself, and so it can easily be overlooked.
One key issue is that time-reversals of stochastic differential equations are not unique \cite{Chetrite2008,Landi2021}, as one has to choose a way of associating a backward trajectory to a given forward trajectory.
This has recently been discussed in detail by \citet{OByrne2024,OByrne2025}.
In particular, one has to specify the action of time-reversal $\Theta$ on phase space: for instance, positions are left invariant $\Theta \bm{r} = \bm{r}$ while velocities change sign, $\Theta \bm{v} = -\bm{v}$).
For instance, the same SDE can have completely different interpretations (namely, describe an equilibrium or a nonequilibrium system) depending on whether the variables are considered to be odd or even under time-reversal \cite{Lucente2025}.
This is not an issue if we start from a (classical or quantum) fully microscopic description, for which the effect of time-reversal is known, but doing so is not always practical\footnote{
\citet{Gaspard2022} reserves the use of the term \enquote{detailed balance} for the full microscopic description, and uses \enquote{reversibility} in partially coarse-grained descriptions; this is a useful distinction, but we will keep with the standard usage of conflating these terms.
}. 
The question of the choice of the backward protocol is also related to the connection between stochastic dynamics and stochastic thermodynamics, which require an extra layer of interpretation specifying how the coupling to the environment is implemented to be meaningful.
If the environment is not explicitly and fully modeled (which is often a daunting task), extra physical assumptions, such as \enquote{local detailed balance}, are required to make a connection between dynamics and thermodynamics
\cite{Gaspard2004,Sekimoto2010,Seifert2010,VanDenBroeck2015,Horowitz2019,Maes2021,Fodor2022,Pachter2024}.

In the context of nonreciprocal systems, measures of irreversibility such as the (informatic) rate of entropy production have been considered \cite{Loos2019,Loos2020,Suchanek2023a,Suchanek2023b,Suchanek2023c,Zhang2023,OByrne2024,OByrne2025} and indeed suggest that non-reciprocity typically leads to irreversibility.

\subsubsection{Markov chains with discrete states}

We first consider the case of a countable set $\mathscr{S}$ of discrete states $x_i$ indexed by integers $i$. 
When the system is Markovian (without memory), the evolution in time of the probability $p_i(t) \equiv p(t,i)$ of state $x_i$ (in short, state $i$) at time $t$ can be described by the equation
\begin{equation}
    \frac{d}{dt} p_i(t)
    = W_{i j} \, p_j(t)
    \equiv 
    \sum_j [\Gamma_{i|j} p_j(t) - \Gamma_{j|i} p_i(t)]
\end{equation}
in which $\Gamma_{ij} \equiv \Gamma_{i|j} \equiv \Gamma_{j \to i}$ is the probability rate of jumping from state $j$ to state $i$.
This equation can be represented in matrix form as $(d/dt) p(t) = \hat{W} p(t)$.
(When the system is not time-translation invariant, $\hat{W}$ may also depend on time, but we do not consider this case here.)
In this context, time reversal acts on the space of states as a map $\Theta : \mathscr{S} \to \mathscr{S}$ with $\Theta^2 = \text{id}$.
Given an integer $i$, we use the symbol $\Theta i$ to represent the integer such that $x_{\Theta i} = \Theta(x_i)$.
We can then define a time-reversal operator $\hat{\Theta}$ acting on probability vectors $p$ as $[\hat{\Theta} p]_i = p_{\Theta i}$.
A stationary distribution $p^{\text{ss}}$ is such that $\hat{W} p^{\text{ss}} = 0$. 

In this context, the detailed balance condition reads \cite{Gardiner2004}
\begin{equation}
    \Gamma_{i|j} \, p^{\text{ss}}_j = \Gamma_{\Theta j | \Theta i} \, p^{\text{ss}}_{\Theta i}.
\end{equation}
By introducing the operator $\hat{p}_{\text{ss}}$ of multiplication by $p^{\text{ss}}$, such that $(\hat{p}_{\text{ss}} v)_i = p^{\text{ss}}_i v_i$, detailed balance can be written in the more compact form
\begin{equation}
    \hat{\Theta} \hat{p}_{\text{ss}} \hat{W}^\dagger \hat{\Theta}^{-1}
    = 
    \hat{W}\hat{p}_{\text{ss}} 
    \label{detailed_balance_mc}
\end{equation}
in which $\hat{W}^\dagger=\hat{W}^T$ as $\hat{W}$ is real-valued.

\paragraph{Example.}
As an example, consider the three-state Markov chain described by the transition rate matrix
\begin{equation}
    W = \gamma \begin{pmatrix}
        -1 &  1 &  0 \\
         0 & -1 &  1 \\
         1 &  0 & -1 
    \end{pmatrix}
\end{equation}
which admits the steady-state $p^{\text{ss}}=(1,1,1)^T/3$,
and consider the two time-reversal operators
\begin{equation}
    \Theta_1 = \text{Id}
    \quad
    \text{and}
    \quad
    \Theta_2 = 
\begin{pmatrix}
 0 & 1 & 0 \\
 1 & 0 & 0 \\
 0 & 0 & 1 
 \end{pmatrix}.
\end{equation}
One can check explicitly that the detailed balance condition Eq.~\eqref{detailed_balance_mc} holds with $\Theta = \Theta_2$, but not with $\Theta = \Theta_1$.

\subsubsection{Stochastic differential equations (SDEs)}
\label{sde_fp}

Consider a set of coupled Itô stochastic differential equations
\begin{equation}
    d X_i = f_i(\vec{X}) dt + \sigma_{i j}(\vec{X}) dW_j
    \label{ito_sde}
\end{equation}
for the random variable $\vec{X}(t) \in \mathbb{R}^d$, in which $W(t) \in \mathbb{R}^d$ is the $d$-dimensional Wiener process.
We refer to \citet{Risken1989,Gardiner2004,Pavliotis2014} for details about stochastic differential equations.
Equation \eqref{ito_sde} may be written in the more familiar form 
$\dot{x}_i = f_i(\vec{x}) + \sigma_{i j} \eta_j(t)$
in which $\eta_i(t) = dW_i/dt$ are uncorrelated Gaussian white noises with unit standard deviation (so $\langle \eta_i(t) \rangle = 0$ and $\langle \eta_i(t) \eta_j(t') \rangle = \delta_{i j} \delta(t-t')$).

The evolution of the probability distribution $p(t, \vec{x})$ of the random variables $\vec{X}(t)$ ruled by Eq.~\eqref{ito_sde} is described by the Fokker-Planck equation~\cite{vanKampen2007,Risken1989,Gardiner2004}
\begin{subequations}
\begin{equation}
\partial_t p = \hat{\WW} p \equiv - \partial_i J_i
\end{equation}
where
\begin{equation}
  J_i(t,\vec{x}) \equiv f_i(\vec{x}) p(t,\vec{x}) - \partial_j [  D_{i j}(\vec{x}) p(t,\vec{x}) ]
\end{equation}
\label{fokker_planck}
\end{subequations}
for an arbitrary distribution $p$, where $\partial_i = \partial/\partial x_i$ and $\bm{D} = \bm{\sigma} \bm{\sigma}^T/2$.
Here, we have used different symbols $\bm{X}$ and $\bm{x}$ to distinguish a point in phase space and a random process, respectively, but we shall conflate these notations from now on.
The stochastic process defined by $\hat{\WW}$ is called stationary if there is a stationary distribution $p_{\text{ss}}(\vec{x})$ such that $\hat{\WW} p_{\text{ss}} = 0$.

The stochastic process defined by $\hat{\mathbb{W}}$ is said to have detailed balance with respect to the stationary distribution $p_{\text{ss}}$ if $\Theta p_{\text{ss}} = p_{\text{ss}}$ and
$\hat{\Theta} \hat{p}_{\text{ss}} \hat{\mathbb{W}}^\dagger \hat{\Theta}^{-1}  = \hat{\mathbb{W}} \hat{p}_{\text{ss}}$
in which $\hat{p}_{\text{ss}}$ is the operator that multiplies a function on phase space with $p_{\text{ss}}$.
Equivalently, detailed balance holds when\footnote{As $p_{\text{ss}}$ is a probability distribution, the operator $\hat{p}_{\text{ss}}$ is positive-semidefinite and has a unique positive-semidefinite square root $\hat{p}_{\text{ss}}^{1/2}$.
It is not necessarily invertible, because there may be points $\bm{x}$ in phase space such that $p_{\text{ss}}(\bm{x}) = 0$, but if it is the case they are irrelevant in the sense that their probability of being observed in strictly zero, so they can be removed from phase space. We will assume that it is the case, so $\hat{p}_{\text{ss}}$ is positive-definite.
\label{pss_positive_definite}
}
\begin{subequations}
\label{db}
\begin{equation}
    \hat{\Theta} \hat{p}_{\text{ss}} \hat{\Theta}^{-1} = \hat{p}_{\text{ss}}
    \quad
    \text{and}
    \quad
    \hat{L} = \hat{\Theta} \hat{L}^\dagger \hat{\Theta}^{-1}
\end{equation}
in which $\dagger$ is the adjoint with respect to the $L^2$ inner product, and
\begin{equation}
    \hat{L} \equiv \hat{p}_{\text{ss}}^{-1/2} \, \hat{\mathbb{W}} \, \hat{p}_{\text{ss}}^{1/2}
    \label{db_op}
\end{equation}
\end{subequations}
In case where all variables are even with respect to time-reversal ($\Theta = \text{Id}$), detailed balance implies that $\mathbb{W}$ is $\hat{p}_{\text{ss}}$-pseudo-Hermitian (with a positive-definite $\hat{p}_{\text{ss}}$, see footnote \ref{pss_positive_definite}).

When $\hat{p}_{\text{ss}}$ commutes with $\Theta$, we define $\mathcal{Q} = \Theta \hat{p}_{\text{ss}} = \hat{p}_{\text{ss}}  \Theta$ so the second part of detailed balance 
\begin{equation}
    \hat{\Theta} \hat{p}_{\text{ss}} \hat{\mathbb{W}}^\dagger
    = 
    \hat{\mathbb{W}} \hat{p}_{\text{ss}} \Theta
\end{equation}
becomes \cite{Kurchan1998}
\begin{equation}
    \hat{\mathbb{W}}^\dagger
    = 
    \mathcal{Q}^{-1}
    \hat{\mathbb{W}} \mathcal{Q}
\end{equation}

When the operator $\mathcal{Q}$ is of the form $\mathcal{Q} = \mathcal{O}  \mathcal{O}^\dagger$ with $\mathcal{O} \in \text{GL}$, then we have proven that $\hat{\mathbb{W}}$ is $\mathcal{O}  \mathcal{O}^\dagger$-pseudo-Hermitian.
This means that one can choose an inner product with respect to which $\hat{\mathbb{W}}$ is Hermitian.
In other words, $\mathbb{W}$ has an exact generalized PT-symmetry (Sec.~\ref{PT_symmetry}), that is \enquote{spontaneously broken} when detailed balance does not hold.

In terms of the function $f$ and $D$ in the Fokker-Planck equation~\eqref{fokker_planck}, detailed balance is encoded in the conditions \cite[\S~6.3.5 p. 145, 4th ed]{Gardiner2004}
\begin{subequations}\label{detailed_balance}
\begin{equation}
    J^{\text{irr}} \equiv \frac{1}{2}
    [\Theta \bm{f}(\Theta \bm{x}) + \bm{f}(\bm{x})] \, p_{\text{ss}}(\bm{x})
    - 
    \nabla \cdot [ \bm{D}(\bm{x}) p_{\text{ss}}(\bm{x}) ] = 0
\end{equation}
and
\begin{equation}
    \bm{D}(\bm{x}) = \Theta \bm{D}(\Theta \bm{x}) \Theta^{-1}
\end{equation}
\end{subequations}
\noindent in which $\nabla \cdot \bm{A} \equiv \partial_j A_{ij}$\footnote{
In components, and assuming that the coordinates have a well-defined TR parity so $\Theta x_i = \varepsilon_i x_i$ these equations read
$J^{\text{irr}}_i \equiv 1/2 [\varepsilon_{i} f_{i}(\Theta \bm{x}) - f_i(\bm{x})] p_{\text{ss}}(\bm{x})
    -\sum_j \partial_j [ D_{ij}(\bm{x}) p_{\text{ss}}(\bm{x}) ]$ and 
$D_{ij}(\bm{x}) = \epsilon_{i} \epsilon_{j} D_{ij}(\Theta \bm{x})$.
}.
The quantity $J^{\text{irr}}$ defined in Eq.~\eqref{detailed_balance} can be seen as the irreversible part of the probability current 
\begin{equation}
    \label{J_irr_rev}
    J = J^{\text{irr}} + J^{\text{rev}}
\end{equation}
while $J^{\text{rev}} = (1/2) [ \bm{f}(\bm{x}) - \Theta \bm{f}(\Theta \bm{x}) ] p(\bm{x})$ is the reversible part.

\paragraph{Relation to non-variational dynamics.}
Let us now assume that all variables are even ($\Theta \bm{x} = \bm{x}$) and that the noise is additive and diagonal, so $D_{ij}(\bm{x}) = D_i \delta_{ij}$, where $D_i$ does not depend on the state $\bm{x}$.
In this case, Eq.~\eqref{detailed_balance} reduces to $f_i p = D_i \partial_i p$, or equivalently $f_i = D_i \partial_i \log p_{\text{ss}}$.
Taking the derivative $\partial_j$ of both sides and reshuffling, we find that \cite{Lan2012}
\begin{equation}
    D_j \partial_j f_i = D_i \partial_i f_j
\end{equation}
is a necessary (but not sufficient) condition for detailed balance.
When all the systems are driven at the same temperature, $D_j = D_i = D$ and having broken detailed balance is equivalent to \eqref{nonsymmetric_jacobian} for the deterministic dynamical system.
In the same way, a system with a symmetric Jacobian but connected to multiple baths at different temperatures is equivalent to a nonreciprocal system. 
Conversely, it is possible to compensate for deterministic nonreciprocity by using baths at different temperatures, but this is only the case when $\partial_j f_i$ and $\partial_i f_j$ have the same sign (else, one would need negative diffusion coefficients).
This mapping between nonreciprocity and multiple baths with different temperatures is analyzed in detailed in \citet{Ivlev2015}, see also \citet{Benois2023,Loos2020}.

\paragraph{Kramers equations and Newton's third law.}
\label{broken_detailed_balance_newton_third_law}

In this paragraph, we analyze the relation between violations of Newton's third law and broken detailed balance.
To do so, we consider inertial particles (not in the overdamped limit) in thermal equilibrium with a bath at temperature $T$.
(Stochastic differential equations describing an underdamped Brownian motion are known as Kramers equations.)
For simplicity, we focus on pairwise interactions, so it is enough to consider two particles with positions and velocities $(x_1, v_1)$ and $(x_2, v_2)$, respectively. 
We assume that these interact through arbitrary forces $F_{1 \to 2}(x_1, x_2)$ and $F_{2 \to 1}(x_1, x_2)$. 
Hence, we have
\begin{align}
    m_1 \ddot{x}_1 + \gamma_1 v_1 
   & = 
    F_{2 \to 1}(x_1, x_2)
    + \sqrt{2 \gamma_1 k_B T} \xi_1(t)
    \\
    m_2 \ddot{x}_2 + \gamma_2 v_1 
    &= 
    F_{2 \to 1}(x_1, x_2)
    + \sqrt{2 \gamma_2 k_B T} \xi_2(t)
    \label{eom_newton_third_law_rel_db}
\end{align}
in which $m_i$ are the masses of the particles, $\gamma_i$ the friction coefficients, $T$ is the bath temperature and
the noises $\xi_1$ and $\xi_2$ are Gaussian white normalized uncorrelated.
Please note that, in writing this equation, we have made the strong assumption that the presence of possibly non-reciprocal forces does not modify the coupling of the particles with their environment. 
This assumption is implicit in the fact that we have assumed that the friction coefficients $\gamma_i$ and the noise strengths $\sqrt{2 \gamma_2 k_B T}$ would satisfy a fluctuation-dissipation relation in the absence of the interactions between the particles.
This is not guaranteed, and whether this is the case depends on the system.
The point of this calculation is to show that, \emph{under this assumption}, interactions that violate Newton's third law lead to a broken detailed balance.
To see that, one can cast Eqs.~\eqref{eom_newton_third_law_rel_db} as first-order equations in the form of Eq.~\eqref{ito_sde}, and assess whether the detailed balance condition \eqref{detailed_balance} is satisfied, including that $p_\text{ss}$ must be a solution of the stationary Fokker-Planck equation, under the assumption that time-reversal acts as $\Theta x_i = x_i$ and $\Theta v_i = - v_i$. 
We end up with
\begin{equation}
    \!\!\!\!
    F_{1 \to 2} = k_B T \partial_{x_2} \log p_{\text{ss}}^0
    \;\;
    \text{and}
    \;\;
    F_{2 \to 1} = k_B T \partial_{x_1} \log p_{\text{ss}}^0
\end{equation}
which means that the force field must be conservative. 
From this, we can also find the necessary (but not sufficient) condition $\partial_{x_1} F_{1 \to 2}
=
\partial_{x_2} F_{2 \to 1}$.

\paragraph{From extended variational representations to inference.}

Nonvariational dynamics and systems violating detailed balance can be represented through extended variational principles, at the price of introducing auxiliary variables. 
These representations are the go-to tool to compute the statistical properties (and in some cases perform simulations) of nonequilibrium systems.
One of the most common of such representations, known as the Martin–Siggia–Rose-Janssen–De Dominicis–Doi-Peliti path integral formalisms which can broadly be interpreted as classical limits of the Schwinger-Keldysh-Kadanoff-Baym (Keldysh) formalism. These approaches are reviewed in \citet{Kamenev2023}, see \citet{Andreanov2006,Hertz2016} for relations and differences between Doi-Peliti and Martin–Siggia–Rose formalisms.
\citet{Shi2025} explored a conceptually related idea where nonreciprocal interactions are represented through a constrained Hamiltonian embedding by a Hamiltonian system equipped with constraints.

In addition, extensions of the inference techniques developed for equilibrium and variational systems have been considered, with the goal of inferring parameters or distributions from data. 
General methods such as maximum caliber \cite{Presse2013,Dixit2018,Ghosh2020} (a generalization to trajectories of the maximum entropy principle) are not always effective with limited data, and several lines of research are ongoing in order to perform this kind of nonequilibrium inference efficiently \cite{Ferretti2020,Chen2023,Schmitt2023,Hempel2024,Colen2024,Yu2025}.

\subsubsection{Open quantum systems}
\label{open_quantum_systems}

In this section, we review the notion of \emph{quantum detailed balance}, which mirrors the classical version, with a focus on the case of Markovian open quantum systems (see \citet{Sieberer2015,Altland2021} for discussions about the general case of non-Markovian dynamics).
In this case, the evolution of the density matrix $\rho$ is governed by a Lindblad master equation
\begin{equation}
    \frac{d \rho}{dt} = \mathcal{L} \rho = - \ii [H,\rho] + \sum_i (2 L_i \rho L_i^\dagger - \{L_i^\dagger L_i, \rho\})
    \label{lindblad}
\end{equation}
in which $H$ is the Hamiltonian and $L_i$ are jump operators representing the coupling to the environment.
The object $\mathcal{L}$ is the Lindbladian superoperator, and it is called this way because it acts on operators like $\rho$.
In this context, detailed balance is encoded in the conditions \cite{Agarwal1973,Majewski1984,Chetrite2012}
\begin{equation}
    \Theta \rho_{\text{ss}} = \rho_{\text{ss}}
    \qquad
    \text{and}
    \qquad
    \mathcal{L}^\dagger = \mathcal{Q}^{-1} \Theta^{-1} \mathcal{L} \Theta \mathcal{Q}
    \label{quantum_db}
\end{equation}
in which $\mathcal{Q}$ is a superoperator acting as multiplication by the steady-state density matrix $\rho_{\text{ss}}$, which at equilibrium reads $\rho_{\text{ss}} = e^{-\beta H}/Z$ ($\beta$ inverse temperature, $Z$ partition function), and $\Theta$ is a superoperator representing time-reversal.
This condition can be compared with the classical version \eqref{db_op} for a Fokker-Planck equation.
One can show that the quantum detailed balance constraint \eqref{quantum_db} holds when an appropriate implementation of time-reversal invariance is imposed on the system plus bath (see \citet{Lieu2022} and references therein).

Contrary to the classical Fokker-Planck equation \eqref{fokker_planck}, the master equation \eqref{lindblad} is not directly written in terms of a current. 
However, a closer link can be established by considering quasiprobability distribution in phase-space \cite{Gardiner2004b,Hillery1984}.
For instance, the time evolution of the Wigner function $W(t, \bm{p}, \bm{x})$ \cite{Wigner1932} can be cast in the form of a continuity equation \cite{Donoso2001,Bauke2011,Steuernagel2013,Braasch2019}
\begin{equation}
    \partial_t W + \nabla \cdot \bm{J}_W = 0
\end{equation}
in which $\bm{J}_W$ is a probability current in phase space.

Note also that in the same way as in classical systems, time-reversal is not uniquely defined \cite{Aurell2015,Roberts2021}.

\subsection{Nonreciprocal responses}
\label{nonreciprocal_responses}

\noindent\emph{\textbf{In a nutshell:} the antisymmetric part of response functions can (sometimes) be seen as nonreciprocal.}
\medskip

In a linear system, the response $\bm{X}$ to an input $\bm{Y}$ can be expressed as 
\begin{equation}
    X_i = R_{i j} Y_j
    \label{linear_response_simple}
\end{equation}
in which $R$ is a matrix of linear response coefficients.
In this context, reciprocity usually refers to the symmetry constraint
\begin{equation}
    R = R^T
    \quad
    \text{i.e.}
    \quad
    R_{ij} = R_{ji}.
\label{linear_response_simple_reciprocity}
\end{equation}
Conversely, reciprocity is usually said to be broken when $R$ is not a symmetric matrix, i.e. $R \neq R^T$.
A constraint on the response matrix $R$ such as Eq.~\eqref{linear_response_simple_reciprocity} can be seen (i) as the consequence of microscopic symmetries and (ii) as the cause of macroscopic behaviors:
\begin{equation}
\begin{gathered}
    \text{physical consequences}
    \\
    \Uparrow \text{(i)}
    \\
    \text{constraint on $R$}
    \\
    \Uparrow \text{(ii)}
    \\
    \text{microscopic symmetries}
\end{gathered}
    \label{nrr_scheme}
\end{equation}
The first arrow (i) in \eqref{nrr_scheme} is often referred to as Lorentz or Maxwell-Betti reciprocity theorem (see \citet{Masoud2019} for a historical discussion), and typically deals with the consequences of having a constraint on $R$ on the solutions of a PDE containing $R$ as a known parameter. As we shall see, such PDEs may include the Navier Stokes equations of fluid dynamics or the equations of elastodynamics with the role of the response function $R$ played by the viscosity and elasticity tensors respectively. The second arrow (ii) in \eqref{nrr_scheme} describes the consequences of microscopic symmetries (such as time-reversal invariance) on the response $R$ of a many-body system in a statistical steady-state. 
It is usually referred to as the Onsager reciprocity theorem.
In any case, the scheme \eqref{nrr_scheme} is not limited to time-reversal invariance or to the constraint \eqref{linear_response_simple_reciprocity}. 
In fact, one can and should derive the constraints on $R$ for every microscopic symmetry present in the system.

\begin{figure}
    \centering
    \includegraphics[width=\linewidth]{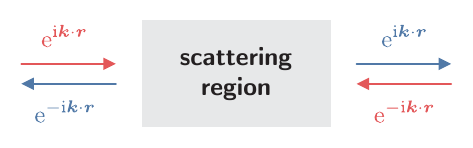}
    \caption{\textbf{Scattering.}
    We send \emph{incoming waves} (in red) onto the scattering region, and receive \emph{outgoing waves} (in blue).
    \label{fig_scattering_simple}
    }
\end{figure}

Section \ref{reciprocity_linear_system} provides a general but abstract definition of reciprocity for linear operators, given in Eq.~\eqref{reciprocal_linear_operator}. 
The direct consequences of this property [arrow (i) in \eqref{nrr_scheme}] are described in Sec.~\ref{reciprocal_theorem}.
Section \ref{scattering_reciprocity} then discusses reciprocity in the context of scattering problems, in which a physical justification for the abstract definition 
of Sec.~\ref{reciprocity_linear_system}
is provided.
Finally, Sec.~\ref{onsager_casimir} discusses the microscopic origin of reciprocity [arrow (ii) in \eqref{nrr_scheme}], and in particular the Onsager-Casimir reciprocity theorem.

\subsubsection{The reciprocal theorem and Green functions}
\label{reciprocal_theorem}

Consider now a linear system
\begin{equation}
    \mathscr{L} \phi = s
    \label{linear_system}
\end{equation}
in which $s$ represents a source and $\phi$ the field resulting from this source (for example, $s$ may represent a charge density and $\phi$ the potential field generated by it).
We assume that $\mathscr{L}$ is reciprocal (with respect to $\mathscr{A}$; see Eq.~\eqref{reciprocal_linear_operator}) and we consider two sources $s_1$ and $s_2$ that we assume to be invariant with respect to the reciprocity operator (namely, $\mathscr{A} s_1 = s_1$ and $\mathscr{A} s_2 = s_2$).
The resulting fields $\phi_1$ and $\phi_2$ satisfy Eq.~\eqref{linear_system}.
Then, we find that
\begin{equation}
\braket{s_1, \phi_2}
=
\braket{s_2, \phi_1}.
\end{equation}
This result is known as the reciprocal theorem.

This result can be rephrased in terms of Green functions.
We can promote the source and the field in the linear system \eqref{linear_system} to be matrices (or operators) instead of vectors and choose for the source the identity matrix $\Id$, leading to
\begin{equation}
    \mathscr{L} G = \Id.
    \label{linear_system_green}
\end{equation}
The solution $G = \mathscr{L}^{-1}$ of this equation is known as the Green function of $\mathscr{L}$.
In terms of the Green function, reciprocity (Eq.~\eqref{reciprocal_linear_operator}) implies that
\begin{equation}
    G^\dagger = \mathscr{A} G \mathscr{A}^{-1}
\end{equation}

When $\mathscr{L}$ is a differential operator, it is more common to write Eq.~\eqref{linear_system_green} in components as 
\begin{equation}
    \mathscr{L} G(\vec{r}, \vec{r'}) = \delta(\vec{r} - \vec{r'}).
\end{equation}
in which $\mathscr{L}$ acts on functions of $\vec{r}$ (not of $\vec{r'}$). 
Assuming that $\mathscr{A}^2 = 1$ and that the basis in which we represent the operators is invariant under $\mathscr{A}$, reciprocity then implies
\begin{equation}
    G(\vec{r_1}, \vec{r_2}) = G^T(\vec{r_2}, \vec{r_1}).
    \label{symmetry_green_function_physical_space}
\end{equation}
When the fields are scalar, the transpose can be dropped and we end up with the even simpler condition $G(\vec{r_1}, \vec{r_2}) = G(\vec{r_2}, \vec{r_1})$. 
This equation illustrates the physical meaning of the reciprocal theorem: the response at point $2$ of a source at point $1$ is identical to the response at point $1$ of a source at point $2$ (in the general case, \enquote{identical} has to be specified precisely, which is the reason for the slightly more complicated general forms of the reciprocal theorem). 
We refer to \citet{Economou2006} for further discussions.

\subsubsection{Scattering and Lorentz reciprocity}
\label{scattering_reciprocity}

In order to probe an object of interest, it is common to send waves upon it and to observe how they are scattered (i.e. transmitted and reflected) by the object (called scatterer). 
The framework used to describe this process is called scattering theory. It is likely that you are currently using light or sound scattering to access this review, but it also applies to other kinds of waves ranging from quantum matter wave to gravitational waves.

The results of a scattering experiment can be summarized in a scattering matrix $S$ that relates input and output through the equation
\begin{equation}
    \vec{\psi^{\text{out}}} = S \vec{\psi^{\text{in}}}.
    \label{Smat}
\end{equation}
Here, $\vec{\psi^{\text{in}}}$ and $\vec{\psi^{\text{out}}}$ are vectors containing the amplitudes of the incoming and outgoing modes.
These are propagating modes (like plane waves) that are respectively moving towards and away from the scatterer, see Fig.~\ref{fig_scattering_simple}.
In this context reciprocity is usually defined as
\begin{equation}
    S = S^T.
\end{equation}
Colloquially, this corresponds to a symmetry between incoming and outgoing mode that has been summarized as: \enquote{if you can see it, it can see you}.

Reciprocity typically arises from the combination of time-reversal invariance on the one hand, and energy conservation (in wave physics) or probability conservation (in quantum mechanics) on the other hand. 
Indeed, energy or probability conservation imposes that $S$ is unitary ($S^\dagger = S^{-1}$).
In addition, time-reversal invariance imposes that $S = V \overline{S^{-1}} Q^{-1}$ in which $V$ and $Q$ are matrices that describe the action of time-reversal on incoming and outgoing modes~\cite{Fulga2012,Beenakker2015}. 
With an appropriate choice of basis for the modes, this can be simplified to $S = \Theta S^{-1} \Theta^{-1}$ where $\Theta$ is the antiunitary time-reversal operator. 
The combination of these two constraint is known as reciprocity.
In an appropriate basis, it is $S = \pm S^T$, where the sign depends on whether $\Theta^2 = \pm 1$ \cite{Beenakker2015}.
These two situations are usually associated with bosons ($\Theta^2 = 1$) and fermions ($\Theta^2 = -1$), see \citet{Sakurai2020}.
(As in Sec.~\ref{reciprocity_linear_system}, the expression is more complicated in an arbitrary basis, and also if the Wigner decomposition of $\Theta$ is composed of different irreducible representations.) 

We emphasize that reciprocity is a symmetry on its own: it is possible to have reciprocity without energy conservation nor time-reversal (e.g. in a passive lossy medium), but any two of the symmetries implies the third~\citet{Carminati2000,Maznev2013,Guo2022}.

\paragraph{The Mahaux-Weidenmuller formula: relating scattering and internal dynamics.}

In this paragraph, we discuss how the scattering matrix can be related to the dynamics of the system of interest.
Let us consider a system with finite-dimensional degrees of freedom, the dynamics of which is described by the linear equation
\begin{equation}
  \dot{a}_m = L_{mn} a_n - \Gamma_{mn} a_n + {W}_{m \alpha} \psi^{\text{in}}_\alpha
  \label{tcm}
\end{equation}
in which $a_n$ is the amplitude of mode $n$, the matrix $L$ describes the behavior of the system of interest when isolated ($H = i L$ would be the Hamiltonian), $\Gamma$ and ${W}$ represent the loss and gain in the system due to exchanges of waves with the channels, respectively.
Equation \ref{tcm} is known as a temporal coupled mode theory in wave physics~\cite{Fan2003,WonjooSuh2004}; in quantum mechanics, it is the Schrödinger equation.
The outgoing modes are given by
\begin{equation}
      \psi^{\text{out}}_\alpha = S^{0}_{\alpha \beta} \psi^{\text{in}}_{\beta} + \tilde{W}_{\alpha n} a_{n}
\end{equation}
in which $\tilde{W}$ represents the emission of waves in the channels from the resonant modes, and $S^{0}$ is the scattering matrix in a trivial reference state where all $a_n = 0$.
Probing the system with monochromatic waves at frequency $\omega$ (eventually) imposes $a(t) = A e^{-i \omega t}$. 
Eliminating the resonant modes $a(t)$, we then find that the effective scattering matrix such that $\psi^{\text{out}} = S^{\text{eff}} \psi^{\text{in}}$ reads
\begin{equation}
  S^{\text{eff}} = S^0 - \tilde{W} \left[ L - \Gamma + i \omega \right]^{-1} {W}
  \label{Seff}
\end{equation}
The generalization to infinite-dimensional systems requires some care, but leads to the same result, which is known as the Mahaux-Weidenmüller formula \cite{mahaux1969shell,Dittes2000}, see also \citet{Kurasov2023} for a version on graphs, \citet{Fan2003,WonjooSuh2004,Zhang2020b,Benzaouia2021} for more details on the coupled mode theory version, \citet{Christiansen2009} for a more formal version, and \citet{Muga2004} for a discussion of the case of non-Hermitian complex potentials.
It is common that $S^{0} = 1$, that the coupling matrices are related through ${W} = - \tilde{W}^\dagger$,  and the self-energy is given by $\Gamma = 1/2 \tilde{W}^\dagger \tilde{W}$ \cite{Fan2003,WonjooSuh2004,Zirnstein2021}.

The Mahaux-Weidenmüller illustrates how the internal symmetries of the system, along with the symmetries of the couplings, translate into symmetries of the scattering matrix. 
These relations are discussed in detail by \citet{Fulga2012,Zijderveld2025}.

\paragraph{Nonreciprocal devices}
\label{nonreciprocal_devices}

\begin{figure}
    \centering
    \includegraphics[width=8cm]{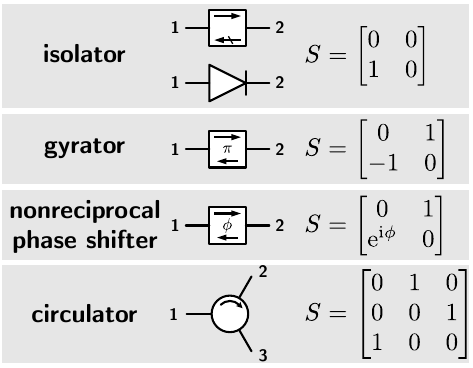}
    \caption{\textbf{Nonreciprocal scattering elements.}
    Standard symbols and canonical scattering matrices for a few basic nonreciprocal scattering elements. 
    In the symbols (loosely following ANSI-Y32/IEC standards), each leg (numbered $1$, $2$, $3$, ...) describes a port with an incoming and an outgoing mode.
    For instance, dipoles correspond to the schematic of Fig.~\ref{fig_scattering_simple}.
    The isolator describes \enquote{amplitude nonreciprocity}.
    The scattering matrix of an isolator (or diode) is a Jordan block of size two (i.e. it is at an exceptional point).
    The gyrator is a particular case of the nonreciprocal phase shifter with differential phase shift $\phi = \pi$, and both describe \enquote{phase nonreciprocity}.
    Finally, a circulator (here with three ports, but it can have any number $n \geq 3$ of ports) describes a \enquote{chiral nonreciprocity} capturing an imbalance between the clockwise and counterclockwise flow of signal or energy in the system.
    \label{nonreciprocal_scattering_elements}
    }
\end{figure}

Nonreciprocal devices have been realized, conceptualized, and standardized in optical, electronic, and high-frequency circuits \cite{Pozar2021,Kord2020} and more recently extended to mechanical waves \cite{Nassar2020}.
Examples of ideal nonreciprocal devices are given in Fig.~\ref{nonreciprocal_scattering_elements}. 
These include the isolator (or diode), which lets signals pass only in one way; the nonreciprocal phase shifter which dephases signals only in one way (the relative dephasing $\phi$ is known as the differential phase shift); the gyrator, a special case where $\phi = \pi$, and the circulator (with three or more ports) where signals flow from one arm to the next in the direction of the arrow, but not to any other arm.

\subsubsection{Transport coefficients and Onsager-Casimir reciprocity}
\label{onsager_casimir}

This section is composed of two parts: first, we describe Onsager-Casimir reciprocity from a purely phenomenological perspective based on thermodynamics.
Second, we describe Onsager-Casimir reciprocity as a theorem resulting from the hypothesis of time-reversal invariance in certain classes of stochastic dynamics.
Both perspectives can be traced to \citet{Onsager1931a,Onsager1931b}. 
The phenomenological perspective, summarized in detail by \citet{DeGrootMazur}, has the advantage of not requiring a microscopic description, but it is fundamentally descriptive.
Some of its aspects have been criticized, in particular by \cite{Truesdell1984}, see \cite{Krommes1993,VanVliet2008} for discussions. 
The modern view, rooted in statistical physics, traces Onsager-Casimir reciprocity to the combination of fluctuation-response relations and detailed balance~\cite{Marconi2008,Campisi2011}.
Within this perspective, generalizations of Onsager reciprocity can be obtained from more general fluctuation relations, at least in the case of (classical or quantum) Gibbs states \cite{Campisi2011}.

\paragraph{Linear irreversible thermodynamics}

The thermodynamics of linear irreversible processes describes transport and relaxation phenomena, such as the diffusion of conserved quantities in spatially extended systems \cite{Gaspard2022,Pottier2010,Beris1994,DeGrootMazur}. The evolution of the densities $x_n(t, \bm{r})$ of extensive thermodynamic quantities $X_i$ follow local balance equations of the form
\begin{equation}
    \partial_t x_n + \bm{\nabla} \cdot (x_n \bm{v} + \bm{J}_{n}) = \sigma_{n}
    \label{lbe_thermo}
\end{equation}
in which $\bm{v}$ is the velocity of the medium, $\bm{J}_{n}$ is the dissipative current of $X_n$ and $\sigma_{n}$ is a local source/sink of $X_n$ (vanishing for conserved quantities).
This equation is complemented by phenomenological laws that give the current and the source as a function of the densities, which can be obtained from thermodynamics when assuming local equilibrium.
Within linear response, the currents are given by $\bm{J}_{m} = L_{m n} \Lambda_n$ and the sources by $\sigma_{m} = L^{\text{r}}_{mn} \Lambda^{\text{r}}_n$ in which $\Lambda_n$ and $\Lambda^{\text{r}}_n$ are called affinities and represent the deviation from equilibrium, while $L$ and $L^{\text{r}}$ are called matrices of phenomenological coefficients (corresponding to transport and relaxation, respectively).
The affinities can be defined phenomenologically as conjugate variables $\Lambda^{\text{r}}_n = \partial s/\partial x_n$ and $\bm{\Lambda}_n = \bm{\nabla}(\partial s/\partial x_n)$ with respect to the entropy density $s$, and  in the case of transport phenomena, $\Lambda_n$ are, up to details, related to gradients $\nabla x_n$ of the corresponding densities.

We encode the behavior under time-reversal of the quantities $X_n$ in an operator $\Theta$ (see Sec.~\ref{harsh}); when $X_n$ is odd or even under time reversal then $\Theta X_n = \pm X_n$, respectively. 
In this context, the symmetry $L^T = \Theta L \Theta^{-1}$ of the matrix of phenomenological coefficients (same for $L^{\text{r}}$) is referred to as Onsager reciprocity \cite{Onsager1931a,Onsager1931b}.
The Onsager relations typically do not hold in systems where time-reversal invariance is broken, for instance where rotation or a magnetic field is present; in this paragraph the symbol $\bm{B}$ collectively refers to all time-reversal-breaking fields.
Instead, a duality relation, called Onsager-Casimir reciprocity, exists between the systems with parameters $\bm{B}$ and $-\bm{B}$, namely $L^T(\bm{B}) = \Theta L(-\bm{B}) \Theta^{-1}$ \cite{Casimir1945}, and the Onsager relation is obtained for $\bm{B} = 0$. 
We emphasize that the Onsager-Casimir relations are only expected to hold when the affinities are chosen in a certain way. 
For perturbations about equilibrium systems, this can be done by taking the affinities as conjugate variables of the $x_i$ with respect to the entropy, as described above, although this requires some care especially for tensorial quantities in the continuum, such as viscosity tensors ~\cite{deGroot1954,Mazur1954,Vlieger1954}.

\paragraph{Onsager reciprocity as a consequence of time-reversal invariance}

From a microscopic perspective, Onsager reciprocity is the manifestation of microscopic time-reversal invariance at the level of response functions in a statistical field theory. 
This relation corresponds to arrow (ii) in Eq.~\eqref{nrr_scheme}.
In its simplest instantiation, it can be seen as the combination of a fluctuation-response relation with detailed balance.
Indeed, the fluctuation-response relates the response of a Markovian system to a perturbation with the temporal correlations in the unperturbed system (which may be out of equilibrium, provided that it has a steady-state with a smooth enough probability density, as discussed by \citet{Agarwal1972,Prost2009,Seifert2010,Baiesi2013}). 
Detailed balance imposes a constraint on time correlations, which directly translates into a constraint on the response function.
More generally, fluctuation theorems can be seen as nonlinear generalizations (i.e., not only at zero forcing) of the Onsager-Casimir reciprocity theorems; in addition to the Onsager-Casimir relations, they relate higher-order nonlinear responses to higher-order correlation functions~\cite{Hanggi1978a,Hanggi1978b,Hanggi1982,Bochkov1977,Andrieux2007,Gallavotti1996,Kurchan1998}.
These can also be applied to quantum systems~\citet{Kurchan2000,Talkner2007,Andrieux2008,Andrieux2009,Chetrite2012}.
We refer to \citet{Esposito2009,Campisi2011,Talkner2020} for further details.

\subparagraph{Linear response theory}

As an example, consider a Markovian master equation
\begin{equation}
    \frac{d}{dt} p = \WW p
\end{equation}
for the probability distribution $p(t)$, such as the Fokker-Planck equation \ref{fokker_planck}. 
We consider the case where the operator $\WW$ depends on an external parameter $h(t)$ as $\WW = \WW_0 + h(t) \WW_1$, and we assume that $\WW_0$ has a steady-state $p^{\text{ss}}$ (so $\WW_0 p^{\text{ss}} = 0$).
The response of an observable $A$ to the perturbation controlled by $h(t)$ is then
\begin{equation}
    \langle A(t)\rangle - \langle A(t)\rangle_0 = \int R_{A,\WW_1}(t-s) h(s) ds
\end{equation}
where the response function $R_{A,\WW_1}$ turns out to be
\begin{subequations}
\label{agarwal_formula}
\begin{equation}
    R_{A,\WW_1}(t) = H(t) \langle A(t) B_{\WW_1}(0) \rangle_{p^{\text{ss}}}
\end{equation}
where $H$ is the Heaviside step function, namely as the temporal correlation function between $A$ and another observable $B_{\WW_1}$ defined as
\begin{equation}
    B_{\WW_1}(x) = \frac{(\WW_1 p^{\text{ss}})(x)}{p^{\text{ss}}(x)}.
\end{equation}
\end{subequations}
This result, known as the Agarwal formula \cite{Agarwal1972}, can be seen as a nonequilibrium generalization of the fluctuation-response theorem~\cite{Seifert2010,Baiesi2013}.
For the manipulations leading to the formulas above to make sense, we need the steady-state to have smooth density~\cite{Baiesi2013,Pavliotis2014}. 
This is not guaranteed out of equilibrium (see Sec.~\ref{transverse_decomposition}); when the density is not smooth, a more general response theory developed by \cite{Ruelle1998,Ruelle1999,Ruelle2009} can still be used.
In the case of Hamiltonian systems at thermal equilibrium, equivalent theorems can be obtained where $\WW$ is replaced by the Liouvillian operator~\cite{Seifert2010,Gaspard2022}.

\subparagraph{Detailed balance and correlations}
\label{symmetry_of_correlations}

One of the key consequences of time-reversal invariance (detailed balance) is that it implies the symmetries of temporal correlations in the steady-state. 
Namely, detailed balance (Eq.~\eqref{db}) implies that
\begin{equation}
    \langle A(t) B(0) \rangle_{p_\text{ss}} = \langle \Theta B(t) \Theta A(0)  \rangle_{p_\text{ss}}
    \label{db_correlations}
\end{equation}
in which correlations are about the steady-state.

Given a list of time-reversal-even observables $A_i$, we can construct a correlation matrix with components $C_{ij}(t) = \langle A_i(t) A_j(0) \rangle$. 
Because of time-translation invariance, we always have $C_{ij}(t) = C_{ji}(-t)$.
Hence, when $\Theta = 1$, the symmetric (antisymmetric) part of $C(t)$ as a matrix is identical to the even (odd) part of $C(t)$ as a function of time. 
In this case, the asymmetry of cross-correlations, which can be related to phase space equivalents of angular momentum \cite{Tomita1974,Tomita2008}, can be constrained by thermodynamic bounds \citet{Ohga2023,Shiraishi2023,Van2024,Gu2024,Aslyamov2025}. 
These are related to conjectured bounds on the thermodynamic cost of maintaining coherent oscillations \cite{Cao2015,Oberreiter2022,Shiraishi2023,Santolin2025}. 
Note that these do not apply as-is when time-reversal-odd variables are present, like in underdamped mechanical systems~\cite{Pietzonka2022}.

\subparagraph{The Onsager reciprocity theorem}

The Onsager reciprocity theorem can be obtained by combining the Agarwal linear response formula \eqref{agarwal_formula} (that assumes a steady-state with a smooth probability density) with the symmetry \eqref{db_correlations} of correlations (that results from time-reversal invariance), leading to
\begin{equation}
    R_{A,B}(t) = R_{\Theta B,\Theta A}(t)
\end{equation}
for any two observables $A$ and $B$. 

\subparagraph{Green-Kubo formulas}

In certain circumstances, the linear response described above can be cast as a Green-Kubo formula, in which transport coefficients are expressed as integrals of correlation functions \cite{Kubo1966,Toda2012,Kubo2012,Gaspard2022,Pottier2010}, see also \citet{Pavliotis2010} for a more mathematical perspective.
For instance, with the appropriate assumptions, the diffusion tensor can be expressed as
\begin{equation}
    D_{ij} = \int_{0}^{\infty} \langle v_i(t) v_j(0) \rangle dt
\end{equation}
in which $\bm{v}$ is the velocity, which is essentially the current of particles.
Similarly, the diffusion tensor can also be expressed as \cite{Pavliotis2010}
\begin{equation}
    D_{ij} = (\psi_i, \mathbb{W}^{-1} \psi_j)
\end{equation}
in which $(\cdot, \cdot)$ is an appropriate inner product.
Generalizations for other transport coefficients, like viscosity or heat diffusion, can be obtained \cite{Balescu1991,Kirkpatrick2025}, see \citet{Fruchart2023} for examples.
When they hold, these expressions show that (i) asymmetric transport coefficients (here, $D_{ij} \neq D_{ji}$) are related to time-asymmetric correlations (here, $\langle v_i(t) v_j(0) \rangle \neq \langle v_i(-t) v_j(0) \rangle $; see Sec.~\ref{symmetry_of_correlations}) and (ii) asymmetric transport coefficients are related to a non-Hermitian Fokker-Planck operator (here, $\mathbb{W} \neq \mathbb{W}^\dagger$; a similar statement can be made in kinetic theory with the linearized collision operator instead, see e.g. \citet{Resibois1970,Lhuillier1982,Fruchart2022}).
A concrete example where a non-symmetric (odd) viscosity (Sec.~\ref{hall_odd_effects}) is obtained from molecular dynamics simulations both from response measurements and time-asymmetric correlations through a Green-Kubo formula can be found in \citet{Han2021}.
Note also that some care is required within overdamped descriptions \cite{Risken1989,Jardat1999,Hoang2023}.

\subsection{Reciprocity in linear operators}
\label{reciprocity_linear_system}

\noindent\emph{\textbf{In a nutshell:} an abstract notion of non-reciprocity can be generically defined for any linear operator.}
\medskip

Consider a linear operator $\mathscr{L}$ acting on a vector space $\mathcal{V}$ equipped with an inner product $\braket{\cdot, \cdot}$.
The linear operator is said to be reciprocal with respect to an antiunitary operator $\mathscr{A}$ on $\mathcal{V}$ such that
\begin{equation}
    \mathscr{A} \mathscr{L} \mathscr{A}^{-1} = \mathscr{L}^\dagger
    \label{reciprocal_linear_operator}
\end{equation}
in which $\mathscr{L}^\dagger$ is the adjoint of $\mathscr{L}$, namely the operator such that $\braket{\mathscr{L}^\dagger \phi, \chi} = \braket{\phi, \mathscr{L} \chi}$ for all $\phi, \chi \in \mathcal{V}$.
As a consequence, we have\footnote{This can be shown from the identity $\braket{\chi, \mathscr{L} \phi} = \braket{\mathscr{A} \phi, (\mathscr{A} \mathscr{L}^\dagger \mathscr{A}^{-1}) \mathscr{A} \chi}$ valid for any linear operator $\mathscr{L}$, antiunitary operator $\mathscr{A}$, and vectors $\phi, \chi \in \mathcal{V}$. } 
\begin{equation}
    \braket{\phi, \mathscr{L} \chi} = \braket{\mathscr{A} \chi, \mathscr{L} \mathscr{A} \phi}
    \label{reciprocity_matrix_elements}
\end{equation}
for any $\phi, \chi \in \mathcal{V}$. 
Equations \eqref{reciprocal_linear_operator} and \eqref{reciprocity_matrix_elements} are the most general expressions of reciprocity for a linear operator~\citet{Bilhorn1964,Deak2012,Sigwarth2022,Schiff1968}.
The antiunitary $\mathscr{A}$ typically represents time-reversal, although \citet{Deak2012} discussed the possibility of using other \enquote{reciprocity operators}.
Please note that \eqref{reciprocal_linear_operator} is not the statement that $\mathscr{L}$ is $\mathscr{A}$-invariant (which would read $\mathscr{A} \mathscr{L} \mathscr{A}^{-1} = \mathscr{L}$, without the adjoint).
\subsubsection{If you can see it, \emph{it} can see you}
The abstract definition \eqref{reciprocal_linear_operator} of non-reciprocity is perhaps most easily motivated from the perspective of scattering (Sec.~\ref{scattering_reciprocity}). 
In short, we want to mathematically encode the idea that \enquote{if you can see it, \emph{it} can see you}. To do so, we need to exchange the incoming and outgoing modes; as these are essentially traveling waves $\ee^{\pm \ii \bm{k}\cdot\bm{r}}$ (Fig.~\ref{fig_scattering_simple}), this is in general performed by an antiunitary operator (because it conjugates the complex exponentials).
We want to also cancel out the effect of a trivial damping of the signal due to dissipation, which is essentially symmetric: this is performed by relating $\mathscr{A} \mathscr{L} \mathscr{A}^{-1}$ to $\mathscr{L}^\dagger$ instead of $\mathscr{L}$.

To make Eq.~\eqref{reciprocal_linear_operator} more concrete, we need to represent the operators in a basis $\phi_i$, so $\mathscr{L} \simeq L$ in which $L$ is a matrix with elements $L_{ij} = \langle \phi_i, \mathscr{L} \phi_j \rangle$, and $\mathscr{A} = U_A \mathcal{K}$ where $\mathcal{K}$ represents complex conjugation.
Equation \eqref{reciprocal_linear_operator} then becomes
\begin{equation}
    L^T = U_A L U_A^{-1}.
    \label{general_A_reciprocity}
\end{equation}
This is the most general expression of $\mathscr{A}$-reciprocity.

In practical cases, it is possible to simplify Eq.~\eqref{general_A_reciprocity} through Wigner's theory of the representation of antiunitary operators~\cite{Wigner1960,Weigert2003}. 
In a nutshell, antiunitary operators can be decomposed into three classes of irreducible representations, corresponding respectively (i) to 1D blocks with $\mathscr{A}^2 = 1$, (ii) to 2D blocks with $\mathscr{A}^2 = - 1$ and (iii) to 2D blocks with $\mathscr{A}^2 = \Omega \neq \Omega^* \in U(1)$. 
A given antiunitary $\mathscr{A}$ may be composed of several of these blocks. As each block is independent of the others, we now describe the possible cases separately.

\paragraph{\texorpdfstring{$\mathscr{A}^2 = 1$}{A²=1}.}
In this case, it is possible to choose a basis of vectors $\phi_i$ such that $\mathscr{A} \phi_i = \phi_i$ (in other words, $U_A$ is the identity).
In this basis, Eq.~\eqref{reciprocity_matrix_elements} reduces to
\begin{equation}
    L = L^T
    \qquad
    \text{(when $\mathscr{A}^2 = 1$).}
    \label{reciprocity_matrix_elements_bosonic}
\end{equation}
This is the most common instance of nonreciprocity in the context of linear systems!
We emphasize that (i) it only applies when the reciprocity operator $\mathscr{A}$ satisfies $\mathscr{A}^2 = 1$ and (ii) it only applies in a basis of $\mathscr{A}$-invariant vectors.
When $\mathscr{A}$ is time-reversal, this means that Eq.~\eqref{reciprocity_matrix_elements_bosonic} applies to bosonic systems (for which $\mathscr{A}^2=1$), in a time-reversal invariant basis, but not to fermionic systems (for which $\mathscr{A}^2 = -1$).

\paragraph{\texorpdfstring{$\mathscr{A}^2 = - 1$}{A²=-1}.} 
When $\mathscr{A}^2 = - 1$, it is not possible to make the same choice of basis as in the first paragraph. Instead, one can choose a basis of vectors $\phi_i^\pm$ such that $\mathscr{A} \phi_i^\pm = \pm \ii \phi_i^\mp$ \footnote{When $\mathscr{A}$ is time-reversal, the vectors $\phi_i^\pm$ are known as a Kramers pair.}. In such a basis, Eq.~\eqref{reciprocity_matrix_elements} becomes
\begin{equation}
    L^T = \sigma_y L \sigma_y^{-1}
    \qquad
    \text{(when $\mathscr{A}^2 = -1$).}
    \label{reciprocity_matrix_elements_fermionic}
\end{equation}
in which $\sigma_y$ is a Pauli matrix. 
This situation is commonly encountered in fermionic systems, because the antiunitary time-reversal operator $\Theta$ acting on fermions satisfies $\Theta^2 = - 1$ \cite{Sakurai2020}.

\paragraph{\texorpdfstring{$\mathscr{A}^2 = \Omega \neq \Omega^*$}{A²=Ω≠Ω*}.} 
In this case, one can choose a basis of vectors $\phi_i^\pm$ such that $\mathscr{A} \phi_i^+ = \omega^{*} \phi_i^-$ and $\mathscr{A} \phi_i^- = \omega \phi_i^+$ in which $\omega \in U(1)$ is such that $\omega^2 = \Omega$, leading to $U_A =\big(\begin{smallmatrix}   0 & \omega^* \\  \omega & 0 \end{smallmatrix}\big)$ in Eq.~\eqref{reciprocity_matrix_elements}.

\subsubsection{Generalized PT-symmetry and pseudo-Hermiticity}
\label{PT_symmetry}

The notions of (generalized) PT-symmetry and pseudo-Hermiticity are ways to encode into an algebraic constraint certain spectral properties of an operator, namely whether its eigenvalues are purely real, complex conjugate, or arbitrary \cite{Bender1998,Bender2007,Mostafazadeh2002,Mostafazadeh2002b,Mostafazadeh2002c,Mostafazadeh2003,Bender2002,Bender2010,Mostafazadeh2015,Ashida2020}.

Consider a linear operator $H \in \text{End}(\mathcal{H})$ acting on a Hilbert space $\mathcal{H}$.

\paragraph{Generalized PT-symmetry}
\label{generalized_PT}

Consider an antilinear operator $X$ (this means that there is a linear operator $M_X$ such that $X \psi = M_X \overline{\psi}$) such as $X^2 = \Id$ (where $\Id$ is the identity matrix). 
The operator $H$ is said to be $X$-symmetric when $[H,X] = 0$ (i.e. when $X H = H X$). 
The $X$-symmetry is called exact (or unbroken) when in addition, there is a complete set of eigenvectors $\psi_n$ of $H$ satisfying $X \psi_n = \psi_n$. 
Else, it is called inexact (or spontaneously broken).
The operator $X$ can be seen as a generalization of the product $PT$ of parity $P$ and time-reversal $T$, and is therefore called a generalized PT symmetry, although it may have nothing to do with the physical parity and time-reversal operations. 

\paragraph{Pseudo-Hermiticity}

Consider an invertible operator $\eta$ on $\mathcal{H}$. The operator $H$ is said to be $\eta$-pseudo-Hermitian when
\begin{equation}
    H^\dagger = \eta H \eta^{-1}.
\end{equation}
The operator is said to be pseudo-Hermitian if there is some $\eta \in \text{GL}(\mathcal{H})$ such that it is $\eta$-pseudo-Hermitian.
The operator is said to be exactly pseudo-Hermitian when it is $\eta$-pseudo-Hermitian with a positive-definite $\eta$ (equivalently, when $\eta = \mathcal{O} \mathcal{O}^\dagger$ with $\mathcal{O} \in \text{GL}(\mathcal{H})$).

\paragraph{Spectral consequences}

Assuming that $H$ is diagonalizable and has a discrete spectrum, the three following statements are equivalent: (i) $H$ is pseudo-Hermitian, (ii) $H$ has an antilinear symmetry, (iii) the eigenvalues of $H$ are either real or come in complex-conjugate pairs. 
In addition, the three following statements are equivalent: (i) $H$ has an exact antilinear symmetry, (ii) $H$ is exactly pseudo-Hermitian, (iii) the eigenvalues of $H$ are real. 

In parallel with the case where PT symmetry is spontaneously broken ($H$ has an inexact antilinear symmetry), we say that PT symmetry is explicitly broken when $H$ does \emph{not} have an antilinear symmetry at all.

\begin{figure}
    \centering
    \includegraphics[width=\linewidth]{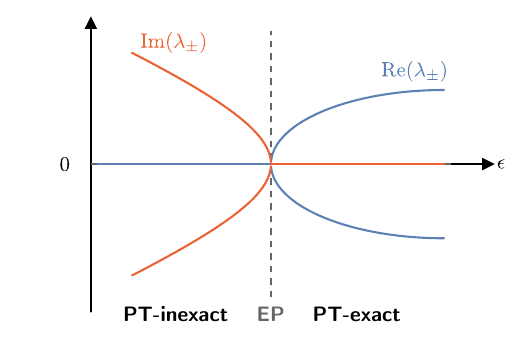}
    \caption{
    \textbf{Exceptional point marking the boundary between PT-exact and PT-inexact regions.}
    The eigenvalue spectrum corresponds to Eq.~\eqref{L_EP_PT} with $\delta = \kappa - \epsilon$ and fixed $\kappa > 0$.
    }
    \label{figure_ep_boundary}
\end{figure}

\paragraph{Exceptional points as boundaries}

Exceptional points can be seen as boundaries between exact and inexact (aka unbroken and spontaneously broken) PT-symmetric phases.
We can see this with the following two-by-two example:
\begin{equation}
    \label{L_EP_PT}
    L = \kappa \sigma_1 + \ii \delta \sigma_2 =  \begin{pmatrix}
        0 & \kappa + \delta \\
        \kappa - \delta & 0
    \end{pmatrix}
\end{equation}
where $\kappa$ and $\delta$ are real numbers, and $\sigma_n$ are Pauli matrices.
The matrix $L$ is not Hermitian whenever $\delta \neq 0$, however it has a generalized PT-symmetry ($X$-symmetry; see Sec.~\ref{generalized_PT}) with
\begin{equation}
    M_X = \begin{pmatrix}
        0 & \eta \\
        \eta^{-1} & 0
    \end{pmatrix}
    \quad
    \text{with}
    \quad
    \eta = \sqrt{\frac{\kappa + \delta}{\kappa - \delta}}.
\end{equation}
Setting $\delta = \kappa - \epsilon$, we find that (i) $L$ has an exceptional point at $\epsilon = 0$ (except when $\kappa = 0$, in which case $L = 0$), (ii) the eigenvalues of $L$ are $\lambda_{\pm} = \pm \sqrt{2 \kappa} \, \sqrt{\epsilon} + \mathcal{O}(\epsilon)$, meaning that the eigenvalues are real when $\epsilon > 0$ and purely imaginary complex conjugate pairs when $\epsilon < 0$.
(An almost identical picture takes place if we consider $L + \omega_0 \, \Id$ ($\omega_0 \in \RR$), except that the complex conjugate pairs are not purely imaginary.)
This illustrates spontaneous PT-symmetry breaking: the generalized PT-symmetry of $M$ is exact when $\epsilon > 0$ and inexact (spontaneously broken) when $\epsilon < 0$.
These regimes are separated by an exceptional point: $L$ is already in Jordan normal form when $\epsilon = 0$, and one can check explicitly that the normalized eigenvectors become collinear as $\epsilon \to 0$.

The shape of the eigenvalue spectrum near the boundary between PT-exact and PT-inexact regions, shown in Fig.~\ref{figure_ep_boundary}, is very distinctive and can easily be recognized in more complicated situations. On the PT-exact side, the imaginary parts are degenerate and the real parts are distinct; while on the PT-inexact part, the imaginary parts are distinct and the real parts degenerate.

\paragraph{Anti–parity-time symmetry}

Note that a related symmetry known as anti–parity-time (APT) symmetry has also been considered \cite{Peng2016,Li2019}.
It is defined as follows: we say that $H$ is $X$-antisymmetric when $X H = - H X$.
PT-symmetry and APT-symmetry may have different physical implications in a given context. 
However, they have the same mathematical content: $H = i L$ is $X$-symmetric if and only if $L$ is $X$-antisymmetric.
Note that PT-exact and PT-inexact are also permuted when considering $L$ instead of $H$.

\paragraph{Application to nonlinear dynamics}

In a non-linear dynamical system $\partial_t \psi = f(\psi)$, the condition of PT-symmetry reads $X f(\psi) = f(X \psi)$, see e.g. Ref.~\cite{Swift1992,Konotop2016}. 
This is particularly convenient when the dynamical system is $U(1)$-equivariant (Sec.~\ref{symmetry_dynamics}) for which travelling wave solutions can be obtained as fixed point of an effective dynamics \cite{Chossat2000}.

\paragraph{An example: the harmonic oscillator.}

\begin{figure}
    \centering
    \includegraphics[width=\linewidth]{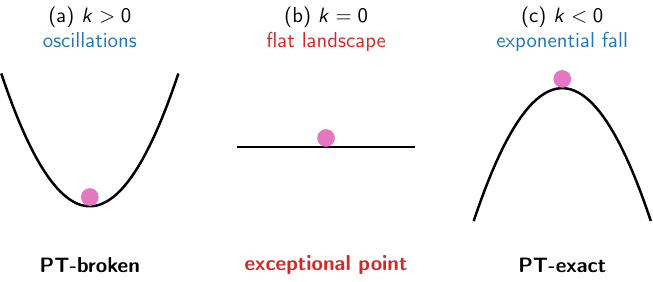}
    \caption{
    \label{figure_ep_harmonic_oscillator}
    \strong{Exceptional points in a harmonic oscillator.}
    The simple harmonic oscillator has an exceptional point when the spring constant $k$ vanishes, separating oscillations (when the potential is confining) from an exponential fall (when the potential is repelling).
    These correspond respectively to the PT-inexact and PT-exact cases.
    Adapted from \citet{Fruchart2021}.
  }
\end{figure}

As an example, let us consider the one-dimensional classical harmonic oscillator $m \ddot{x} = - k x$, where $x$ is the position of a particle of mass $m > 0$ in a harmonic potential with stiffness $k$.
This second-order differential equation is equivalent to the first order Hamilton equations
\begin{equation}
	\label{harmonic_oscillator}
	\partial_t \begin{pmatrix} x \\ p \end{pmatrix}
	=
	\begin{pmatrix} 0 & 1/m \\ -k & 0 \end{pmatrix}
	\begin{pmatrix} x \\ p \end{pmatrix}
\end{equation}
in which $p = m \dot{x}$.
The eigenvalues of the PT-symmetric matrix in Eq.~\eqref{harmonic_oscillator} are $\pm \ii \sqrt{k/m}$.
When $k > 0$ (Fig.~\ref{figure_ep_harmonic_oscillator}a), the eigenvalues are complex conjugate imaginary numbers and the particle oscillates in the potential. This corresponds to the PT-inexact case.
When $k < 0$ (Fig.~\ref{figure_ep_harmonic_oscillator}c), the eigenvalues are real, and the particle falls off exponentially with a characteristic time $\tau = \sqrt{|m/k|}$ from the top of the inverted parabola. This correspond to the PT-exact case.
The critical case $k = 0$ (Fig.~\ref{figure_ep_harmonic_oscillator}b) separating these regumes corresponds to a free particle (the potential is completely flat), and in this case the matrix in Eq.~\eqref{harmonic_oscillator} is not diagonalizable: it is a Jordan block of size $2$ with eigenvalue zero, so the free particle $k = 0$ is an exceptional point.

Another simple simple example of exceptional point is given by the damped harmonic oscillator, in which an exceptional point occurs at the critical damping which separates the underdamped and overdamped regimes.

\subsubsection{Normal and non-normal matrices}
\label{non_normal_matrices}

A matrix $H \in \mathcal{M}_n(\CC)$ is said to be normal when it commutes with its adjoint ($H H^\dagger = H^\dagger H$) \footnote{For simplicity we focus on matrices but similar statements can be made for infinite dimensional operators, with proper care.}
All (anti)symmetric, (anti)Hermitian, unitary matrices are normal. 
Perhaps most crucially for our purposes, a matrix is normal if an only if it can be diagonalized by a unitary matrix in $\CC$, or in other words if an only if there is an orthonormal basis of $\CC^n$ consisting of eigenvectors of $H$.
Several other equivalent statements can be found in \citet{Grone1987,Elsner1998}.
From the standpoint of physics, the eigenvectors may correspond to normal modes.
When the corresponding matrix is not normal, then the \enquote{normal modes} cannot be taken to be orthogonal to each other anymore \citet{Trefethen1993,Trefethen2005} with striking consequences for the stability analysis of noisy systems, see the later Sec. \ref{non_normal_amplification_and_noise}.

Matrices that cannot be diagonalized (called defective, nonsemisimple, or nondiagonalizable) are the most extreme case of non-normal matrices. 
Following \citet{Kato1984}, the points in parameter space (or in the space of matrices) where this happens are called exceptional points.
Despite their name, they are ubiquitous in nonequilibrium systems including nonreciprocal ones.
These matrices exhibit non-trivial Jordan blocks (i.e. of size larger than one) of the form
\begin{equation}
    \begin{pmatrix}
\lambda & 1       & 0             & \cdots & 0             \\
0           & \lambda & 1             & \cdots & 0             \\
0           & 0       & \lambda       & \cdots & 0             \\
\vdots   & \vdots  & \vdots     & \ddots & \vdots   \\
0           & 0       & 0             & \cdots & \lambda      \\
\end{pmatrix}.
\end{equation}

\section{Where to find non-reciprocity?}
\label{where}

In this section, we discuss a selection of concrete situations in which non-reciprocity may be encountered, in order to illustrate the different ways it can arise, and the variety of domains where it can happen.
Our take-home message is the following: non-reciprocity is the rule rather than the exception.

\subsection{Nonvariational dynamics}
\label{where_nonvariational_dynamics}

\begin{figure*}
    \centering
    \includegraphics[width=\linewidth]{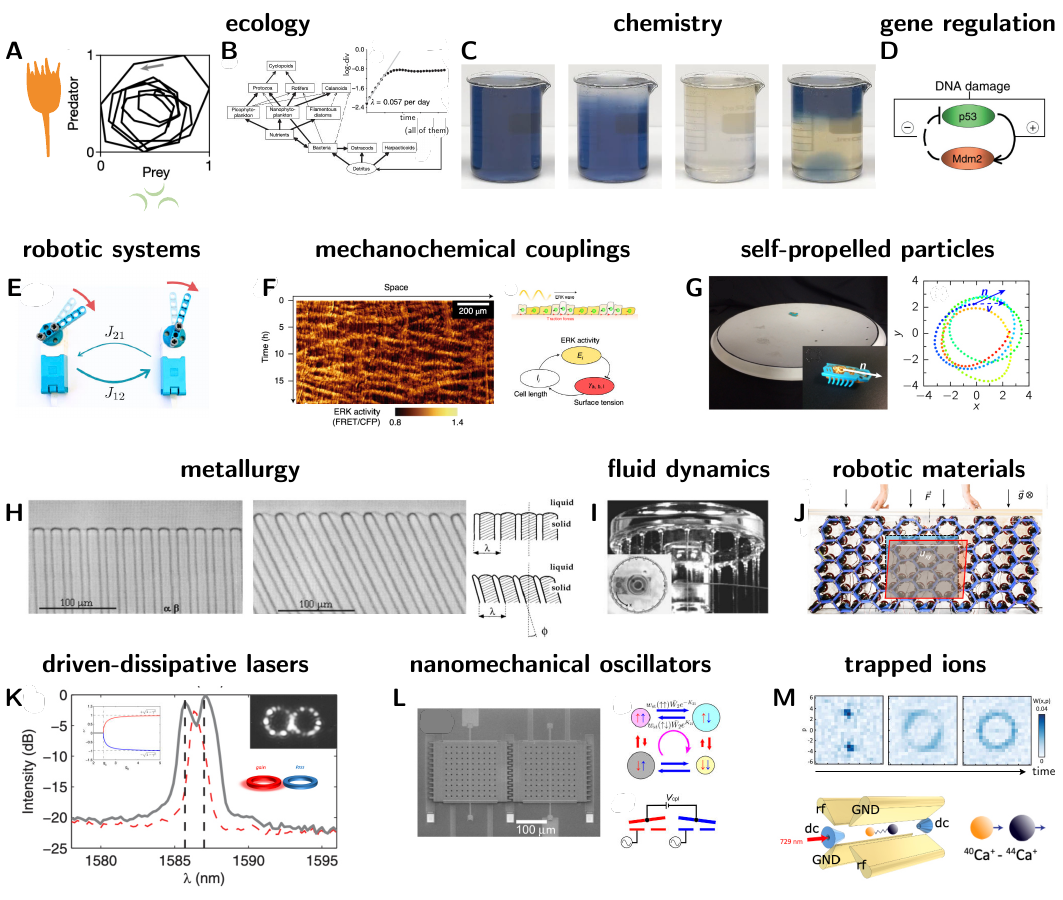}
    \caption{
    \label{examples_nonvariational}
    \textbf{Nonvariational dynamics.}
	(A) Oscillations in a planktonic prey-predator system (respectively \emph{Monoraphidium minutum} and \emph{B. calyciflorus}) observed in a year-long experiment (data shown corresponds to about 45 days).
    Adapted from \citet{Blasius2019}.
	(B) Complex food web isolated from the Baltic Sea, consisting of bacteria, phytoplankton species, herbivorous and predatory zooplankton species, and detritivores. The inset shows the experimentally computed Lyapunov exponent for one of the species, which is positive, hinting at a chaotic dynamics.
    Adapted from \citet{Beninca2008}.
	(C) Oscillations in a Briggs–Rauscher oscillating chemical reaction. The solution is not mixed externally during the experiment, meaning that spatial homogeneisation is mostly diffusive.
    Adapted from \citet{Braun2017}.
	(D) The tumor suppressor p53 transcriptionally activates Mdm2, which in turn targets p53 for degradation. After strong DNA damage, a change in the strength of regulation and in the concentrations leads to damped oscillations in p53 and Mdm2.
    Adapted from \citet{Lahav2004}.
	(E) Nonreciprocal XY spins interactions (alignement and antialignment) can be implemented in robotic systems, here using programmable LEGO toys. Adapted from \citet{Mandal2024b}.
	(F) Spatiotemporal pattern arising from mechanochemical feedback loops in cell monolayers. Extracellular signal-regulated kinase (ERK) activity causes mechanical changes, which in turn affect ERK activity. Adapted from \citet{Boocock2020}.
	(G) Self-propelled and self-aligning particle (hexbug) in a dish antenna.
    The nonreciprocal coupling between the velocity and the robot's orientation leads to a limit cycle.
    Adapted from \citet{Dauchot2019}.
	(H) Directional solidification fronts in the lamellar eutectic alloy \ce{CBr4-C2Cl6}. In this nonequilibrium system, a nonreciprocal interaction between different harmonics of the field representing the interface occur, and this leads to a parity-breaking instability captured by a drift-pitchfork bifurcation. Adapted from \citet{Ginibre1997}.
	(I) Parity-breaking instability (similar to panel H) in a liquid column pattern with periodic boundary conditions formed at the side of an overflowing dish. Adapted from \citet{Brunet2001}.
	(J) Robotic metamaterial exhibiting nonreciprocal odd elasticity, which arises from the nonreciprocal (but momentum-preserving) programmed interactions between constituents. Adapted from \citet{Veenstra2025}.
	(K) Drift-pitchfork bifurcation in driven-dissipative ring resonator lasers, leading to a two-peak emission curve due to thermal noise. Adapted from \citet{Hassan2015}.
	(L) Implementation of two nonreciprocal Ising spins using driven nanomechanical oscillators. Adapted from \citet{Han2024}.
	(M) Synchronization between two quantum van der Pol oscillators made from ion crystals in a Paul trap. Adapted from \citet{Liu2025b}.
    }
\end{figure*}

One of the most iconic instance of nonreciprocity is given by preys and predators.
These involve at least two different kinds of nonreciprocity. 
First, a cat chasing a mouse may be described by effective social forces that do not satisfy Newton's third law: this is discussed in the later Sec.~\ref{violations_newton_third_law}.
On longer time scales, the presence of prey promotes the presence of predators, while the presence of predators represses the presence of preys.
This ecological dynamics is captured by so-called prey-predator or consumer-resource models~\citet{Murray2011a,Murray2011b,Turchin2013,Maynard1978}.
One of the simplest such model (that ignores the run-and chase motion in real space) are the Lotka–Volterra equations
\begin{equation}
    \label{glv}
    \frac{d N_i}{dt} = N_{i} \left( \alpha_i + \sum_{i=1}^{S} \alpha_{ij} N_j \right)
\end{equation}
describing the population densities $N_i$ of $S$ species $i=1,\dots,S$ with growth rates $\alpha_i$ and interaction strengths $\alpha_{ij}$ which encode the influence of species $i$ on species $j$.
In general, $\alpha_{ij}$ and $\alpha_{ji}$ can have different values and different signs, encoding the nonreciprocity of the system.
Here, the nonreciprocity is, in a sense, minimal: there are two degrees of freedom that have different roles.
Generalizations of Eq.~\eqref{glv} with random couplings between many species, often including effects of noise, have been used for modeling ecosystems \cite{Ros2022,Hu2022,vanDenBerg2022,Brenig1988,Nutku1990,Altieri2021,Biroli2018}.

Although Eq.~\eqref{glv} reproduces certain qualitative features of population dynamics, it is not entirely realistic~\cite{Murray2011a}. 
Notably, the Lotka–Volterra equations have a Hamiltonian structure~\cite{Kerner1964,Kerner1997}
and therefore cannot exhibit, for instance, limit cycles.
In this sense, it can be seen as \enquote{purely nonreciprocal}.
More realistic models have been developed~\citet{Murray2011a,Murray2011b,Turchin2013,Maynard1978} that cure this disease.
For instance, the Rosenzweig-MacArthur model~\cite{Rosenzweig1963} models the evolution of isolated prey through a logistic growth (instead of exponential) by replacing $\alpha_i$ with $\alpha_i^0 (N_0 - N_i)$ when $i$ is a prey, and includes a saturation of the interaction, called functional response \cite{EdelsteinKeshet2005}.
In the case of a single predator and a single prey, this model is contained in the coupled differential equations
\begin{equation}
    \begin{split}
    \dot{u} &= r (u_0 - u) u - k \frac{u v}{u_1 + u} \\
    \dot{v} &= \tilde{k} \frac{u v}{u_1 + u} - s v
    \end{split}
	\label{rma_two}
\end{equation}
where $u$ and $v$ are the numbers (or concentrations) of prey and predators, respectively, $r$ is the natural growth of prey, $u_0$ is the carrying capacity of the environment, $s$ is the natural decay rate of predators, while $k$ and $\tilde{k}$ measure predation, which reduces the number of prey while increasing the number of predators (not necessarily in the same amount), and $u_1$ describes the saturation of this coupling.
Crucially, Eq.~\eqref{rma_two} is neither variational nor volume-preserving, and therefore can exhibit limit cycles (Sec.~\ref{nonvariational_dynamics}), which it does.
In fact, Eq.~\eqref{rma_two} exhibits a bifurcation between a fixed point where prey and predator concentrations are finite and do not change in time and a limit cycle where both oscillate periodically, for instance as one increases the carrying capacity $u_0$. 
One can show that this bifurcation is a Hopf bifurcation, and so after a nonlinear change of variable, Eq.~\ref{rma_two} is locally equivalent to the normal form
\begin{equation}
    \dot{z} = (r + \ii \omega) z - \beta |z|^2 z + \mathcal{O}(z^4)
    \label{normal_form_hopf_predator_prey}
\end{equation}
in which $z(t)$ is a complex variable encoding $u_1(t)$ and $v_1(t)$, and $r = [u_0 - u_0^{\text{c}}]/u_0^{\text{c}}$ where $u_0^{\text{c}}$ is the critical carrying capacity where the bifurcation occurs.

Additional ingredients may be required to quantitatively capture experimental data; for instance, Fig.~\ref{examples_nonvariational}a shows predator-prey cycles persisting for approximately 50 cycles and 300 predator generations were observed in a lab experiment on a planktonic predator–prey system which have been modeled through a stochastic time-delayed model in addition to the saturation mechanisms included in the Rosenzweig-MacArthur model \cite{Blasius2019}.
(Time delays, which are discussed in Sec.~\ref{time_delayed}, are another way of inducing nonreciprocity.)
With more species, like in Fig.~\ref{examples_nonvariational}b, the dynamics of coupled populations can lead to chaos \cite{Beninca2008}, as famously predicted by \citet{May1976}.

\subsubsection{Couplings between different fields of the same nature}
\label{coupling_same_kind_of_fields}

The example of preys and predators fits in a broader class where nonreciprocity takes place between two different fields at the macroscopic level.

As you see in Fig.~\ref{examples_nonvariational}c, oscillations can occur in spatially extended systems.
This figure shows snapshots of the Briggs–Rauscher reaction (see \citet{Kim2002} and references therein for details about this particular reaction), an oscillating chemical reaction in which the concentrations in various chemicals changes periodically in time.
The same kind of behavior found in consumer-resources models can be observed in simpler chemical systems, which can exhibit behaviors ranging from limit cycle oscillations \cite{Higgins1967,Nicolis1973,Field1989,Schnakenberg1979} to chemical chaos~\citet{Argoul1987,Scott1991}.
In fact, most of the early work on dissipative structures has been done in the context of chemical networks, see \citet{Nicolis1973,Epstein1998,Kuramoto1984,Haken1977,Manneville2014} for introductions.
Strictly speaking, limit cycles or chaotic attractors only occur when the system is maintained out of equilibrium, for instance with a chemostat, and this is what leads to a nonvariational dynamics. 
Even though the system eventually reaches equilibrium when the system is closed (like in Fig.~\ref{examples_nonvariational}c), this nonvariational effective dynamics often proves more instructive than the full one.

In this class of systems, nonreciprocity takes place between two different fields that have the same nature (the concentrations in two different chemical species).
The associated field theory is a spatially extended version of the local dynamics (which also follows the normal form Eq.~\eqref{normal_form_hopf_predator_prey}).
In this example, advection can be neglected, and the spatially extended system is described by a PDE of the form
\begin{equation}
    \partial_t \psi = (r + \ii \omega) \psi - \beta |\psi|^2 \psi + D \Delta \psi
    \label{normal_form_hopf_predator_prey_spatial}
\end{equation}
in which $\psi(t,\bm{r})$ is a complex field, basically containing the two nonreciprocally coupled chemical concentrations fields as real and imaginary parts.
Equation \eqref{normal_form_hopf_predator_prey_spatial} is often known as the complex Ginzburg-Landau equation \citet{Aranson2002,Kuramoto1984}.
Here, the system remains homogeneous on average, but this is not necessarily the case, as we discuss in Sec.~\ref{classification_instabilities}.

This also applies to generalized reaction networks including chemical reaction networks, biochemical regulation networks (Fig.~\ref{examples_nonvariational}D), but also purely physical systems and social opinion dynamics.
In all of these examples, the interactions correspond to repression or enhancement (depending on their sign).

It is also possible to have different kinds of interactions.
As an example, Fig.~\ref{examples_nonvariational}E shows a robotic realization of two nonreciprocally coupled XY spins, in which the interaction corresponds to alignment or antialignment, depending on the sign. 
The spins are programmed to approximate the equation of motion
\begin{equation}
    \partial_t \theta_a = \sum_j J_{ab} \sin(\theta_b - \theta_a)
\end{equation}
in which $\theta_a(t)$ is the angle made by the XY spin $a=1,2$ with a fixed direction, and where the coupling $J_{ab}$ may be asymmetric ($J_{ab} \neq J_{ba}$), manifesting nonreciprocity.
Mathematically, (anti)alignment may also describe synchronization and simple models of social dynamics.

\subsubsection{Couplings between fields of different kinds}
\label{coupling_different_kinds_of_fields}

A slightly different situation from Sec.~\ref{coupling_same_kind_of_fields} when a nonreciprocal coupling arises between fields that describe different kinds of quantities, for instance a scalar density and a vector velocity. 
In this case, one has to take care that the fields may have different physical dimensions and behaviors under time-reversal in order to assess the degree of symmetry of their coupling and whether it is compatible with detailed balance.

As an example, Fig.~\ref{examples_nonvariational}F shows a system where the mechanical properties of a biological tissue are coupled to activity of a signaling pathway known as ERK that regulates the cell cycle. This nonequilibrium mechanochemical coupling leads to active waves in the tissue that are believed to participate in biological function.
Similarly, self-propelled \cite{Marchetti2013} and self-aligning particles \cite{Baconnier2025} can exhibit nonreciprocal couplings between different fields.

For instance, consider a one-dimensional active Ornstein-Uhlenbeck particle (AOUP), described by the equations \cite{Fodor2016}
\begin{align}
    \dot{x} &= f_x(x,v) \equiv  - \mu \partial_x U + \ v \\
    \tau \dot{v} &= f_v(x,v) \equiv - v + \sqrt{2D} \eta(t)
\end{align}
in which $\mu$ is a mobility, $U$ an external potential, $\tau$ a time scale, and $\eta(t)$ a standard Gaussian white noise.
The dynamics is not volume-preserving ($f_v$ describe the relaxation to the fixed point $v=0$), and as $\partial_v f_x \neq \partial_x f_v$, the deterministic dynamics is not variational.
Computing the Jacobian
\begin{equation}
    J = \begin{pmatrix}
        - \mu U'' & 1 \\
        0 & -1/\tau
    \end{pmatrix}
\end{equation}
of the deterministic dynamics, we find that not only it is not symmetric, but that it is generically non-normal and that a careful choice of potential (such that $\tau \mu U'' = 1$) makes it reach an exceptional point.
Similar features can be found in other models of active particles~\cite{Romanczuk2012,Bechinger2016,Weis2025}, which also emerge as effective descriptions of phase transitions involving dynamic phases \cite{Suchanek2023a,Suchanek2023b}.

Similarly, consider an overdamped self-aligning particle (Fig.~\ref{examples_nonvariational}) in a harmonic trap described by the equations \cite{Dauchot2019,Baconnier2025}
\begin{subequations}
\label{self_align_potential}
\begin{align}
    \dot{x} &= \alpha \cos \phi - \beta x \\
    \dot{y} &= \alpha \sin \phi - \beta y  \\
    \dot{\phi} &= \gamma[ x \sin \phi - y \cos \phi ] 
\end{align}
\end{subequations}
in which $(x,y)$ is the displacement of the particle with respect to the bottom of the potential and $\phi$ the angle that the particle makes with a fixed axis.
In addition to self-propelling (with strength $\alpha$), the particle tends to align with its velocity with a self-alignment strength $\gamma$, while $\beta$ measures the stiffness of the potential. 
Again, Eq.~\eqref{self_align_potential} is non-variational and also not volume preserving. Contrary to the AOUP, it admits limit cycle solutions \cite{Dauchot2019}.
The Jacobian at a fixed point ($x=y=0$)
\begin{equation}
    J_0 = \begin{pmatrix}
        - \beta & 0 & 0 \\
        0 & - \beta & \alpha \\
        0 & - \gamma & 0
    \end{pmatrix}
\end{equation}
shows how self-propulsion ($\alpha$) and self-alignment ($\gamma$) can be seen as the reciprocal of each other.

\subsubsection{Couplings between different modes of the same field}
\label{modes_same_field}

It is not necessary to have multiple \emph{fields} to have nonreciprocal couplings. 
Nonreciprocity at the level of a single field can arise between different components of the same field (Sec.~\ref{components_same_field}), but it can also happen even with a scalar field through the coupling of different modes or harmonics of the field.

Perhaps the simplest example consists in the advection of a real-valued field $\phi(t,x)$ by a constant uniform velocity field $v_0$, described by the advection-diffusion equation
\begin{equation}
    \partial_t \phi = D \partial_x^2 \phi + v_0 \partial_x \phi
\end{equation}
in which $D$ is a diffusion coefficient. 
This equation, which also describes the macroscopic behavior of a biased random walk, is not variational.
Expanding $\phi(t,x) = \sum_k a_k(t) \sin(kx) + b_k(t) \cos(k x)$, we find that
\begin{equation}
\frac{d}{dt} \begin{pmatrix} a_k \\ b_k \end{pmatrix}
= 
\begin{pmatrix}
- D k^2 & - v_0 \\
v_0 & - D k^2
\end{pmatrix}
\begin{pmatrix} a_k \\ b_k \end{pmatrix}
\end{equation}
where the non-reciprocal coupling between $a_k$ and $b_k$, proportional to $v_0$, is apparent. 
Physically, this non-reciprocity manifests itself as an asymmetry in the evolution of $\phi$, where disturbances are propagated preferentially towards the right or towards the left, depending on the sign of $v_0$.
Mathematically, this phenomenon is related to the so-called non-Hermitian skin effect \citet{Ashida2020,Okuma2020,Bergholtz2021}.

Figure~\ref{examples_nonvariational}H shows an instance of nonreciprocal coupling between different harmonics that arises in metallurgy, in the context of directional interface growth.
In the experiment of Fig.~\ref{examples_nonvariational}H, a sample of liquid eutectic alloy \ce{CBr4-C2Cl6} is moved at constant speed in an imposed temperature gradient, leading to the solidification of the liquid~\cite{Ginibre1997}. 
Depending on the parameters, the interface may stay still (left panel) or move at constant velocity (right panel), either to the left or to the right, spontaneously breaking parity.
The shape of the liquid-solid interface in the comoving frame is described by single scalar field $u(t,x)$ decomposed as
\begin{equation}
    u(t, x) = A_1(t,x) \ee^{\ii (q_{\text{c}} x - \omega t)} + A_2(t,x) \ee^{\ii (2 q_{\text{c}} x - \omega t)} + \text{c.c.}
\end{equation}
which follows the amplitude equation \cite{Malomed1984,Douady1989,Coullet1989,Coullet1990,Fauve1991}
\begin{equation}
\label{amplitude_equation_interface_growth}
\begin{split}
	\partial_t A_1 &= \mu_1 A_1 - \overline{A_1} A_2 - \alpha |A_1|^2 A_1 - \beta |A_2|^2 A_1 \\
	\partial_t A_2 &= \mu_2 A_1 + \epsilon A_1^2 - \gamma |A_1|^2 A_2 - \delta |A_2|^2 A_2
\end{split}
\end{equation}
The coefficient of $\overline{A_1} A_2$ is set to $-1$ by rescaling, and the non-reciprocity between the different Fourier components is captured by the coefficient $\epsilon$, which must be positive for traveling patterns (left panel) to appear~\cite{Douady1989}.
The bifurcation between static and moving patterns can be identified as a drift-pitchfork bifurcation, for which the Jacobian is not diagonalizable at the bifurcation point (Sec.~\ref{bifurcations_eps}).
The same mechanism arises in completely different systems such as overflowing fountains, as illustrated on Fig.~\ref{examples_nonvariational}I: at the bifurcation, the liquid columns start moving either clockwise or counterclockwise (at random) along the edge of the fountain \citet{Counillon1997,Brunet2001}.

\subsubsection{Couplings between different components of the same field}
\label{components_same_field}

Nonreciprocity can also arise with a single field when it has multiple components; for instance, there can be a nonreciprocal coupling between two components of the velocity field $\bm{v}(t,x)$ when space dimension $d > 1$.
At first glance, this is similar to a coupling between two fields of the same kind (Sec.~\ref{coupling_same_kind_of_fields}). 
A distinction emerges when we take into account how the fields transform under change of coordinates: in this sense, a vector field or a tensor field is a single object with different components (this section), in contrast with a composite field made of several scalar fields concatenated (Sec.~\ref{coupling_same_kind_of_fields}).

As an example, the dynamics of the robotic solid in Fig.~\ref{examples_nonvariational}J is described by the odd elastodynamics equations \cite{Scheibner2020,Fruchart2023}
\begin{equation}
    \label{odd_elastic_dynamics}
    \partial_t \bm{u} = (\mu \bm{\delta} + \mu_{\text{o}} \bm{\epsilon}) \Delta \bm{u} + (B \bm{\delta} + B_{\text{o}} \bm{\epsilon}) \nabla(\nabla \cdot \bm{u})
\end{equation}
in which $\bm{u}(t,\bm{r})$ is the deformation field of the elastic solid, $\bm{\delta}$ is the identity and $\bm{\epsilon}$ the fully antisymmetric tensor, $\mu$ and $B$ are standard elastic moduli, while $\mu_{\text{o}}$ and $B_{\text{o}}$ are odd elastic moduli that arise because it is chiral and active --- they encode the violation of Maxwell-Betti reciprocity (Sec.~\ref{nonreciprocal_responses}) that make the dynamics \eqref{odd_elastic_dynamics} nonvariational.

\subsubsection{Driven-dissipative systems and nonreciprocity}
\label{driven_dissipative}

Nonreciprocity can be induced by gain and loss, a property that arises in systems ranging from hydrodynamic instabilities to open quantum systems.

This can be seen in the example of Fig.~\ref{examples_nonvariational}K, where two coupled ring-shaped laser cavities are considered, with one lossy cavity (blue) and one cavity with gain (red) \cite{Hassan2015}.
The evolution of the (nondimensionalized) complex amplitudes of light in each cavity $A_1$ and $A_2$ is described by
\begin{subequations}
\label{laser_equations}
\begin{align}
	\partial_t A_1 &= - \gamma A_1 + g_0 \, \frac{A_1}{1+|A_1|^2} + \ii A_2 \\
	\partial_t A_2 &= - \gamma A_2 - f_0 \, \frac{A_1}{1+|A_2|^2} + \ii A_1
\end{align}
\end{subequations}
The lasing transition occurs when the complex amplitudes $A_a$ acquire a finite value. 
This primary bifurcation corresponds to the spontaneous breaking of the $U(1)$ symmetry $A_a \to \ee^{\ii \theta} A_a$ of Eq.~\eqref{laser_equations} (see e.g. Ref.~\cite{DeGiorgio1970,Graham1970,Haken1975,Gartner2019}).
The resulting lasing state is a fixed point of Eq.~\eqref{laser_equations} corresponding to oscillations at a frequency $\omega_0$ included in the modulated electric field (not the amplitudes).
A secondary drift-pitchfork bifurcation then occurs from this state, leading to two branches of limit cycles, that describe oscillations at frequencies $\omega_0 \pm \Delta \omega$, where $\Delta \omega$ undergoes a pitchfork-like behavior through the bifurcation.
Experimentally, this corresponds to a transition from one to two peaks in the emission spectrum of the laser, as shown in Fig.~\ref{examples_nonvariational}K \cite{Hassan2015}. 
Note that both peaks appear at the same time because noise continuously makes the system jump between the two limit cycles, effectively restoring the broken chiral symmetry on average.

Another example is shown in Fig.~\ref{examples_nonvariational}L in which coupled parametric micromechanical oscillators implement asymmetrically coupled Ising spins \cite{Han2024}. 

\begin{figure}
    \centering
    \includegraphics[width=\linewidth]{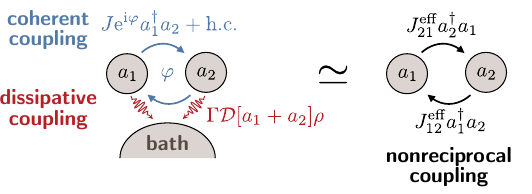}
    \caption{\textbf{Simple model of quantum nonreciprocity.}
    Two bosonic modes with annihilation operators $a_1$ and $a_2$ are coupled through a coherent (Hamiltonian) coupling $H = J \ee^{\ii \varphi} a_1^\dagger a_2$ (in blue) and are also connected to a common bath, modeled through a Lindbladian dissipative coupling with rate $\Gamma$ (in red).
    The phase $\varphi$ can be thought as a Peierls phase due to a magnetic field, and it breaks time-reversal invariance when nonzero.
    This system is effectively equivalent to a nonreciprocal coupling when $\varphi,\Gamma \neq 0$.
    Adapted from \citet{Metelmann2015,SuarezForero2025}.
    }
    \label{figure_quantum_nr}
\end{figure}

\subsubsection{Quantum nonreciprocity}

The same strategy also applies to quantum systems, with some additional subtleties. For instance, \citet{Clerk2022} reviews how nonreciprocity emerges from the balance between coherent (Hamiltonian) interactions and incoherent (dissipative) interactions. 
In order to properly take these into account, it is required to describe the system through a density matrix evolved by a quantum master equation (Sec.~\ref{open_quantum_systems}). 
Nevertheless, the classical limit of these systems tends to produce nonvariational dynamics similar to that of classical nonreciprocal systems.

This can be illustrated with a simple model from \citet{Metelmann2015}, illustrated in Fig.~\ref{figure_quantum_nr}, in which two bosonic modes $a_1$ and $a_2$ are coupled through a Hamiltonian coupling $H = J \ee^{\ii \varphi} a_1^\dagger a_2$ and are also coupled to a common bath through a dissipative coupling implemented $\Gamma \mathcal{L}[a_1 + a_2] \rho$ where $\mathcal{D}[L]$ is the standard dissipative superoperator associated to the jump operator $L$, defined by $\mathcal{D}[L] \rho = L \rho L^\dagger - 1/2[L L^\dagger \rho + \rho L L^\dagger]$ and entering the quantum master equation.
The equations of motion for the average values $\langle a_1\rangle$ and $\langle a_2\rangle$ are then given by \cite{Metelmann2015,SuarezForero2025}
\begin{subequations}
\label{quantum_nr_toy_model}
\begin{align}
    \frac{d \langle a_1\rangle}{dt} &= - \frac{\Gamma}{2} \langle a_1\rangle - J^{\text{eff}}_{12} \langle a_2\rangle
    \\
    \frac{d \langle a_1\rangle}{dt} &= - \frac{\Gamma}{2} \langle a_2\rangle - J^{\text{eff}}_{21} \langle a_1\rangle
\end{align}
where
\begin{align}
J^{\text{eff}}_{12} = \frac{\Gamma}{2} + \ii J \ee^{\ii \varphi}
\qquad
\text{and}
\qquad
J^{\text{eff}}_{21} = \frac{\Gamma}{2} + \ii J \ee^{-\ii \varphi}
\end{align}
\end{subequations}
which are in general different. 
In particular, by setting $\varphi=\pi/2$ and $J=\Gamma/2$, we can have $J^{\text{eff}}_{12} = 0$ and $J^{\text{eff}}_{21} = \Gamma \neq 0$, which is manifestly nonreciprocal (unidirectional in the classification of Fig.~\ref{classes_of_nonreciprocity}; the other classes can be obtained by changing the parameters). 
This particular case is related to the notion of cascaded quantum systems, that are by definition essentially one-way.
In Eq.~\eqref{quantum_nr_toy_model}, breaking time-reversal invariance is crucial to break reciprocity, but other schemes that do not require time-reversal to be broken have been proposed \citet{Wanjura2023,Wang2023}.
We refer to \citet{Clerk2022} for details, to \citet{SuarezForero2025} for a perspective from chiral quantum optics. 
See also \cite{Brighi2025,Soares2025} for the case of fermions.
A review of recent progress on quantum nonreciprocity is provided by \citet{Barzanjeh2025}.

In this context, the notion of quantum limit cycles and of quantum synchronization, conceived as quantum analogues of the classical notions, have emerged and are under investigation \citet{Lee2013,Lee2014,Walter2014,Walter2014b,Roulet2018,Buca2022,Nadolny2023,Dutta2025,Nadolny2025b}.
For instance, Fig.~\ref{examples_nonvariational}M shows the Wigner function $W(x,p)$ at different times (top row) as a quantum van der Pol oscillator realized with trapped ions evolves towards a limit cycle state \cite{Liu2025b}. 
This system is roughly described by the master equation (in a rotating frame) \cite{Lee2013,Walter2014}
\begin{equation}
    \frac{d \rho}{dt} = - \ii [H,\rho] + \gamma_1 \mathcal{D}[b^\dagger] \rho + \gamma_2 \mathcal{D}[b^2] \rho
\end{equation}
for a bosonic mode $b$, where $H = - \gamma b^\dagger b + \ii \Omega (b + b^\dagger)$. 
To a first approximation, this system can be seen as a noisy classical limit-cycle oscillator: the classical dynamical system describing the average $z(t) = \langle b \rangle$ is
\begin{equation}
    \dot{z} = (\gamma_1/2 + i \Delta) z - \gamma_2 |z|^2 z  - \Omega
\end{equation}
in which we recognize the normal form \eqref{hopf_normal_form} of a Hopf oscillator, up to a rescaling and a constant term $-\Omega$ due to the rotating frame. 
Note that here, the nonreciprocity takes place within a single mode.
Going beyond this classical picture, \citet{Liu2025b} also implement mutual synchronization between two quantum oscillators, and report that the synchronized dynamics can be observed through a joint measurement of both oscillators, but not from any local measurement.
Many-body effects are discussed in Sec.~\ref{what_are_consequences}, see \citet{Fazio2025,Sieberer2025} for reviews.

\subsubsection{Time-delayed interactions and causality}
\label{time_delayed}

In certain situations, the dynamics does not follow an equation of the form \eqref{ds} [or \eqref{coupled_langevin}], but instead includes time-delayed interactions~\cite{Zwanzig1961,Zwanzig2001,Smith2010}.
A system with this kind of coupling is called non-Markovian.
In a deterministic system, the ODEs in Eq.~\eqref{ds} are replaced with delay differential equations of the form
\begin{equation}
    \frac{d \vec{x}(t)}{d t} = \vec{f}\left[ \{x(t-\tau)\}_{\tau \geq 0} \right]
    \label{delay_eom}
\end{equation} 
in which the rate of change at time $t$ can depend on the entire past trajectory of the system.
Because of causality, it does not depend on the future, an asymmetry that can be exploited to produce nonreciprocity.
Delays can similarly be introduced in a Fokker-Planck equation \cite{Guillouzic1999,Frank2005,Frank2005b}.
Non-Markovian systems typically arise when the system maintains some memory about its past~\cite{Zwanzig1961,Keim2019}, which is stored in unmonitored (hidden) degrees of freedom.
From a mathematical perspective, it means that we have integrated out some degrees of freedom from a more fundamental Markovian description. 
Conversely, it is possible to \enquote{Markovianize} a non-Markovian system by introducing auxiliary degrees~\cite{Vogel1965,Fargue1974,Smith2010,Kondrashov2015}.

\citet{Loos2019,Loos2020} discusses the relations between non-reciprocal and delayed interactions as well as their thermodynamic implications.
In particular, they show that in linear systems, non-reciprocal couplings can lead to non-monotonic memory kernels when some degrees of freedom are traced out.

To see how delays can induce non-reciprocity, let us consider the underdamped dynamics of two degrees of freedom $X=(x,y)$ in the potential
\begin{equation}
    V(x,y) = \alpha \frac{x^2}{2} + \gamma x y + \beta \frac{y^2}{2}
\end{equation}
and let us artificially introduce a delay $\tau \geq 0$ in the couplings, leading to
\begin{subequations}
\begin{align}
    \dot{x} &= - \alpha x - \gamma y(t-\tau) \\
    \dot{y} &= - \beta y - \gamma x(t-\tau).
\end{align}
\end{subequations}
When $\tau = 0$, we recover the usual overdamped dynamics.
In the limit $\tau \to 0$, we can expand at first order
\begin{equation}
    X(t-\tau) = [e^{- \tau (d/dt)} X](t) \simeq X(t) - \tau \dot{X}(t)
\end{equation}
and reorganize the equations to get
\begin{subequations}
\begin{align}
    \dot{X} = 
    \frac{1}{\gamma ^2 \tau ^2-1}
    \begin{pmatrix}
         \alpha +\gamma ^2 \tau  & \beta  \gamma  \tau +\gamma  \\
 \alpha  \gamma  \tau +\gamma  & \beta +\gamma ^2 \tau  \\
    \end{pmatrix}
    X
\end{align}
\end{subequations}
in which the coupling matrix is indeed asymmetric.

\subsubsection{Systems violating Newton third law}

Mechanical systems violating Newton's third law, that are discussed in Sec.~\ref{violations_newton_third_law}, are usually nonvariational.
To see that, let us consider the overdamped limit of Eq.~\eqref{example_nr_newton}, that we can formally obtain by setting $m = 0$ in the equation, leading to
\begin{subequations}
\label{example_nr_newton_overdamped}
\begin{align}
    \dot{x}_1 &= f_1(x_1, x_2) \equiv k_{12}/\gamma \, (x_2 - x_1) \\
    \dot{x}_2 &= f_2(x_1, x_2) \equiv k_{21}\gamma \, (x_1 - x_2)
\end{align}
\end{subequations}
which is not variational as $\partial_{x_1} f_2 \neq \partial_{x_2} f_1$ when $k_{12} \neq k_{21}$.
This tends to induce run and chase dynamics (that may or may not be limit cycles) as well as oscillations in the overdamped regime, as shown in Fig.~\ref{violations_newton_third_law}E,G,I,I',K,K'.

\subsection{Violations of Newton third law}
\label{violations_newton_third_law}

Examples of interactions that violate Newton third law abund in nature and include virtually all field-mediated interactions, such as
\begin{itemize}[nosep,left=0pt,label=--]
    \item electromagnetic interactions between moving charges \cite{Goldstein2002} \item hydrodynamic interactions between particles in a fluid \cite{Happel1983,Kim1991} and \citet{Beatus2006,Guillet2025}
    \item acoustic interactions  \cite{Wu2025,King2025,StClair2023,Morrell2025}
    \item catalytically active chemotactic colloids \cite{Soto2014,Saha2019,Niu2018,Sengupta2011,Usta2008,Wu2021} or light-activated active colloids \cite{Schmidt2019,Codina2022}  \item vapour-mediated interactions between water droplets \cite{Cira2015} or micelle-mediated interactions between oil droplets \citet{Meredith2020}
\item wake-mediated interactions in complex plasma \citet{Chaudhuri2011,Ivlev2015,Hebner2003,Melzer1999} \item Casimir and depletion forces \cite{Dzubiella2003,Buenzli2008}
    \item optics \cite{Sukhov2015,LuisHita2022,Rieser2022,Reisenbauer2024,Yifat2018,Peterson2019}
\end{itemize}
They also include more complex interactions mediated by an non-equilibrium environment such as social forces \cite{Helbing2001} in humans or animals (pigeons \cite{Nagy2010} or penguins \citet{Zampetaki2021}).
Some of the fundamental consequences of interactions violating Newton's third law are discussed by \citet{Poncet2022,Ivlev2015,Kryuchkov2018,Saha2020,You2020}.

The effective lack of linear momentum conservation (that is, the fact that the momentum of the particles is not conserved) is possible because some momentum has been exchanged with the field mediating the interaction.
The overall conservation of momentum always holds when the momentum of the field itself is taken into account; in practice, doing so requires some care and we refer to \citet{Pfeifer2007,Bliokh2025,Stone2002,Maugin2015,Maugin1993} for details and references.

\begin{figure*}
    \centering
    \vspace{-0.75cm}
    \includegraphics[width=\linewidth]{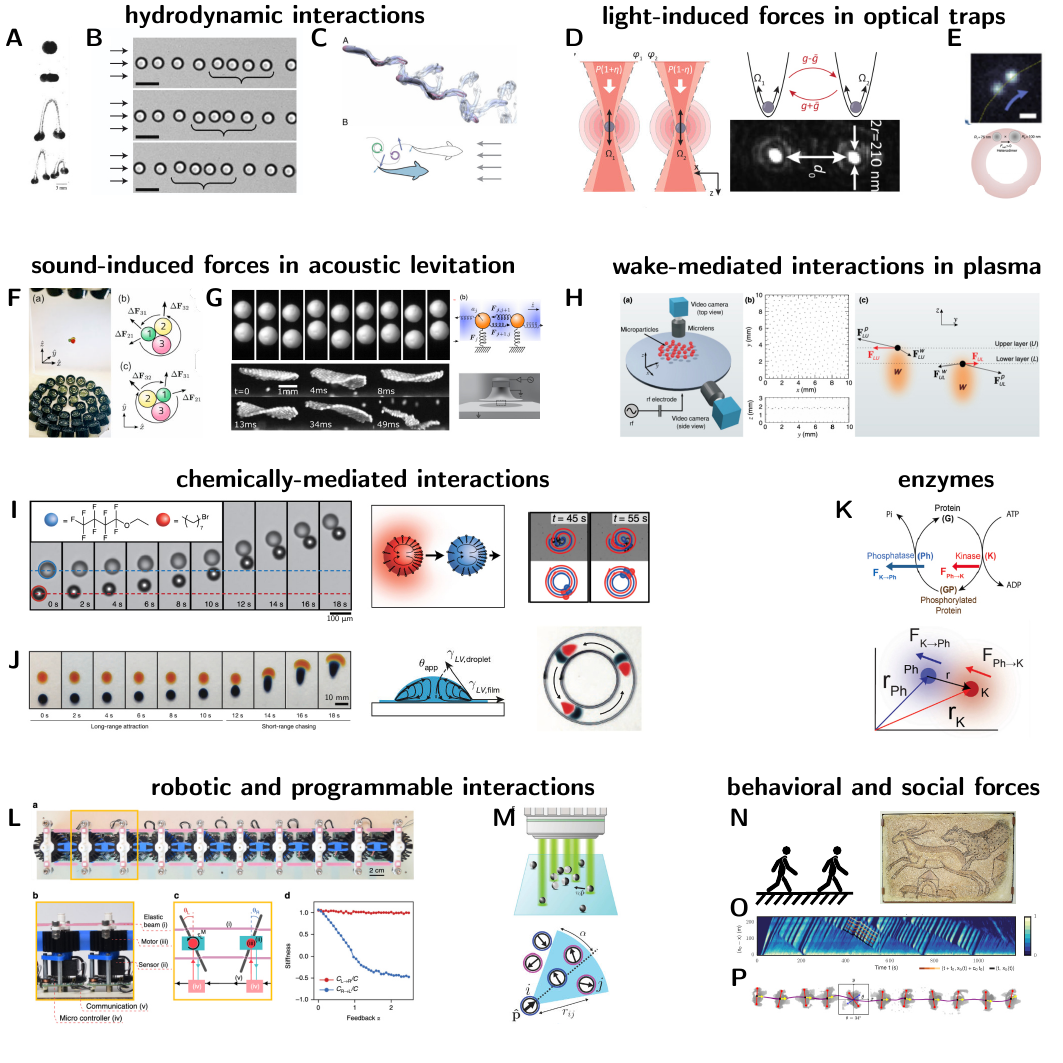}
    \vspace{-1.25cm}
    \caption{
    \label{examples_newton_third_law}
    \textbf{Violations of Newton's third law.}
	(A) Sedimentation of a cloud of glass beads in silicon oil, exhibiting complex instabilities. The interaction between the different particles is not reciprocal. Adapted from \citet{Metzger2007}.
	(B) The interaction between droplets of immiscible fluid into a liquid-filled microfluidic channel is not reciprocal because of the underlying mean flow (arrows). This leads to asymmetric wave propagation in microfluidic crystals made of many droplets, as well as instabilities. Adapted from \citet{Beatus2006}.
	(C) The interaction between fishes swimming in water, mediated by the fluid flow, is not reciprocal: the front fish affects the back fish more than the converse. This interaction is mostly mediated by vortices, as shown in the figure. Adapted from \citet{Verma2018,Li2020b}.
	(D) Transverse light-induced forces between optically levitated nanoparticles are not reciprocal. Adapted from \citet{Rieser2022}.
	(E) Nonreciprocal light-induced forces in a ring-shaped optical trap leads to run-and-chase motion. Adapted from \citet{Yifat2018}.
	(F-G) Interactions between acoustically levitated millimeter-scale (expanded polystyrene or polyethylene) particles is nonreciprocal. 
    The interaction can be multi-body (panels F and G, bottom). In the two-sphere system of panel G, a stable limit cycle was reported.
    Adapted from \citet{King2025} (F), \citet{Morrell2025} (G, top), \citet{Lim2022} (G, bottom).
	(H) In complex dusty plasma, wake-mediated interactions are nonreciprocal. The wake (orange) coming from the ion wind is directional, so particles on the bottom layer are influenced more by particles on the upper layer than conversely. Adapted from \citet{Ivlev2015}.
	(I) The micelle-mediated interactions between droplets made from different oils in water is nonreciprocal, because of the solubility difference between the oils. The red oil droplet chases the blue oil droplet. Clusters from local sticky interactions may undergo circular motion.
Adapted from \citet{Meredith2020}.
	(J) The vapor-mediated interactions between droplets made with different water-propylene mixtures is nonreciprocal. This can lead to run-and-chase motion. Adapted from \citet{Cira2015}.
	(K) The interaction between enzymes such that the substrate of one enzyme is the product of the other enzyme and vice-versa is nonreciprocal. The mechanism is similar to that of panels I-J. Adapted from \citet{Mandal2024}.
	(L) Robotic mechanical interactions can be designed to be nonreciprocal. Here, a chain of active rotors implements a nonreciprocal elastic solid. Adapted from \citet{Brandenbourger2019}.
	(M) Other programmable interactions, here implementing a cone of vision, here in carbon-coated silica Janus particles, can be nonreciprocal. Adapted from \citet{Lavergne2019}.
	(N-P) Social and behavioral forces are typically nonreciprocal. This is for instance the case between a predator and a prey, but also between identical pedestrians (because of a vision cone; panel P), and so on.
	Consequences can be seen in a kymograph of the longitudinal velocity (panel O), of a crowd in the Chicago Marathon, showing directional velocity waves.
    Adapted from \citet{Gu2025} (N, left); Mosaic (Roman, Homs, Syria, 450-462 AD), Chazen Museum of Art, photography Daderot on Wikimedia under license CC0 (N, right),
\citet{Bain2019} (O),
     \citet{Willems2020} (P).
}
\end{figure*}

\subsubsection{Electromagnetism}
\label{EM}

It is a testament to physics curricula that Newton's third law is so often taken as a universal truth even though a cursory glance at \citet{Feynman1989}'s lectures (\S~26) provides a counterexample.
Let us follow Feynman and consider two charged particles moving with fixed velocities $\bm{v}_1$ and $\bm{v}_2$ at right angle, positioned so that they don't collide, as shown in the following schematic:
\begin{center}
    \includegraphics[width=0.9\columnwidth]{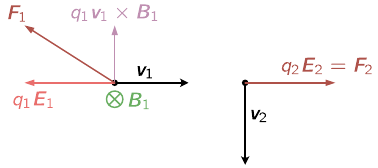}
    \label{EMschematic}
\end{center}
If a particle $i$ is not too fast with respect to the speed of light, the electric field $\bm{E_i}$ it generates is given by Coulomb's law and the magnetic field by $\bm{B_i} = \bm{v_i} \times \bm{E_i}/c^2$.
While the electric parts of the Lorentz force $\bm{F} = q(\bm{E} + \bm{v} \times \bm{B})$ are equal and opposite, this is not the case of the magnetic forces, as shown on the schematic.

One could hope to \enquote{fix} this issue by using a more accurate description, but this would only make things worse: the reason why the principle of action and reaction is broken (as explained by \citet{Feynman1989} in \S~27) is that the electromagnetic field carries momentum. As mentioned above, this requires some care, especially in matter, where the existence of subtleties is referred to as the Abraham–Minkowski controversy \cite{Pfeifer2007}.

\subsubsection{Hydrodynamic interactions}
\label{hydroint}

Consider a set of identical particles sedimenting in a viscous fluid under the action of gravity.
All particles are subject to the same external force $\bm{F}_\text{ext} = - m g \hat{z}$ where $m$ is the mass of a particle and $g$ the local acceleration of gravity.
At low Reynolds numbers, and in the comoving frame falling at the settling velocity $\bm{v}_0 = \bm{F}_\text{ext}/[6 \pi \eta a]$ of a single isolated particle with radius $a$ in the fluid with kinematic viscosity $\nu$, the dynamics of the positions $\bm{x}_i$ of the particles $i$ is described by
\begin{equation}
    \frac{d \bm{x}_i}{dt} = \sum_{j \neq i}
    \bm{f}_{j \to i}(\{\bm{x}_\ell\}_\ell)
\label{stokesian_dynamics}
\end{equation}
in which $\bm{f}_{j \to i} = \bm{f}(\bm{x}_i - \bm{x}_j) \equiv 1/[8\pi \eta] G(\bm{x}_i, \bm{x}_j) \bm{F}_\text{ext}$ where
$G(\bm{x}_i, \bm{x}_j) = G(\bm{x}_i - \bm{x}_j)$ is the Green function of the Stokes equation (called the Oseen tensor), given by
\begin{equation}
    G_{ij}(\bm{r}) = \frac{1}{r} \delta_{ij} + \frac{1}{r^3} r_i r_j
\end{equation}
in a standard fluid, where $r = \lVert \bm{r} \rVert$.
From this expression, you can see that even for two particles $\bm{f}(\bm{r}) \neq - \bm{f}(-\bm{r})$, and so $\bm{f}_{j \to i} \neq \bm{f}_{i \to j}$, namely the hydrodynamic interaction does not satisfy Newton's third law, except when $\bm{F}_\text{ext} = 0$.

Physically, this arises from the fact that (i) linear momentum is exchanged between the particles and the fluid, (ii) the rate of exchange depends on the configuration of the particles.
Linear momentum conservation holds for the system composed of the particles plus the fluid, but not for the effective dynamics \eqref{stokesian_dynamics}. This is manifested both in the fact that the dynamics is overdamped (this corresponds to i), and by the fact that interactions do not satisfy Newton's third law (due to i and ii).
In addition, it is crucial that there is an external drive ($\bm{F}_\text{ext}$) that puts the system out of equilibrium.

The complexity of hydrodynamic interactions, which can already be illustrated by Eq. \eqref{stokesian_dynamics} (for instance, three particles exhibit a chaotic saddle \cite{Janosi1997}) underlies the complex statistical physics of sedimentation and dense suspensions \cite{Ramaswamy2001,Ness2022} which ranges from reversible-irreversible transitions  \cite{Pine2005,Menon2009} to clumping instabilities \cite{Crowley1971,Lahiri1997,Lahiri2000}.

For instance, Fig.~\ref{examples_newton_third_law}A shows the experimental observation of the breakup of a cloud of sedimenting particles, that is reproduced by Eq. \eqref{stokesian_dynamics}, as discussed by 
\citet{Metzger2007}.
Additional subtleties arise when the particles are not spherical: in the case of disks, for instance, the orientational degrees of freedom have to be taken into account and instability from non-normal amplification of perturbations \cite{Chajwa2020} (see Fig.~\ref{figure_nr_particle_systems}B).

As an aside, note that the Oseen tensor of a standard fluid satisfies Eq.~\eqref{symmetry_green_function_physical_space}, which can be traced to the symmetry of the viscosity tensor. 
This is usually known as a manifestation of \enquote{Lorentz reciprocity}, as discussed in detail by \citet{Masoud2019}.
This illustrates that there is no systematic relation between Lorentz reciprocity and Newton's third law.

Other flows may yield different interactions.
Figure~\ref{examples_newton_third_law}B shows water
droplets in an oil-filled channel that interact through non-reciprocal dipolar interactions stemming from the externally imposed flow in the channel, eventually leading to asymmetric wave propagation in the system \cite{Beatus2006}.

Hydrodynamic interactions can lead to non-reciprocal effective forces in a variety of contexts also beyond Stokes flows; for instance, fishes following each other are believed to interact through vortices induced by their swimming  \cite{Verma2018}, as illustrated in Fig.~\ref{examples_newton_third_law}C.
In active fluids, more elaborate fluid-mediated interactions can also emerge \cite{Baek2018,Khain2024}.

\subsubsection{Light-induced forces}

Optical fields can induce forces, known as optical binding forces, between microscopic or nanoscopic particles~\cite{Dholakia2010}.
For instance, this arises for particles in optical traps.
It turns out that these light-induced forces can be non-reciprocal \cite{Sukhov2015}.
Consider for instance two spherical particles in optical traps as in \citet{Rieser2022,Reisenbauer2024}.
In addition to the trapping force and radiation pressure, which arise only from the trapping field, the optical binding force on particle $i$ is then given by
\begin{equation}
    \label{optical_force}
    \bm{F}_i = \text{Re} \sum_{j \neq i} \nabla_{\bm{r}_i}
    \left(
    \frac{\alpha_i \alpha_j}{2 \epsilon_0}
    \bm{E}_0^*(\bm{r}_i)
    G(\bm{r}_i - \bm{r}_j)
    \bm{E}_0(\bm{r}_j)
    \right)
\end{equation}
in which $\bm{r}_j$ is the positions of particle $j$, $\alpha_j$ its polarizability, $\epsilon_0$ the polarizability of vacuum, $G(\bm{r}) = G^T(\bm{r})$ is the (tensor) Green function of the transverse Helmholtz equation, and $\bm{E}_0$ is the external electric field. 
Equation \eqref{optical_force} suggests that a non-reciprocal interaction between $i$ and $j$ may arise when $\alpha_{i} \neq \alpha_{j}$, like with particles of different sizes \cite{Sukhov2015,Karasek2017} or when the background field is not the same at $\bm{r}_i$ and $\bm{r}_j$.
In the experimental setup of \citet{Rieser2022} shown in Fig.~\ref{examples_newton_third_law}D, the two identical particles are also trapped by the optical traps, and the linearized equations of motion about their equilibrium position turns out to be
\begin{subequations}
\begin{align}
    \ddot{z}_1 + \gamma \dot{z}_1 &= - (\Omega_1^2 + \kappa_+ + \kappa_-) z_1 + (\kappa_+ + \kappa_-) z_2
    \\
    \ddot{z}_2 + \gamma \dot{z}_2 &= - (\Omega_2^2 + \kappa_+ - \kappa_-) z_2 + (\kappa_+ - \kappa_-) z_1
\end{align}
\end{subequations}
in which $z_i(t)$ is the vertical position of the particle $i$, $\gamma$ represents damping, $\Omega_i$ are the eigenfrequencies of the isolated harmonic traps, and $\kappa_{\pm}$ respectively represent the conservative and nonconservative parts of the couplings arising from the transverse optical binding force.
Here, the nonreciprocity comes from the difference in optical phase at the positions of the two particles.

As shown in Fig.~\ref{examples_newton_third_law}E, light-mediated interactions can also lead to a run and chase dynamics when the particles are not trapped (here, they are confined into an annulus) \citet{Yifat2018,Peterson2019}.
Similarly, light can induce nonreciprocal interactions between different parts of a  nanostructure \cite{Zhao2010,Zhang2014}, and this can lead to synchronized limit-cycle oscillations in nanoscale mechanical oscillators \cite{Liu2023,Liu2025}.
See also \citet{Parker2025} for nonpairwise forces (Sec.~\ref{beyond_pairwise}).
\citet{LuisHita2022} discusses the case of random electromagnetic fields, and show theoretically that in an homogeneous and isotropic random electromagnetic field, the interaction between two particles with different absorption cross-sections can have nonreciprocal interactions.

\subsubsection{Sound-induced forces}

In the same way as light, sound can induce nonreciprocal forces.
Figure~\ref{examples_newton_third_law}F shows and experimental setup from \citet{King2025} where three expanded polystyrene particles levitated in an acoustic trap exhibit three-body nonreciprocal interactions (Sec.~\ref{beyond_pairwise}).
A different setup involving similar levitated particles shown in Fig.~\ref{examples_newton_third_law}G shows oscillations due to two-body nonreciprocal interactions.

\subsubsection{Complex and dusty plasma}

Figure \ref{examples_newton_third_law}H shows a complex plasma in which microparticles levitate above a flat electrode. 
The particles may levitate at two different heights, and so the plasma can be seen as a binary mixture. In addition to direct interactions between these particles, the ion flow due to the electrode leads to a wake downstream of each particles (orange clouds in the figure), leading to wake-mediated interactions that are not reciprocal \cite{Chaudhuri2011,Hebner2003,Melzer1999}. \citet{Ivlev2015} analyzed theoretically and experimentally the consequences of such interactions on the statistical mechanics of collections of particles.

\subsubsection{Chemical signaling}

Figure \ref{examples_newton_third_law}I shows the behavior of two droplets made with different oils (red and blue) and suspended in water: the interaction between the droplets is nonreciprocal, and the red droplet chases the blue one \cite{Meredith2020}.
In short, this arises from the coupling micelle-mediated oil
transport with Marangoni flows: the red oil is preferentially solubilized by micelles in water (with respect to the blue one). The resulting asymmetry in the solubilized oil leads to a gradient of surface tension that in turn produces flows inside the droplet through the Marangoni effect, leading to self-propulsion \cite{Michelin2023}.

This illustrates a more general mechanism delineated by \citet{Soto2014,Golestanian2007}, who focused on phoretic interactions between catalytically active colloids.
In short, two ingredients are at play.
First, the particles produce or consume a scalar field (for instance, catalytic colloids produce or consume a chemical), leading to a gradient of the scalar field.
Second, the particle move in response to the gradient of the scalar field through phoretic effects (see \citet{Anderson1989,Moran2017,Illien2017,Michelin2023} for reviews).
In the setup considered by \citet{Soto2014}, this leads to the overdamped equations of motion
\begin{equation}
   \frac{d\bm{r}_{i}}{dt} = \sum_{k\neq i} \frac{\alpha_{k}\mu_{i}R^{2}}{6\pi D} \frac{\bm{r}_{ki}}{|\bm{r}_{ki}|^{3}}
   + \bm{f}_{\text{rep},i}
   + \bm{\xi}(t) 
\end{equation}
for the positions $\bm{r}_{i}(t)$ of the particles or radius $R$, in which the surface activity $\alpha_i$ describes the production/consumption of the chemical, the surface mobility $\mu_i$ describes the strength and direction of the phoretic effect, while $D$ is the effective diffusion coefficient of the chemical, $\bm{f}_{\text{rep},i}$ is a repulsive force (satisfying Newton's third law) and $\bm{\xi}$ a noise.
Notably, $\alpha_i$ and $\mu_i$ can have arbitrary signs.
The physical consequences of this kind of interactions are discussed in \cite{Soto2014,Saha2019,Niu2018,Sengupta2011,Usta2008,Wu2021}, see also \cite{Schmidt2019,Codina2022} for light-activated active colloids.
Many-body effects are reviewed in Sec.~\ref{what_are_consequences}.

Figure \ref{examples_newton_third_law}J shows a similar effect at a different scale where droplets made of water/propylene glycol mixes with different proportions chase each other through vapour-mediated interactions between water droplets \cite{Cira2015}.
Similar effects are expected in model biomolecular condensates \cite{JambonPuillet2024}.
On yet another scale, Fig.~\ref{examples_newton_third_law}K illustrates a similar mechanism at the scale of single molecules described by \citet{Mandal2024}, which rests on parallels between the behavior of enzymes and active matter at larger scales \cite{Ghosh2021}.  \citet{Mandal2024} theoretically consider two enzymes $E_1$ and $E_2$ in solution (\ce{Ph} and \ce{K} in the figure), such that the product of the reaction catalyzed by $E_2$ is the substrate of the reaction catalyzed by $E_1$, and conversely. 
These are put out of equilibrium by a mechanism fixing the concentrations of certain chemicals (here ATP, ADP, and P).
This yields to a gradient in the catalyzed chemicals, which have been observed to induce chemotactic motion, although the underlying mechanisms are still debated \cite{Feng2020,Zhang2021,Mandal2023}, and, for the same reason as before, this leads to interactions violating Newton's third law.

\subsubsection{Robotic and programmable interactions}

Perhaps the most direct way to experience non-reciprocal forces at human scales are through robotic systems where the forces are the product of sensing, computation, and actuation.
Figure \ref{examples_newton_third_law}L shows an example where a one-dimensional chain of particles of mechanically coupled units composed of a sensor, a microcontroller, and a motor implement nonreciprocal interactions \citet{Brandenbourger2019}.
In this work, the behavior of the chain is described by the mass-and-spring model
\begin{equation}
    m \frac{d^2 u_i}{dt^2} = F_{i-1 \to i} + F_{i+1 \to i}
\end{equation}
in which $u_i(t)$ is the displacement of the particle $i$ along the direction of the chain, while
\begin{equation}
    F_{i \pm 1 \to i} = k(1\mp \varepsilon) [u_i - u_{i\pm1}].
\end{equation} 
In the continuum, this leads to the continuum equation $\partial_t^2 u - c^2 \partial_x^2 u + 2 \epsilon c^2/a \partial_x u$ for the deformation field $u(t,x)$, in which $c = a \sqrt{k/m}$ and $a$ is the lattice spacing. 
This equation, which exhibits a coupling between different modes (Sec.~\ref{modes_same_field}), shows that waves going to the left are amplified while waves going to the right are damped (or conversely, depending on the sign of $\varepsilon$), a phenomenon known as the non-Hermitian skin effect \citet{Brandenbourger2019} that is analogous to our discussion of biased diffusion in Sec.~\ref{modes_same_field}, but here the density of diffusing particles is replaced by linear momentum.

Other kinds of programmable interactions can be used to generate nonreciprocal forces: for instance, \citet{Lavergne2019} controlled the motilities of light-activated active particles with an external feedback loop in order to implement vision-cone interactions, as shown in Fig.~\ref{examples_newton_third_law}M.

\subsubsection{From traffic to behavioral and social forces}

As they move in space, humans, animals, and other similarly complex systems have no reason to satisfy Newton's third law, because their motion is the combination of self-propulsion with behavior (Fig.~\ref{examples_newton_third_law}N). 
For instance, if we were to model a predator chasing a prey by an interaction law, it would not be reciprocal.
In practice, animal behavior (including humans) may be arbitrarily complex, so a picture in terms of physics-like forces is simply not suitable. 
(\citet{Riiska2024} reviews animal behavior from the perspective of physics.)
Nevertheless, in select situations, an effective description in terms of forces can be fruitful.
For instance, Fig.~\ref{examples_newton_third_law}O shows  a kymograph of the longitudinal velocity of runners in the Bank of America Chicago Marathon, where velocity waves propagate upstream at the same speed \cite{Bain2019} while Fig.~\ref{examples_newton_third_law}P shows the orientation dynamics of pedestrians in a lab environment \cite{Willems2020}. 
In both cases, the humans can only see in front of them, and hence any interaction is nonreciprocal.

More details can be found in the following places: \citet{Helbing2001} reviews the dynamics of car and pedestrian traffic and similar situation, \citet{Jusup2022} reviews social dynamics including small and large scale motion, and \citet{Corbetta2023} reviews the physics of human crowds.

\begin{figure*}
    \centering
    \includegraphics[width=0.95\linewidth]{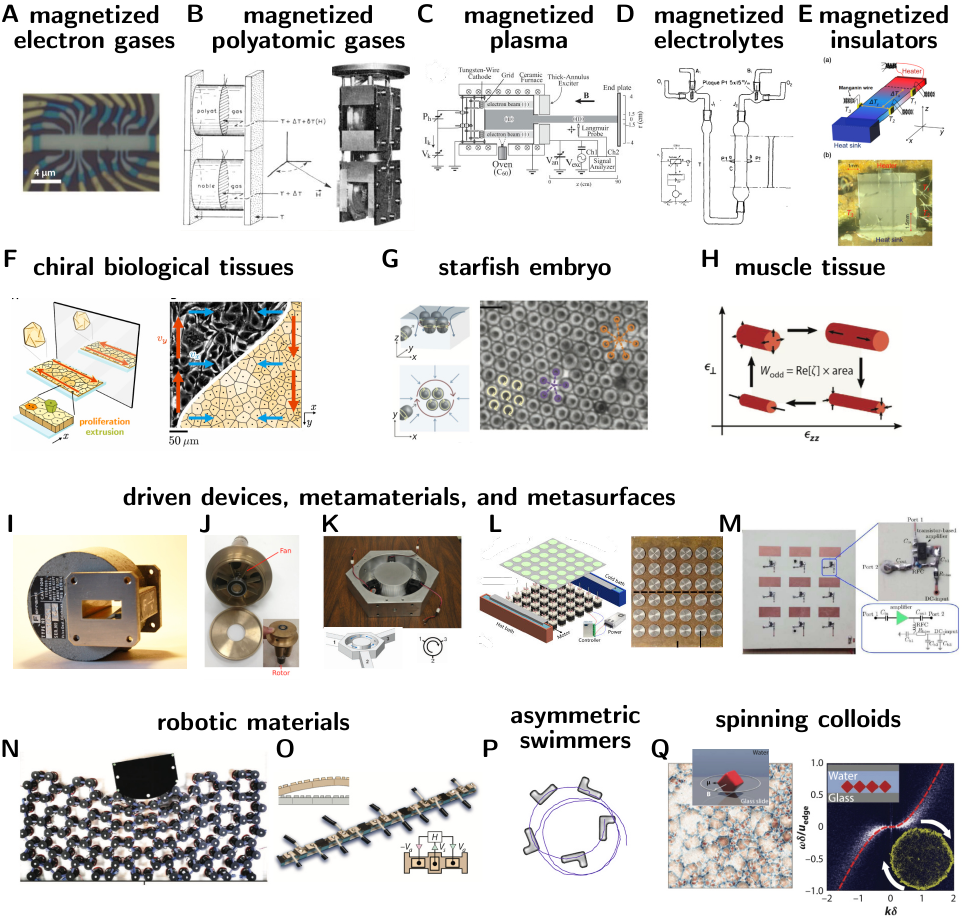}
    \caption{
    \label{figure_responses}
    \textbf{Nonreciprocal responses.}
    (A) Two-dimensional electron gas (in graphene) under magnetic field exhibiting Hall conductivity and viscosity. Adapted from \citet{Berdyugin2019}.
(B) Magnetized polyatomic gases exhibit odd/Hall viscosity and thermal conductivity (called Senftleben–Beenakker effect in this context). Here, a schematic and photograph of the measurement device for the thermal conductivity. Adapted from \citet{Hermans1970}.
(C) Magnetized plasma (here a pair-ion plasma) exhibit Hall responses. Adapted from \citet{Oohara2005}.
(D) Electrolytes (acqueous solutions and molten salts) under magnetic field exhibit Hall conductivity. Here, we show the setup used to measure it. Adapted from \citet{LaforgueKantzer1965}.
(E) Nonmagnetic insulator (\ce{SrTiO3}) under magnetic field, exhibiting a thermal Hall effect. Adapted from \citet{Li2020c}.
(F) Chiral biological tissues such as Human fibrosarcoma cells (HT1080) exhibit chiral responses captured by odd viscoelasticity in the continuum. Adapted from \citet{Chen2025b}.
(G) Starfish embryo have been reported to exhibit odd elasticity. Adapted from \citet{Tan2022}.
(H) Muscles have been reported to exhibit nonreciprocal responses coupling shear and volume changes of the muscle. Adapted from \citet{Shankar2024}.
(I) Ferrite resonance absorption microwave isolator, letting power pass only in one direction.  Adapted from \citet{IsolatorWP}.
(J) Device containing a fan realizing nonreciprocal acoustic Willis coupling (an example of coupling between different kinds of fields, here monopole and dipole). Adapted from \citet{Quan2021}.
(K) Acoustic circulator working by using fans to circulate air in a ring cavity. Adapted from \citet{Fleury2014}.
(L) Large magnet-free thermal Hall effect at room temperature in an active thermal metamaterial made of a stationary solid matrix plus rotating solid particles. Adapted from \citet{Xu2023}.
(M) Metasurface implementing nonreciprocal electromagnetic response without magnets and Faraday couplings, using instead a surface-circuit-surface (SCS) coupling to transistors for breaking time-reversal symmetry. Adapted from \citet{Taravati2017}. 
(N) Robotic metamaterial with programmed interactions exhibiting odd elasticity. Adapted from \citet{Veenstra2025}.
(O) Beam with piezoelectric patches that couple bending and shearing degrees of freedom, leading to odd micropolar elasticity.  Adapted from \citet{Chen2021}.
(P) Asymmetric L-shaped swimmers self-propelled by surface chemical reactions. The particles swim in circles, and at long times are expected to exhibit odd diffusion. Adapted from \citet{Kummel2013}. 
(Q) Two-dimensional fluid in which are suspended small cubes with a permanent magnetic dipole moment under a rotating magnetic. It exhibits odd viscosity, and possible other odd responses. Adapted from \citet{Soni2019}. 
    }
\end{figure*}

\subsection{Nonreciprocal responses}
\label{nrresponse}

Most of the examples of Secs.~\ref{violations_newton_third_law} and \ref{where_nonvariational_dynamics} lead to nonreciprocal responses, as explained in Secs.~\ref{nonreciprocal_responses} and \ref{onsager_casimir}.
A few examples are discussed in this section.

\subsubsection{Nonreciprocity in materials}
\label{mat}

Nonreciprocal responses, both at the level of bulk local responses and at the device level, arise in virtually all domains of physics. 
\citet{Asadchy2020} and \citet{Caloz2018} review nonreciprocity in optics and electromagnetism.
They identify a set of conditions under which the response of a material to an electromagnetic field can be expected to be reciprocal: (i) the response is linear, (ii) the material is at equilibrium, and (iii) the material is time-reversal invariant.
In this context, obtaining nonreciprocal responses then entails breaking at least one of these conditions.
For instance, a standard way of obtaining nonreciprocal responses in optics consists in using the Faraday effect, a magneto-optic effect that rotates the polarization plane of light as it propagates in an optically active media under an external magnetic field, which breaks time-reversal invariance.
\citet{Nassar2020} reviews nonreciprocity in acoustic and elastic materials, where similar ideas apply, see also \citet{Yves2025} for an overview based on symmetries.
In this case, it is often not practical to utilize an external magnetic field, because the couplings to deformations are often weak. 
Instead, various strategies ranging from active components to nonlinear responses are typically used.
We also note that reciprocity, when it is present, can greatly simplify the solution of certain problems: this is discussed in detail by \cite{Achenbach2004} in elastodynamics and by \citet{Masoud2019} in fluid dynamics and transport phenomena.

\subsubsection{Nonreciprocal devices}

Nonreciprocal devices (Sec.~\ref{nonreciprocal_devices}) such has isolators, circulators, and so on have been realized for many wave systems. They can be used as building blocks for realizing material-level responses, and conversely they can (but not not have to) be made of materials with the desired response.
One of the most common example is the electronic diode, made of an appropriate arrangement of doped superconductors.
Some other examples are shown in Fig.~\ref{nonreciprocal_devices}, including microwave isolators (panel I) that can be made with ferrites and magnets, circulators for acoustic waves using fans to circulate air (panel K), or more complex devices such as the system in panel J that realizes the equivalent of a nonreciprocal acoustic Willis coupling, involving two different fields.

\subsubsection{Topology and response}

Topological responses (as in topological insulators) can be seen as an intermediate level between materials and devices: they are rooted in the behevior of the bulk, but manifest themselves, for instance through robust edge states, at the level of the entire system. 
There is a complex relationship between nonreciprocity and topology.
First, bulk nonreciprocal responses (such as Hall responses, Sec.~\ref{hall_odd_effects}) are related to topological states (such as the \enquote{quantum} Hall effect), although there is no one-to-one correspondence. 
Second, the presence of unidirectional edge modes makes a system nonreciprocal, in the sense that it behaves like a circulator. 
It is however not acting as an isolator when the underlying system is non-dissipative (e.g. Hermitian quantum systems and equivalent classical systems that have broken time-reversal invariance).
We refer to \citet{Zhang2021b,Nelson2024} for discussions on these aspects.

\subsubsection{Hall/odd effects}
\label{hall_odd_effects}

One of the most iconic example of a nonreciprocal response is the Hall effect, most commonly encountered in electron gases in metals or semiconductors (Fig.~\ref{figure_responses}A) but also present, for instance, in electrolytes (panel D). 
In an isotropic two-dimensional electron gas under an external magnetic field $B$ (orthogonal to the plane), the local version of Ohm's law giving the relation between the electric field $\bm{E}$ and the electric current $\bm{j}$ reads
\begin{equation}
    \bm{j} = [\sigma \bm{\delta} + \sigma_{\text{H}} \bm{\epsilon}] \bm{E}
\end{equation}
in which $\sigma$ is the parallel (usual) conductivity, while $\sigma_{\text{H}}$ is known as the Hall conductivity, and describe a chiral transverse current in response to the applied field. 
Crucially, $\sigma_{\text{H}}$ is an odd function of the imposed magnetic field ($\sigma_{\text{H}}(B) = - \sigma_{\text{H}}(-B)$ and so it vanishes when there is no magnetic field), while $\sigma$ is an even function.
The presence of a finite $\sigma_{\text{H}}(B)$ is a violation of reciprocity. I
This bulk local nonreciprocal response leads to nonreciprocal devices such as gyrators and circulators (Sec.~\ref{nonreciprocal_devices}), as reviewed by \citet{Viola2014}.

The Hall conductivity is an instance of a larger class of nonreciprocal responses, known as Hall or odd responses, which are typical of systems that break chirality and time-reversal invariance \cite{Barabanov2015}, see Fig.~\ref{figure_responses}A-G, N, and P-Q for various examples.
They have for instance been discussed for particle diffusion (under the name of odd diffusion) in chiral active matter \cite{Larralde1997,Hargus2021,Kalz2022,Caprini2025} and in chiral porous media \cite{Koch1987,Auriault2010} (Fig.~\ref{figure_responses}P gives an example with self-propelled circle swimmers), for light diffusion \cite{Rikken1996,vanTiggelen1995}, for the thermal conductivity (under the name of Righi-Leduc effect or thermal Hall
effect), both in fluids (like polyatomic gases under magnetic field, Fig.~\ref{figure_responses}B) and solids (like insulators under magnetic field, Fig.~\ref{figure_responses}E) as well as in robotically active metamaterials made of rotating particles on top of a fixed matrix (Fig.~\ref{figure_responses}L).
Equivalents for elastic and viscous responses in chiral fluids and solids have been considered \cite{McCourt1990,Fruchart2023} (under various names such as Hall/odd/gyro viscosity and elasticity), see Fig.~\ref{figure_responses}B for magnetized polyatomic fluids (see \citet{McCourt1990}), Fig.~\ref{figure_responses}C for magnetized plasma (see \citet{Braginskii1965} as well as \citet{Stacey2006,Bae2013} for a discussion in the context of tokamaks and \citet{Kono2015} for pair-ion plasma), Fig.~\ref{figure_responses}F for biological tissues made of chiral cells and Fig.~\ref{figure_responses}G for assemblies of starfish embryo; as well as in the dynamics of muscles \cite{Zahalak1996,Shankar2022}, see Fig.~\ref{figure_responses}H.
It has also been proposed for active cholestetics \cite{Kole2021}.

Let us illustrate how it works in the context of mechanics (see \citet{Fruchart2023} for a review): in the continuum, the stress tensor
\begin{align}
\label{stress_overview} 
    \sigma_{ij} = C_{ijk\ell} \partial_\ell u_k \, + \, \eta_{ijk\ell} \partial_\ell \dot u_k \, + \, \cdots 
\end{align}
that summarizes the surface forces $f_i= \partial_j \sigma_{ij}$ between material elements. 
Here, the elasticity tensor $C_{ijk\ell}$ is the proportionality coefficient between the stress tensor and the displacement gradient tensor $\partial_\ell u_k$, while the viscosity tensor $\eta_{ijk\ell}$ is the proportionality coefficient between the stress and the velocity gradient tensor $\partial_i \dot u_j$.
In equilibrium systems, which are time-reversal invariant and where energy is conserved, these coefficients can be expressed as
\begin{equation*}
    C^{\text{eq}}_{i j k \ell} = \frac{\delta^2 F}{\delta (\partial_j u_i) \delta (\partial_\ell u_k ) } 
    \quad
    \text{and}
    \quad
    \eta^{\text{eq}}_{i j k \ell} = \, \frac{T\delta^2 \dot{S}}{\delta ( \partial_j \dot u_i) \delta (\partial_\ell \dot u_k)}
\end{equation*}
where $F$ is the free energy of the elastic medium, $\dot S$ the rate of entropy production of the fluid, and $T$ is the temperature. 
As a consequence, they satisfy the symmetries $C^{\text{eq}}_{i j k \ell} = C^{\text{eq}}_{k \ell i j}$  and $\eta^{\text{eq}}_{i j k \ell} = \eta^{\text{eq}}_{k \ell i j}$.
However, this needs not be the case in nonequilibrium media, where one can have
\begin{equation}
    C_{i j k \ell} \neq C_{k \ell i j}
    \qquad
    \text{and}
    \qquad
    \eta_{i j k \ell} \neq \eta_{k \ell i j}
\end{equation}
respectively violating Onsager reciprocity and Maxwell-Betti reciprocity. 
The antisymmetric part of these tensors is referred as odd viscosity and odd elasticity~\cite{Avron1998,Scheibner2020}.
As illustrated in Sec.~\ref{components_same_field} in the case of odd elastodynamics, nonreciprocal response functions can induce non-variational dynamics.

Note that some subtleties arise in these systems, that require some care to handle. 
For instance, in the case of diffusion of a scalar (e.g. particles or heat), the antisymmetric part of the diffusion tensor $D_{ij}$ (giving the current $J_{i} = - D_{ij} \partial_j \rho$ as a function of gradients of the scalar density $\rho$) drops out of the diffusion equation (when it is uniform), as
\begin{equation}
    \partial_t \rho = - \partial_i J_i = D_{ij} \partial_i \partial_j \rho.
\end{equation}
This can be easily seen in an isotropic system in 2D, where $D_{ij} = D^{\text{e}} \delta_{ij} + D^{\text{o}} \epsilon_{ij}$.
As a consequence, the effects of a uniform $D^{\text{o}}$ are only observable when one can measure the flux directly, or when the flux enters boundary conditions of the diffusion equation. 
The presence of a nonuniform $D^{\text{o}}$ leads however to observable consequences even in the bulk.
Similar effects are also present for part of all response tensors, see \citet{Rao2020,Rao2023,Fruchart2022} for a discussion in the case of viscosity.

One minimal situation where odd diffusion can be derived from microscopic are Brownian particles subject to a magnetic Lorentz force.
This apparently simple situation is rather subtle, because the low mass limit is singular, and distinct from the large friction limit, and results in stochastic differential equation with a nonwhite noise whose correlation matrix has antisymmetric components \cite{Chun2018}. 
Starting from the inertial dynamics for a two-dimensional particle
\begin{subequations}
\begin{align}
    \bm{\dot{x}} &= \bm{v} \\
    \bm{\dot{v}} &= \bm{f}(\bm{x}) - G \bm{v} + \sqrt{2 \gamma k_{\text{B}} T} \bm{\xi}
\end{align}
\end{subequations}
in which $G = \gamma \, \Id - q \bm{B} \times$ where $\bm{B}$ is the magnetic field and with a standard Gaussian white noise $\bm{\xi}$, \citet{Chun2018} show that in the small mass limit, this reduces to the overdamped dynamics
\begin{equation}
    \bm{\dot{x}} = G^{-1} \bm{f}(\bm{x}) + \bm{\eta}
\end{equation}
in which $\bm{\eta}(t)$ is a Gaussian nonwhite noise with asymmetric correlations
\begin{equation}
    \langle\bm{\eta}(t) \otimes \bm{\eta}(t')\rangle = \lim_{m \to 0} \frac{k_{\text{B}} T}{m} e^{-\gamma/m |t-t'|} 
    R_{\omega_B (t-t')}
\end{equation}
where $R_\theta$ is the rotation matrix with angle $\theta$ and $\omega_B=qB/m$.
For small but finite $m$, this noise can be seen as a chiral Ornstein-Uhlenbeck process.
From the master equation perspective, this leads to nondiffusive fluxes, sometimes known as Lorentz fluxes \cite{Vuijk2019,Abdoli2020,Vuijk2020}.

Odd responses also occur in turbulence: for instance, the eddy diffusivity tensor is asymmetric in the presence of mean vorticity \cite{Durbin2011}.
The transport of a passive scalar by a turbulent flow can be captured by the equation
\begin{equation}
    \partial_t C + U_i \partial_i C = \alpha \nabla^2 C - \partial_i \Pi_i
\end{equation}
in which $C$ is the mean concentration field, $\bm{U}$ is the mean velocity, $\alpha$ is the bare diffusivity (without turbulence), and $\Pi_i = \overline{c u_i}$ is called the scalar flux due to turbulence, in which the overbar represents average over turbulence, while $c$ and $\bm{u}$ are the fluctuating parts of the concentration and velocity fields. 
A simple model (closure) for the scalar flux is given by tensor diffusivity models, in which
\begin{equation}
    \Pi_i = - \alpha_{ij} \frac{\partial C}{\partial x_j}
\end{equation}
where $\alpha_{ij}$ is the eddy diffusivity tensor, which, under certain hypotheses, can be modeled as \citet{Durbin2011}
\begin{equation}
    \label{closure_turbulent_eddy_diffusivity}
    \bm{\alpha} = \left( \frac{\epsilon}{C_c K} \bm{1} + \bm{\dot{e}}\right)^{-1} \bm{\overline{uu}}
\end{equation}
in which $\bm{\overline{uu}}$ is the tensor with components $\overline{u_i u_j}$, ${\dot{e}}_{ij}=\partial_j U_i$ are mean velocity gradients, $\epsilon$ is the rate of injection of energy, $K= \overline{u^2}/2$ is the mean turbulent kinetic energy, while $C_c$ is a dimensionless number of order unity.
Equation \eqref{closure_turbulent_eddy_diffusivity} shows that the presence of mean vorticity (antisymmetric part of ${\dot{e}}_{ij}$) leads to an asymmetric eddy diffusivity tensor.

\subsubsection{Cross-field couplings}

In thermodynamics, one of the striking manifestations of reciprocity are relations between cross-coupling phenomena.
As an example, thermoelectric effects include the Seebeck effect, whereby temperature differences give rise to an electrical tension, and the Peltier effect, whereby an electrical current leads to a heat flow.
As suggested by their names, these two effects were discovered separately.
At equilibrium, it turns out that the corresponding response coefficients are related by Onsager reciprocity. 
Examples are reviewed in \cite{Gaspard2022,Pottier2010,Beris1994,DeGrootMazur}.
Also in cross-field couplings, the presence of time-reversal breaking fields can remove reciprocity. For instance, this arises for thermoelectric effects from the presence of a supercurrent in mesoscopic structures \citet{Giazotto2006}.
Similar effects arise in nonreciprocal bianisotropic electromagnetic media \cite{Kong1972,Cui2024} or Willis phononic metamaterials \cite{Nassar2020,Yves2025}.
Other examples are reviewed in \citet{Lakes2025}.

\subsubsection{Response to temporal sequences}

In this section, we discuss a slightly different kind of nonreciprocity, whereby a system reacts (on average) to a nonreciprocal temporal sequence.
The archetypal example is swimming at low Reynolds number, where the average self-propulsion velocity can be seen as the response to the sequence of deformations of a deformable body \citet{Taylor1951,Purcell1977,Shapere1987,Shapere1989,Shapere1989b,Lauga2009}. 
In this case, \citet{Purcell1977}'s scallop theorem states that, under certain conditions (e.g. in a Stokes flow in a simple fluid), a swimming stroke (the family of shapes taken by the objects parameterized by time) that is invariant under time-reversal leads to a vanishing self-propulsion velocity. 
In contrast, time-reversal breaking strokes like corkscrew or undulatory motions can lead to self-propulsion on average.
Note that here, it is the reciprocity (or lack thereof) of the input that produces a constraint on the self-propulsion.
However, the behavior of the medium under time-reversal also plays a part in the theory, as reviewed in \citet{Lauga2009,Lauga2011}. In particular, inertia or other memory effects break the theorem. See also \citet{Lapa2014} for the (lack of) effect of odd viscosity.

\subsubsection{Economics and game theory}

Nonreciprocal responses can also arise in situations where the microscopic description is less tractable, ranging from economics to sociological behavior.

For instance, in economics, a demand function is a rule giving the quantity $q_i$ of a good $i$ that is bought as a function of the unit prices $p_j$ of all goods, plus other variables (we follow \citet{Deaton1980}).
Marshallian (uncompensated) demand functions $q_i = g_i(p, b)$ depend only on the prices $p$ and on the total budget $B$ of the consumer (who have to choose how much to buy of all the good they need constrained by their budget, so that $b = \sum_i q_i p_i$). 
In contrast, Hicksian (compensated) demands functions $q_i = h_i(p, u)$ depend only on the prices and on the utility $u = \psi(p, x)$ associated to the purchases.
Crucially, Hicksian demand functions can be expressed as
\begin{equation}
h_i = \frac{\partial c}{\partial p_i}
\end{equation}
in which the cost function $c(p, u)$ [such that $b = c(p,u)$] is the inverse of the utility function $\psi$.
As a consequence, Hicksian demands satisfy the symmetry relation
\begin{equation}
    \label{slutsky}
    \frac{\partial h_i}{\partial p_j} = \frac{\partial h_j}{\partial p_i}
\end{equation}
because of the symmetry of second derivatives. 
The matrix with elements $s_{ij} = {\partial h_i}/{\partial p_j}$ (which is therefore symmetric) is known as the or Slutsky matrix of compensated price responses \cite{Deaton1980}.

In practice, producers are interested in how much a change in prices will affect the demand. This is quantified by a quantity known as the elasticity of demand
\begin{equation}
    E_{i j} = \frac{\partial \ln s_i}{\partial \ln X_j} = \frac{X_j}{s_i} \frac{\partial s_i}{\partial X_j}.
\end{equation}
which measures how much a market share $s_i$ of a good $i$ changes when the unit price $X_j$ of another good $j$ is modified ($E_{ii}$ is the (self-)elasticity while $E_{i j}$ with $i \neq j$ are called cross-elasticities).
\citet{Bonfrer2006} reviews the possible asymmetries in the cross-elasticities. 
In particular, the Slutsky symmetry \eqref{slutsky} produces constraints (but not symmetry) on the part of the cross-elasticities corresponding to the compensated demand, but not on the full matrix. As a consequence, they argue and show with an empirical analysis of observable data that the price-elasticity matrix $E_{ij}$ can not only have magnitude asymmetry with $E_{ij} \neq E_{ji}$ but even sign asymmetry where the two components have opposite signs. 
Here, we have two products A and B such that increasing the price of A decreases the demand for B ($E_{B A} < 0$), while increasing the price of B increases the demand for A ($E_{A B} > 0$), meaning that A is a substitute for B, while B is a complement to A \cite{Bonfrer2006}.
From the perspective of statistical physics, \citet{GarnierBrun2023} derived a fluctuation-response formula for the Slutsky matrix, relating it to spontaneous
fluctuations of consumption (not response to changing prices and budgets), and discuss its asymmetry in relation to rationality and herding effects in a model where agents influence each other's choices.

In game theory, something a bit similar to Hicksian demand functions arises in potential games \cite{Monderer1996,Candogan2011}. 
In short, a game is called potential if there is a real-valued function (called the potential) on the space of all possible strategies, such that the change in utility of a player between two strategies is equal to the change in the potential between these strategies (perhaps with appropriate weighting). 
This means that the players of a potential game are, effectively jointly maximizing the potential (although they are only trying to maximize their utility).
In the case of differential games (games in continuous time), this leads to nonvariational dynamics (Sec.~\ref{nonvariational_dynamics}), as discussed in \citet{FonsecaMorales2017}. 
For instance, such an emergent potential structure can be found in models of social segregation, in which this leads to stationary steady-states \cite{Grauwin2009} while non-potential decision structures tend to lead to dynamic states \cite{Zakine2024,GarnierBrun2025,Lama2025,Seara2025}.
Separately, notions of reciprocity have been considered in economy and sociology \cite{Stephens2002,Roberts1998,Malmendier2014}, defined as \enquote{the tendency to reciprocate
kind acts with kindness and unkind acts with spite} \cite{Malmendier2014}. These can be encoded mathematically, for instance in the framework of game theory \cite{Hofbauer1998}.

\section{What are the consequences of non-reciprocity?}
\label{what_are_consequences}

One of the recurring features of many-body nonreciprocal systems is their ability to exhibit dynamic steady-states.
You can see an example of this in Fig.~\ref{examples_nonvariational}C, which shows collective oscillations of chemical species in a spatially extended system.
These temporal phases of matter include many-body limit cycles or tori, also known as time crystals or quasicrystals (with various more or less restrictive definitions), as well as more complex spatiotemporal phenomena.
As we illustrate in this section, the phenomena observed in a variety of different nonreciprocal systems share common features, raising the question of how to generalize to these systems the notions of phases, phase transitions, and universality classes familiar from equilibrium statistical physics.

\subsection{Nonequilibrium phase transitions}
\label{neqpt}

In equilibrium, a dynamical perspective on phase transitions and critical phenomena was put forward by \citet{Hohenberg1977}, see \citet{Folk2006,Kamenev2023} for a modern perspective. 
In short, the dynamics of a (hopefully small) number of fields $a_i(t,\bm{r})$ is described by stochastic PDEs that are derived from the effective free energy of the system augmented with kinetic coefficients, both of which are constrained by equilibrium constraints. 
Different dynamics can be constructed from the same free energy functional, depending on the constraints imposed on the system (e.g. whether a quantity is conserved or not).
In many-body nonreciprocal systems, the dynamics may be novariational and the system out of equilibrium, so it is not possible to separate dynamics and energetics; instead, we only have dynamics at our disposal.
From a phenomenological perspective, this entails directly working at the level of the (stochastic) dynamics, rather than the effective free energy. 
Techniques from equivariant dynamical systems (Sec.~\ref{symmetry_dynamics}) can be used for the deterministic part, and progress is ongoing to classify the symmetries of stochastic dynamics from a more general perspective \cite{Sa2025,Sa2023,Kawabata2023,Altland2021,Lieu2020,Kawasaki2022}.
From the perspective of coarse-graining, taking dynamics into account entails working directly with the stochastic actions where it is encoded.

In spite of some fundamental differences, the notion of universality in critical phenomena originally developed in equilibrium systems has been successfully extended to many classes of nonequilibrium systems \cite{Vicsek1992,Odor2004,Odor2008,Kardar2007,Henkel2008,Henkel2010,Munoz2018,Dupuis2021,Sieberer2025}, see Sec.~\ref{nonvariational_renormalization_group} for more discussions and references.
As explained in Sec.~\ref{hh_and_beyond}, the perspective we discuss here relates this approach to bifurcation theory through generalizations of the Hohenberg-Halperin approach to dynamic critical phenomena \cite{Hohenberg1977}.

One extra challenge is that, in many situations of practical interest, only finite size, finite time, finite number observations exist.
This is in contrast with usual equilibrium phase transition which occur in systems composed of several Avogadro numbers of constituents. 
In nonreciprocal many-body systems, the number of constituents may be large, but not sufficiently large for the standard thermodynamic limit to be appropriate, because it may wash out important distinctions \footnote{The question \enquote{Is X a real phase transition?} may sometimes be irrelevant in practice despite its fundamental interest. For instance, \citet{Schutz2003,Rajewsky2000} provide an example in which a very sharp (but smooth) change in an observable when a parameter is varied by a tiny amount. This results in a very large (but finite) correlation length, which would require a number of the order of $10^{70}$ sites to resolve. Such a sharp crossover is not practically distinguishable from a genuine phase transition. }.
This challenge is compounded by the fact that noise plays an important role, and so the fully deterministic limit is also not appropriate.
Efforts to develop classification schemes that handle all of these aspects are ongoing, in which the thermodynamic limit is be replaced by a long-time limit, a scaling of the system parameters, or any other way of making the theory singular.
These include approaches in the spirit of macroscopic fluctuation theory \cite{Bertini2015,KourbaneHoussene2018,Agranov2023,Agranov2021,Martin2025}, dynamical phase transitions (where a long-time limit is taken) \cite{Garrahan2007,Garrahan2009,Tsobgni2016,Heyl2018,Jack2020}, or finite-component phase transitions (where a rescaling of the system parameters is considered) \cite{Hwang2015,Hwang2016,Ashhab2013,Felicetti2020}.
Note also that sharp changes in behavior can also occur as the number of components is changed \cite{Calvert2025}.

\subsubsection{The Hohenberg-Halperin classification and beyond}
\label{hh_and_beyond}

At equilibrium, phase transitions are usually described in terms of a free energy. 
The minimum of this free energy corresponds to the current phase, and a phase transition occurs when it ceases to be a global minimum, or a minimum at all.
This static landscape picture relies on an underlying dynamics that leads the system towards the global minimum~\cite{Landau1954,Hohenberg1977,Cross1993,Cugliandolo1995,Keim2019}. 

Phase transitions can be related to bifurcations of dynamical systems~\cite{Lagues2003,Muoz2018,Sornette2000}.
For example, the ferromagnet/paramagnet transition in the equilibrium Ising model corresponds to a pitchfork bifurcation: in mean-field, the dynamics of the scalar order parameter $\phi$ (the rescaled magnetization) is described by
\begin{equation}
    \frac{{\rm d} \phi}{{\rm d}t} = \alpha \phi - \phi^3
\end{equation}
in which $\alpha \propto T - T_{\text{c}}$ is proportional to the distance from the critical temperature.
Conversely, the thermal field theory corresponding to locally-coupled pitchfork bifurcations where \begin{equation}
    \partial_t \phi = \alpha \phi - \phi^3 + D \Delta \phi + \eta(t,r)
\end{equation} 
where $\eta$ is a Gaussian white noise, leads upon renormalization to the Ising critical point~\cite{Tauber2014}.

This dynamics is an instance of the \enquote{model A} dynamics in the classification of \cite{Hohenberg1977}.
It describes the relaxational dynamics of order parameter fields $\phi$ that do not satisfy a conservation law.
In this case, the dynamical system that describes the time evolution of the order parameter~$\phi$ near its equilibrium value reads~\cite{Landau1954,VanHove1954,Hohenberg1977,Cross1993,Hohenberg2015}
\begin{equation}
    \partial_t \bm{\phi}(t,\bm{r}) = - \Gamma \frac{\delta \mathcal{F}}{\delta \bm{\phi}(t,\bm{r})} +\bm{\eta}(t,\bm{r})
    \label{model_A}
\end{equation}
for a system described by the Ginzburg-Landau free energy $F[\phi]$ (or a Ginzburg-Landau-Wilson effective Hamiltonian \cite{Wilson1974,Hohenberg2015}), $\Gamma$ a kinetic coefficient, and $\eta(t,\bm{r})$ a Gaussian white noise with $\langle\eta(t,\bm{r})\rangle = 0$ and $\langle\eta^{m}(t,\bm{r}) \eta^{n}(t',\bm{r'})\rangle = 2 \Gamma k T \delta(t-t') \delta(\bm{r}-\bm{r'}) \delta^{mn}$ \footnote{In general, the kinetic coefficient $\Gamma$ may be replaced by a matrix, which is related to Onsager coefficients at equilibrium. This is described in detail by \citet{Folk2006}.}.
In the example of the Ising model, $F$ is a $\phi^4$ free energy.

In order to encompass nonreciprocal systems, we instead consider a nonreciprocal model A of the form
\begin{equation}
    \partial_t \bm{\phi}(t,\bm{r}) = \bm{f}[\bm{\phi}] + \bm{\eta}(t,\bm{r})
\end{equation}
in which the functional $\bm{f}$ may be nonvariational, namely $\delta f_a(\bm{r})/\delta \phi_b(t,\bm{r'}) \neq \delta f_b(\bm{r'})/\delta \phi_a(t,\bm{r})$.
Hence, bifurcations with no analogue at equilibrium can occur in the mean-field dynamics, in particular those leading to limit cycles.
In the context of a many-body system, these may correspond to phase transitions that are forbidden at equilibrium.
In addition, $\bm{f}$ is not necessarily related to $\Gamma$.

For instance, a non-reciprocal version of the Ising model with two asymmetrically coupled species may be described by a non-reciprocal $\phi^4$ theory of the form \cite{Avni2025b}
\begin{equation}
    \partial_t \phi_i = -\phi_i + J_{ij} \phi_j - \frac{1}{3} [J_{ij} \phi_j]^3 + D \nabla^2 \phi_i + \eta_i
\end{equation}
in which $\phi_i$ ($i=1,2$) are scalar fields representing the magnetization of each species, while $J_{ij}$ is a matrix of couplings with $J_{ij} \neq J_{ji}$.
With appropriate values of the parameters, this can be reduced to the Hopf field theory  where limit cycles appear (Sec.~\ref{hopf_field_theory}).

As in Landau theory, the general form of $\bm{f}$ can be constructed from symmetries (Sec.~\ref{symmetry_dynamics}).
As an example, we can consider $N$ species $a=1\dots,N$ each described by a vector field $\bm{\phi}_a$ with vector components  $\phi_a^\mu$ ($\mu=1,\dots,d$ in space dimension $d=2$). 
Then, imposing equivariance with respect to the diagonal action of $O(2)$ imposes that \cite{Fruchart2021}
\begin{equation}
    \label{general_O2}
    \partial_t \bm{\phi}_a(t,\bm{r}) = \mathscr{A}_{ab} \bm{\phi}_b + \mathscr{B}_{abcd} ( \bm{\phi}_b \cdot \bm{\phi}_c )\bm{\phi}_d
    + \mathcal{O}(\nabla,\bm{\phi}^4)
\end{equation}
See Sec.~\ref{where_nonvariational_dynamics} for other examples.

Similarly, the dynamics of a conserved order parameter is, at equilibrium, described by the model B dynamics
\begin{equation}
    \partial_t \rho = - \nabla \cdot \left( - \Gamma \nabla \mu + \bm{\eta}(t,\bm{r}) \right)
    \quad
    \text{with}
    \quad
    \mu = \frac{\delta \mathcal{F}}{\delta \rho}
\end{equation}
in which the chemical potential $\mu$ is obtained as a variation of the free energy, and $\Gamma$ is a kinetic coefficient (mobility) constrained by detailed balance (not the same as in model A; see \citet{Folk2006,Cates2019} for details).

In the nonreciprocal Cahn-Hilliard models introduced in \citet{Saha2019,You2020}, this is generalized as 
\begin{subequations}
\label{nrch}
\begin{equation}
    \partial_t \rho_a = - \nabla \cdot \left( - \Gamma_{ab} \nabla \mu_b + \bm{\eta}_a(t,\bm{r}) \right)
\end{equation}
where the chemical potentials $\mu_a[\phi_1,\phi_2,\dots]$ are not constrained to be variational, and the mobility $\Gamma_{ab}$ is not necessarily constrained by detailed balance. \citet{Saha2019,You2020} focused on the influence of nonvariational character of the chemical potentials, taking $\Gamma_{ab} = \delta_{ab}$ and
\begin{equation}
    \mu_a = \mu^{\text{eq}}_a + \kappa_{ab} \rho_b
\end{equation}
\end{subequations}
in which $\mu^{\text{eq}}_a$ is an equilibrium chemical potential, while $\kappa_{ab} \neq \kappa_{ba}$.
For a particular choice of the equilibrium part, this leads to 
\begin{equation}
    \partial_t \rho_a = \nabla \cdot [(\chi_a+\phi_a^2-\gamma_i \nabla^2)\nabla \phi_a + \kappa_{ab} \nabla 
    \phi_b +\eta_a]
\end{equation}
corresponding to asymmetric cross-diffusions.
This equation can arise, for instance, when coarse-graining two populations of particles with interspecies interactions violating Newton's third law, but also in mixtures of active and passive particles \cite{You2020}.
It turns out that a generalized version of the nonreciprocal Cahn-Hilliard model emerges as normal form (or universal amplitude equation) for pattern formation in nonequilibrium scalar mixtures with more than one
conservation law \citet{FrohoffHulsmann2023b}.
\citet{Sahoo2025} discusses the consequences of having mobilities with different equilibrium-like functional dependencies on $\rho_i$.
To the best of our knowledge, the consequences of having a non-equilibrium mobility, and whether it is identical to having a non-variational chemical potential, are not yet known.

A nonvariational and/or detailed-balance-breaking structure can arise in all other classes of models (with conserved, nonconserved, or mixed dynamics). For instance, \citet{Cates2019} reviews nonequilibrium field theories in soft matter with a focus on systems with self-propulsion, while \citet{Cates2025} focuses on phase separation in active systems.

\paragraph{Nonequilibrium noises.}

In the examples discussed up to now in this section, the deterministic dynamics is combined with an additive Gaussian white noise, and the fact the system is out of equilibrium is mainly encoded in the nonvariational character of the functional $f$.
This makes the nonequilibrium character of the system apparent even in a mean-field description where noise is neglected.
This can be compared and constrasted with other archetypal non-equilibrium phase transitions such as directed percolation~\cite{Henkel2008}, in which the non-equilibrium character is encoded in the noise.
Indeed, in directed percolation, the order parameter $\rho(t, r)$ (density of occupied states) satisfies~\cite{Henkel2008}
\begin{equation}
    \label{dp_pde}
	\partial_t \rho = - a \rho - b \rho^2 + c \nabla^2 \rho + \sqrt{\rho} \, \xi(t, r)
\end{equation}
in which $\xi(t, r)$ is a white noise satisfying $\braket{\xi(t,r)} = 0$ and $\braket{\xi(t,r) \xi(t',r')} = \kappa \delta(t-t') \delta(r-r')$.
The deterministic part of this equation is variational, with an effective free energy $\mathcal{F} \propto (a/2) \rho^2 + (b/3) \rho^3 + (c/2) (\nabla \rho)^2$ and the non-equilibrium character is entirely contained in the noise.

The directed percolation universality class is a paradigmatic example of absorbing phase transition \cite{Odor2004,Odor2008,Henkel2008} in which the microscopic irreversibility manifests itself at macroscopic scales by the existence of an \enquote{absorbing phase} from which it is impossible to escape even when noise is present. 
This is expected to happen in a variety of physical contexts, for instance in the case of directed percolation, which is believed to describe the transition to turbulence in pipes~\cite{Lemoult2016,Barkley2016,Avila2023}. 
Absorbing phase transitions do have a manifest nonreciprocal character in the sense that it is possible to go from the fluctuating phase to the absorbing phase, but not the other way around. 
This large-scale nonreciprocity is a manifestation of an asymmetry in the microscopic rates of change between states, which amounts to broken detailed balance.  
Indeed, \citet{Klamser2025} reported that non-reciprocal interactions (in the sense of Eq.~\ref{newton_third_violated}) between particles can lead to an absorbing phase transition with exponents compatible with directed percolation (Fig.~\ref{figure_nr_particle_systems}E).

In general, a combination of nonvariational dynamics and nonequilibrium noises may be required to adequately describe a given system. 
Examples can be found in \citet{Odor2004,Tauber2014,Dupuis2021,Hinrichsen2000,Janssen2005}.

\begin{figure}
    \centering
    \includegraphics[width=8.5cm]{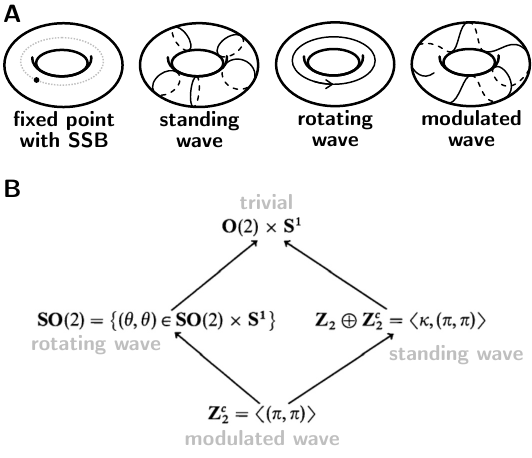}
    \caption{
    \label{figure_symmetries}
    \textbf{Classification of dynamic states with symmetries.}
    (a) Different classes of behavior in symmetric systems include (i) the trivial state (not represented) with full symmetry, (ii) static ordered states that spontaneously break spatial and/or internal symmetries, (iii) standing waves that do not have mixed symmetries [involving both time and other symmetries], (iv) rotating waves that move along a group orbit and (v) modulated waves that are a mix of travelling and rotating motions, as well as (vi) other more complicated kinds of motion (not pictured).
    (b) A precise analysis can be made using equivariant bifurcation theory when a particular group action is specified. 
    Here, the isotropy lattice of a particular action of $O(2) \times S^1$ on $\CC^2$, relevant in the case of $O(2)$-symmetric Hopf bifurcations, is pictured (adapted from \citet{Golubitsky1988}).
    }
\end{figure}

\paragraph{Classifying dynamic phases and phase transitions.}

At equilibrium, phases of matter can, to a large extent, be classified by their patterns of spontaneous symmetry breaking \footnote{This idea, originally formalized in \citet{Landau1937}, is known as the Landau paradigm (see \citet{Aron2020} for an application to nonequilibrium steady-states, \citet{Meibohm2023} for an application to dynamical phase transitions, and \citet{Huang2023} for an effective field theory perspective on nonequilibrium many-body physics).
In the last few decades, certain (equilibrium) phases of matter such as those with topological order have been described as being \enquote{beyond Landau paradigm} \cite{Senthil2004,Wen2017}. More recently, the discovery of \enquote{generalized symmetries} \cite{Gaiotto2015,McGreevy2023,Shao2023,SchaferNameki2024} opened the way to incorporate these phases into the Landau paradigm \cite{Moradi2023,Bhardwaj2024,Chen2025}.}.
The same strategy applies to dynamic phases, but one has to also consider temporal symmetries (Sec.~\ref{symmetry_dynamics}), in addition to the group $G$ of spatial and internal symmetries of the system, we have to consider the action of time translations $T_\tau$ which act on an order parameter $\bm{\phi}(t)$ as $[T_\tau \bm{\phi}](t) = \bm{\phi}(t-\tau)$. 
The group of time translation is $\RR \ni \tau$ and so the full symmetry group to consider is $G \times \RR$. 

\subparagraph{Classes of dynamic states.}
\label{classes_dynamic_states}

A first meta-classification (Fig.~\ref{figure_symmetries}A) distinguishes \cite{Chossat2000,Knobloch1990}
\begin{itemize}[nosep,left=0pt,label=--]
    \item the \emph{trivial state} $\bm{\phi} = 0$, with full symmetry $G \times \RR$ \footnote{We have assumed that there is a unique fully symmetric state that can be represented by $\bm{\phi} = 0$, so $\bm{\phi}$ is an order parameter in the usual sense of the term.}
    \item \emph{static ordered states} $\bm{\phi} \neq 0$ that spontaneously break the symmetry, but do not depend on time; they have symmetry $H \times \RR$ where $H$ is a strict subgroup of $G$
    \item \emph{rotating waves} are time-dependent (limit cycle) states $\bm{\phi}(t)$ that correspond to motion along group orbits of $G$.
    Depending on the context, these are also called travelling waves or more generally relative equilibria.
    They are invariant under the action of a diagonal $S^1$ subgroup of $G \times \RR$ that combines a rotation subgroup of $G$ with time translation.
    Namely, this action is defined for $\varphi \in S^1$ as $\bm{\phi}(t) \mapsto R_{\varphi} \bm{\phi}(t-2\pi\varphi/\omega)$ in which $\omega$ is the frequency of oscillation and $R_{\varphi} \in G$.
As the temporal phase shift has the same effect as a rotation of the state, they can compensate each other so some of the symmetry of the state is effectively restored, on average, at the price of time translation invariance.
    A scalar example is $z(t) = z_0 \ee^{\ii \omega t}$. 
    \item \emph{standing waves}, in contrast, only have purely temporal and purely spatial/internal symmetries, with no mixed action. They correspond to motion orthogonal to the group orbits of $G$ in the sense that the symmetry of the state does not change as a function of time.
    For instance, this would be the case of $z(t) = \cos(\omega t) \ee^{\ii \theta_0}$.
    The symmetry elements of these states are of the form $(g,\text{id}_\RR)$ with $g \in G$ or $(\text{id}_G, \tau)$ for $\tau \in G$.
    \item \emph{modulated waves}, that are a combination of both rotating and standing motions, and are generically quasiperiodic
    \item other more complicated behaviors.
\end{itemize}

\subparagraph{Phases of the nonreciprocal $O(2)$ model.}

\begin{figure*}
    \centering
    \vspace{-1.5cm}
    \includegraphics[width=155mm]{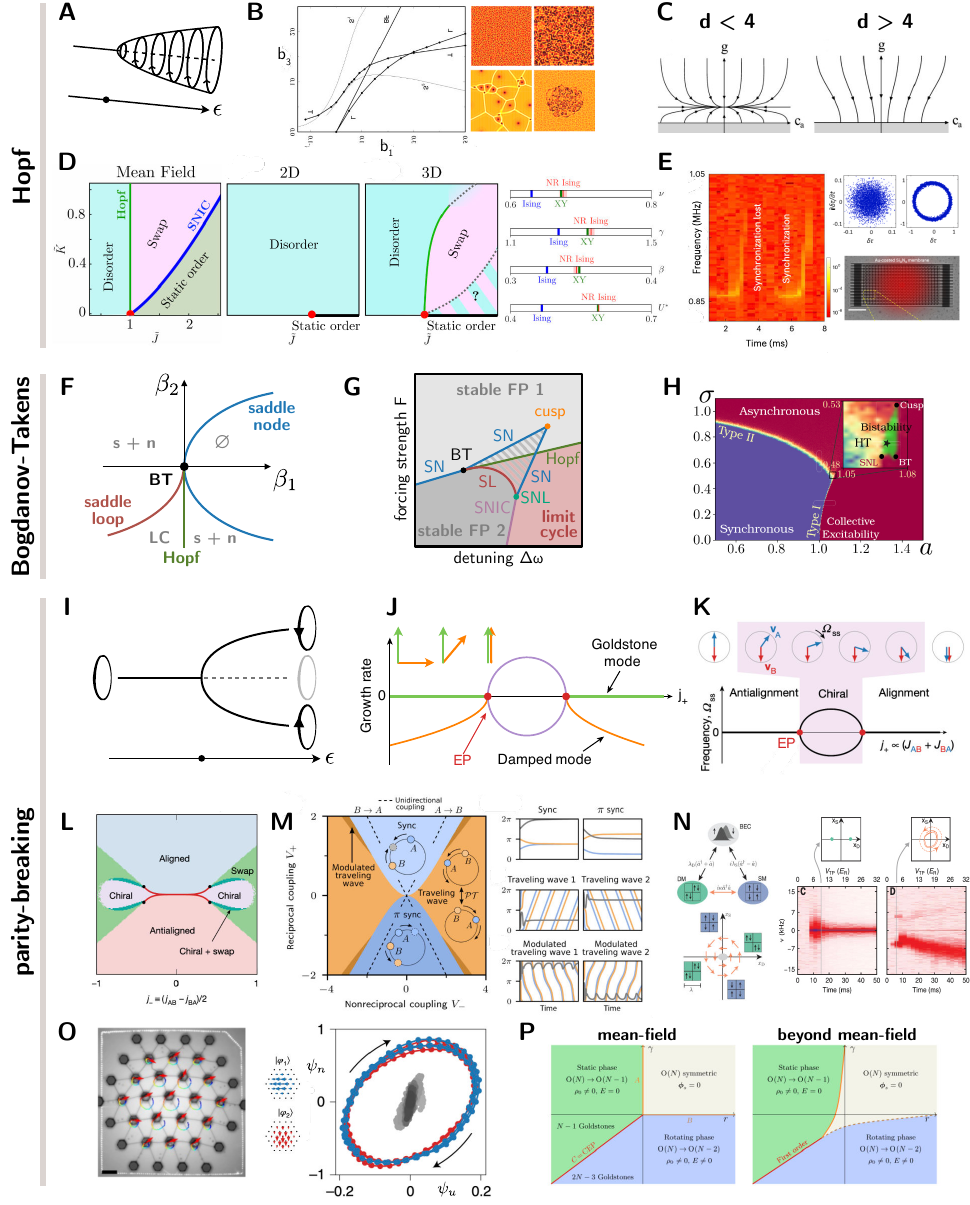}
    \vspace{-0.25cm}
    \caption{\textbf{Many-body bifurcations.}
(A) Supercritical Hopf bifurcation: a stable fixed point becomes unstable when $\epsilon=0$ and gives rise to a stable limit cycle.
	(B) Finite-size phase diagram of the complex Ginzburg-Landau equation, exhibiting regimes with stable plane waves, phase turbulence, defect turbulence, frozen states. Snapshots are given on the right. Adapted from \citet{Chate1996}.
	(C) Renormalization group flow for the spatially extended Hopf bifurcation. In $d < 4$, a nontrivial fixed point formally related to the XY universality class is found. Adapted from \citet{risler2005universal}.
	(D) Schematic phase diagram of the nonreciprocal Ising model with two species summarizing kinetic Monte-Carlo simulations. A many-body limit cycle (swap phase) is present in mean-field and in 3D. Critical exponents for the 3D disorder/swap transition are compatible with XY values. Adapted from \citet{Avni2025b}.
	(E) Experimental observation of a many-body limit cycle in photonic metamaterials. Adapted from \citet{Liu2023,Liu2025}.
	(F) Local bifurcation diagram of the normal form of the Bogdanov-Takens (BT) bifurcation (black point). The BT bifurcation (black dot at $\beta_1=\beta_2=0$) can be seen as the intersection of saddle-node (SN), Hopf, and saddle-loop (SL) bifurcation lines.
	(G) Phase diagram of a driven Kuramoto model exhibiting a BT bifurcation as an organizing center (along with a saddle-node-loop (SNL) and a cusp codim two bifurcations). Adapted from \citet{Childs2008}.
	(H) Numerical simulations of coupled excitable oscillators exhibiting a BT point as an organizing center.
    Adapted from \citet{Buendia2021}.
	(I) Schematic of a drift-pitchfork bifurcation with bifurcation parameter $\epsilon$. A stable circle of fixed points becomes unstable and gives rise to two counter-rotating stable limit cycles when $\epsilon = 0$.
	(J-K) This bifurcation is marked by the coalescence (alignment) of a damped (orange) and a Goldstone (green) eigenmodes at an exceptional point (EP, red circle). Here, two of such bifurcations are shown, with a dynamic phase (purple) in between (J). 
    The corresponding phase diagram in a $O(2)$ model with two species is given in panel K: the (anti)alignement states are nontrivial fixed points, while chiral state corresponds to the two limit cycles.
    Adapted from \citet{Fruchart2021}.
	(L) Slice of the mean-field phase diagram of the nonreciprocal $O(2)$ model \eqref{general_O2} also described in panels J-K. This slice corresponds to nonreciprocal XY and flocking models.
    Adapted from \citet{Fruchart2021}.
	(M) Phase diagram in the thermodynamic limit for a nonreciprocal model with two species of quantum spins. 
    Adapted from \citet{Nadolny2025}.
	(N) Experimental observation of a chiral dynamical instability in a spinor Bose gas interacting with an optical resonator.
    Adapted from \citet{Dogra2019}.
	(O) Active solids made of self-propelled and self-aligning units connected in a lattice undergo collective actuation, as shown by the limit cycle on the right. Adapted from \citet{Baconnier2022}. (P) Phase diagram of nonequilibrium $O(N)$ models obtained by renormalization group analysis. Adapted from \citet{Zelle2024}.
    \label{figure_many_body_bifurcations}
    }
\end{figure*}

As an example, the phases observed in the nonreciprocal $O(2)$ model \eqref{general_O2} with two species are
\begin{itemize}[nosep,left=0pt,label=--]
    \item the trivial disordered state where $\bm{\phi}_1 = \bm{\phi}_2 = 0$
    \item \emph{static ordered states} aligned and antialigned
    \item \emph{rotating waves} (chiral, two of them) 
    \item \emph{standing waves}, swap, a circle of them
    \item \emph{modulated waves}, chiral+swap
\end{itemize}
and these are found in multiple models with the same symmetry, see Fig.~\ref{figure_many_body_bifurcations} (panels D,L,M,P).

\subparagraph{Classification of branches of solutions using equivariant dynamical system theory.}

A more specific classification of the possible symmetry-breaking patterns can be done on a case-by-case basis using the methods of equivariant dynamical systems (Sec.~\ref{symmetry_dynamics}), which depends on the particular action of the symmetry group, and of the bifurcation under consideration. 
As an example, consider a Hopf bifurcation in a $O(2)$ symmetric system \cite{Crawford1991,Golubitsky1988,Chossat2000}. 
Here, a key point is that the normal form of a (nonsymmetric) Hopf bifurcation in polar coordinates has an emergent phase-shift symmetry $z \to z \ee^{\ii \phi}$ (this can directly be checked on the normal form \eqref{hopf_normal_form}). 
This extra symmetry, denoted by $S^1$, distinct from any spatial or internal symmetry, can be traced to the time-translation invariance of the equations, and implies that the normal form should be equivariant under the action of $O(2) \times S^1$.
The simplest nontrivial setup for a $O(2)$-equivariant Hopf bifurcation involves two complex mode amplitudes $z_1(t)$ and $z_2(t)$ and it turns out that around such a bifurcation, it is always possible to choose the coordinate system such that the action of $O(2) \times S^1$ on $\CC^2$ as (i) $\theta \cdot (z_1, z_2) = (\ee^{\ii \theta} z_1, \ee^{\ii \theta} z_2)$ for $\theta \in S^1$, (ii) $\varphi \cdot (z_1, z_2) = (\ee^{-\ii \varphi} z_1, \ee^{-\ii \phi} z_2)$ for $\varphi \in SO(2)$, (iii) $\kappa \cdot (z_1, z_2) = (z_2, z_1)$ for $\kappa \in \ZZ_2$ (where $O(2) = SO(2) \ltimes \ZZ_2$) \cite[XVII \S 1]{Golubitsky1988}.
The isotropy lattice of this action of $O(2) \times S^1$ is shown in Fig.~\ref{figure_symmetries}B and contains the trivial state, a rotating wave, a standing wave, and modulated waves. 
Note that there is no static ordered state appearing in this isotropy lattice, because it is not encompassed in the normal form of the $O(2)$ Hopf bifurcation.

Other symmetries may be analyzed in a similar fashion \cite{Crawford1991,Golubitsky1988,Chossat2000}, see for instance \citet{Ma2025b} for a discussion of permutation symmetry among species in nonreciprocal systems.

\medskip

A path forward to classify phase transitions towards and between dynamical states then consists in considering an ensemble of noisy coupled degrees of freedom, each following the normal forms of a given bifurcation. 
One of the simplest way to do so, for locally coupled systems, consists in promoting the normal form of a bifurcation to a stochastic field theory by adding diffusive couplings and a noise.
This top-down approach, in the spirit of Landau's theory, is dual to a bottom-up approach in which microscopic degrees of freedom are coarsed-grained into a field theory.

\subsubsection{The Hopf field theory}
\label{hopf_field_theory}

\paragraph{The Hopf bifurcation}

The Hopf bifurcation describes the appearance of a limit cycle from a fixed point \cite{Strogatz2018,Kuznetsov2023}.
This bifurcation is captured by the equation
\begin{equation}
\label{hopf_normal_form}
\dot{z} = (\epsilon + i \omega) z - |z|^2 z   
\end{equation}
for the complex numbers $z$ and takes place at $\epsilon = 0$.
This equation is a \enquote{normal form} for the (supercritical) Hopf bifurcation: any system undergoing this bifurcation can be locally transformed into Eq.~\eqref{hopf_normal_form}.
The term $i \omega$ in Eq.~\eqref{hopf_normal_form} encodes its nonreciprocal character; this can perhaps be visualized more easily in terms of the vector $X = (x, y)$ where $z = x + i y$, for which
\begin{equation}
\label{hopf_normal_form_real}
\frac{d}{dt} \begin{pmatrix}
    x \\ y
\end{pmatrix} = \begin{pmatrix}
    \epsilon - [x^2+y^2] & - \omega \\
    \omega & \epsilon - [x^2+y^2]
\end{pmatrix}
\begin{pmatrix}
    x \\ y
\end{pmatrix}.
\end{equation}
As it has limit cycle solutions, Eq.~\eqref{hopf_normal_form} is indeed nonvariational.
At the bifurcation point, the Jacobian of this dynamical system reads
\begin{equation}
    J = \omega \begin{pmatrix}
        0 & -1 \\
        1 & 0
    \end{pmatrix}
\end{equation}
which is purely antisymmetric, betraying the underlying nonreciprocity (\enquote{antagonistic} in the classification of Fig.~\ref{classes_of_nonreciprocity}).

\paragraph{The complex Ginzburg-Landau equation}

In spatially extended systems, $z(t)$ is replaced by a complex field $\psi({\vec r},t)$ with diffusive couplings, and the corresponding equation is known as the complex Ginzburg-Landau equation~\cite{Aranson2002,garcia2012complex,kuramoto1984chemical}
\begin{equation}
    \label{CGLE}
    \partial_t \psi = a \psi+b \nabla^{2}\psi
  + c|\psi|^2 \psi + \eta(\bm{r},t)
\end{equation}
in which $a$, $b$ and $c$ are complex numbers and $\eta$ a complex-valued random noise.
The complex Ginzburg-Landau equation (CGLE) exhibits a number of phenomena related to its nonreciprocal character, as reviewed in \citet{Aranson2002}.
\citet{Chate1996} provide a phase diagram describing the behavior of the CGLE in a two-dimensional finite system from numerical simulations (see also \citet{Aranson2002}). 
The phase diagram however depends on the system size and on the observation time, and the fate of the system in the thermodynamic limit is not entirely known.

The behavior of Eq.~\eqref{CGLE} at the Gaussian level is determined by linearizing the equation about a steady-state. In the case of the trivial state $\psi = 0$, this leads to a linearized equation $\partial_t \delta \Psi_k = L_k \delta \Psi_k + \eta$ for the Fourier modes $\delta \Psi_k$ with
\begin{equation}
    L_k = \begin{pmatrix}
        \epsilon_k & -\omega \\
        \omega & \epsilon_k
    \end{pmatrix}
\end{equation}
in which $\epsilon_k = \epsilon_0 - b k^2$, where $\epsilon_0$ is the distance to the Hopf bifurcation.
This leads to a steady-state correlation matrix of the form
\begin{equation}
    \langle \delta \Psi_k(t+\tau) \otimes \delta \Psi_k(t) \rangle
    =
    -\frac{e^{\epsilon_k |\tau|}}{2\epsilon_k}
    \begin{pmatrix}
    \cos \omega\tau & \sin\omega\tau  \\
    -\sin \omega\tau & \cos \omega\tau
    \end{pmatrix}
\end{equation}
in which the cross-correlations and the oscillations are consequences of the nonreciprocity.

\paragraph{The Hopf universality class.}

\citet{risler2004universal,risler2005universal} have analyzed the fate of the phase transition generalizing the Hopf bifurcation using perturbative renormalization group calculations ($d=4-\epsilon$ expansion). 
It turns out that this phase transition can be formally related to the RG fixed point of model A dynamics with $O(2)$ symmetry (i.e., to the XY universality class).
Using this analogy, \citet{risler2004universal,risler2005universal} predict that this phase transition should exist and lead to an oscillating phase with long-range order for space dimension $d>2$, while no coherent oscillations are expected in $d<2$.
These results were later reproduced and generalized by \citet{daviet2024nonequilibrium,Zelle2024} who also addressed $O(N)$ models.
Note that even though this transition is in indeed predicted to be in the XY universality class, \citet{risler2004universal} also show that additional universal exponents  characteristic of a Hopf bifurcation are present because of the nonequilibrium character of the model (see \citet{Caballero2020} for a similar situation for active models in the Ising universality class).

Extensive numerical simulations have up to now confirmed these predictions, including the existence of an oscillating phase with long-range order and of critical exponents compatible with the XY universality class. 
\citet{wood2006critical,wood2006universality} performed numerical simulations of coupled three-state oscillators for lattices of size $L^d$ going up to $L=80$ in $d=3$, in which a comparison of the rescaled finite-size scaling data with Ising and XY exponents suggested a better collapse with XY exponents.
\citet{Avni2025a,Avni2025b} performed large-scale simulations of a nonreciprocal Ising model exhibiting this transition with lattices of sizes up to $L=320$ in $d=3$, and obtain values of the critical exponents that are compatible with XY exponents but not with Ising exponents.
In addition, these numerical simulations show that the oscillating phase can be seen as a classical time crystal (Sec.~\ref{time_crystals}).
Note that discontinuous (subcritical) transitions to oscillatory phases can also happen \cite{Guislain2024c}.

Things are less clear in space dimension $d=2$. \citet{risler2004universal,risler2005universal} conjectured that quasi-long-range may be present. 
More recent studies based on a mapping to a version of the Kardar-Parisi-Zhang (KPZ) equation with compact variables suggest that the phase with algebraic order does not persist in the limit of large systems (nor an equivalent of the Berezinskii-Kosterlitz-Thouless (BKT) transition) but are replaced with a stretched exponential decay of correlations corresponding to the rough phase of the KPZ equation \cite{altman2015two,wachtel2016electrodynamic,Deligiannis2022,Daviet2025}, see Sec.~\ref{time_crystals} for more details on this mapping.
This can be interpreted in the light of earlier results from \citet{Aranson1998b}, who argue that the modified effective interaction between spiral defects (compared to that between XY vortices) allows for a finite density of defects at any temperature, potentially suppressing the algebraic order.

\subsubsection{Exceptional phase transitions}
\label{ept}

\paragraph{Bifurcations and exceptional points.}
\label{bifurcations_eps}

Exceptional points correspond to matrices that are non-diagonalizable (Sec.~\ref{non_normal_matrices}) and encode a \enquote{unidirectional} kind of nonreciprocity in the classification of Fig.~\ref{classes_of_nonreciprocity}.
Several bifurcations are associated with exceptional points.
One of the simplest is the Bogdanov–Takens bifurcation which has the normal form \citet{Kuznetsov2023}
\begin{subequations}
\begin{align}
    \dot{x}_1 &= x_2
    \\
    \dot{x}_2 &= \beta_1 + \beta_2 x_1 + x_1^2 \pm x_1 x_2
\end{align}
\end{subequations}
Equilibria satisfy $x_2 = 0$ and $\beta_1 + \beta_2 x_1 + x_1^2 = 0$. 
At the point $\beta_1 = \beta_2 = 0$ in parameter space, the Jacobian at the only equilibrium point $(x_1, x_2) = (0,0)$ is not diagonalizable, and takes the form
\begin{equation}
    J = \begin{pmatrix}
    0 & 1 \\
    0 & 0
    \end{pmatrix}.
\end{equation}
The Bogdanov–Takens bifurcation has codimension two, meaning that one generically has to tune two parameters to reach it. 
Nevertheless, it often arises in systems ranging from neuroscience \cite{Buendia2021,Cowan2016,Cowan2014,Buice2007} to ecology \cite{Touboul2018}.
For instance, \citet{Childs2008} considered a periodically forced Kuramoto model.
In a comoving frame, this mean-field model can be described (on the so-called Ott-Antonsen manifold) by the two-dimensional dynamical system
\begin{equation}
    \label{forced_kuramoto_oa}
    \dot{z} = \frac{1}{2} [ (K z + F) - (K z^* + F) z^2]
    - [\Delta + \ii \Delta \omega] z.
\end{equation}
in which $z$ is the Kuramoto complex order parameter, $K$ is the coupling strength, $F$ is the forcing strength, $\Delta \omega$ is the detuning between the forcing frequency and the center frequency of the Lorentzian distribution of frequencies in the Kuramoto model, which has size $\Delta$.
Figure \ref{figure_many_body_bifurcations}G shows the bifurcation structure of Eq.~\eqref{forced_kuramoto_oa} computed by \citet{Childs2008}, which exhibits multiple codimension-two bifurcations (points in the 2D phase diagram), including a Bogdanov-Takens point (BT, in black).
Similarly, it arises in the phase diagram of coupled excitable oscillators (\ref{figure_many_body_bifurcations}H) analyzed in \citet{Buendia2021}. 
See also \citet{Hesse2017} for an analogy between the BT point seen as a transition between fold and Hopf bifurcations, and the saddle-node loop bifurcation seen as a transition between the saddle loop (aka saddle homoclinic) and a SNIC bifurcation.

Another class of bifurcations associated with exceptional points are drift-pitchfork bifurcations (also known as parity-breaking bifurcations), that correspond to the normal form~\cite{Kness1992,Knobloch1995} 
\begin{subequations}
\label{parity_breaking}
\begin{align}
    \dot{x} &= y \label{driftpf_x} \\
    \dot{y} &= \epsilon \, y - y^{3}.
\end{align}
\end{subequations}
The variable $x$ may either be a unbounded real number $x \in \RR$ or a compact variable $x \in S^1$ such as an angle or the position in a space with periodic boundary conditions (in which case the first line may be replaced with $\dot{x} = \sin y$ at lowest order).
When $\epsilon < 0$, the system simply has a line or a circle of fixed points corresponding to all the possible values of $x$.
As the variable $y$ undergoes a pitchfork bifurcation, the full dynamical system undergoes a drift-pitchfork bifurcation and $x$ drifts at a speed controlled by $y$. 
The spontaneous $\ZZ_2$ symmetry breaking associated with the pitchfork bifurcation corresponds to the appearance of counter-rotating limit cycles with angular frequencies $y_\infty^\pm = \pm \sqrt{\epsilon}$ (when $x$ is an angle) or to steadily moving states with a drift speed $y_\infty^\pm$ (when $x$ is unbounded).
Hence, the period of oscillation diverges at the bifurcation (from the limit cycle side), contrary to a Hopf bifurcation where it remains finite.
The sign of the angular frequency/drift speed is determined by the sign of the initial condition $\operatorname{sgn}[y(t=0)]$. 
Hence, the sense of rotation (clockwise or counterclockwise) of the limit cycle is not determined in advance: it depends on the initial conditions, in contrast with the case of a Hopf bifurcation.

The drift-pitchfork bifurcation can be seen as a degenerate case of a pitchfork of limit cycles, in which Eq.~\eqref{driftpf_x} is replaced with $\dot{x} = \omega_0 + y$ (or a more complicated version of this, see \citet{Kuznetsov2023,Guckenheimer2013} for details). 
Similar classes of bifurcations can arise where the pitchfork is replaced by another bifurcation such as a fold bifurcation \cite{Kuznetsov2023}. 
In all of these cases, the nonreciprocity and exceptional points, when they are present, are easy to see in highly symmetric situations such as the normal forms we consider, because one can reduce the analysis to that of fixed points. 
This does not hold in the general case, where the same limit cycles or other attractors can be arbitrary; instead, one can consider covariant Lyapunov vectors to generalize the notion of exceptional points \cite{Weis2025}.

\paragraph{Critical exceptional points.}
\label{cep}
In many-body systems, these bifurcations correspond to nonreciprocal phase transitions in which the relevant fields are coupled nonreciprocally, with unidirectional nonreciprocity in the classification of Fig.~\ref{classes_of_nonreciprocity} \cite{Fruchart2021}.
The corresponding critical points, known as critical exceptional points, have been analyzed in detail by
\citet{Zelle2024} in the case of nonequilibrium $O(N)$ models and by \citet{Hanai2020}.
In these models, spontaneous symmetry breaking leads to the presence of neutral modes (Goldstone modes) in the symmetry-broken phase, that effectively reduce the codimension of critical exceptional points \cite{Fruchart2021}.
\citet{Zelle2024} consider the field theory
\begin{align}
\label{ON_inertial_field_th}
\begin{split}
\partial_t^2 \bm{\phi} &+ (2\gamma + u\|\bm{\phi}\|^2 - Z\nabla^2)\partial_t \bm{\phi} 
\\
&+ r \bm{\phi}
+ \lambda\|\bm{\phi}\|^2 \bm{\phi} - v^2\nabla^2 \bm{\phi} + \bm{\xi} = 0
\end{split}
\end{align}
in which $\bm{\phi}(t,\bm{r})$ is a $N$ component order field and $\bm{\xi}$ are independent Gaussian white noise with zero mean and variance $D_0$.
They show that this field theory exhibits critical exceptional points, around which a linear dispersion relation of the form
\begin{equation}
    \omega_\pm(\bm{q}) = - \frac{\ii}{2}(D \lVert \bm{q} \rVert^2 + \epsilon) \pm v \lVert \bm{q} \rVert
\end{equation}
in which $\epsilon$ represents the distance to threshold.
In addition, giant fluctuations where the mode occupation
\begin{equation}
    \langle \bm{\phi}(-\bm{q},t) \bm{\phi}(\bm{q},t) \rangle \sim \frac{D_0}{\lVert \bm{q} \rVert^4}
\end{equation}
is superthermal, i.e. it is enhanced with respect to the equilibrium scaling $\sim 1/\lVert \bm{q} \rVert^2$.
These enhanced irreversible fluctuations, which are a manifestation of the out of equilibrium nature of critical exceptional points \cite{Zelle2024}, are analyzed in \citet{Suchanek2023a,Suchanek2023b,Suchanek2023c} through the mesoscopic entropy production around CEPs.

Both of these features, associated with the linearized dynamics, stem from its non-normal character (Sec.~\ref{non_normal_amplification_and_noise}).
They can be understood from the simple linearized field theory 
\begin{equation}
\partial_t \bm{\psi}
=
    \begin{pmatrix}
        \epsilon & 1
        \\
        0 & \epsilon
    \end{pmatrix}
    \bm{\psi}
    +
    \begin{pmatrix}
        D_0 + \Delta & D_{12}
        \\
        D_{21} & D_0 - \Delta
    \end{pmatrix}
    \nabla^2 \bm{\psi}
    + \bm{\xi}
\end{equation}
in which we coupled two fields $\psi_1$ and $\psi_2$ (gathered in a vector $\bm{\psi}=(\phi_1,\phi_2)$).
Here, the homogeneous dynamics has an exceptional point, that becomes critical when $\epsilon$ vanishes, and we have added an arbitrary diffusive coupling.
The corresponding dispersion relation reads
\begin{equation}
    s_{\pm} = \epsilon - D_0 q^2 \pm \ii v |q| + \mathcal{O}(q^3)
\end{equation}
in which $v = \sqrt{D_{21}}$ is (up to a time scale) a velocity.
In other words, a critical exceptional point can convert cross-diffusive couplings into propagative modes, aka sound modes. 
This feature has been identified in \citet{Coullet1985b,Coullet1985c} in the context of pattern forming instabilities where the presence of the critical exceptional point can be traced to Galilean invariance that produces a neutral (Goldstone) mode, see \citet{Fauve2020} for a recent review.
In terms of dynamical exponents, this amounts to converting $z=2$ to $z=1$, but in fact there is no unique dynamics as both a diffusive and a propagative timescale that scale differently with system size ($\tau_{\text{d}} \sim q^-2$ and $\tau_{\text{p}} \sim q^{-1}$) coexist \cite{Hanai2020,Zelle2024}.

In mean-field, nonreciprocal phase transitions with critical exceptional points have been predicted (and in certain cases, observed) in many-body systems spanning the natural sciences \cite{Fruchart2021}.
These include coupled oscillators~\cite{Hong2011,Hong2011b,Hong2014},
driven-dissipative quantum condensates \citet{Hanai2019,Belyansky2025,Zou2025} and spin systems \cite{Chiacchio2023,Nadolny2025,Hanai2025,Jachinowski2025}, and active matter~\cite{Fruchart2021,Saha2020,You2020}, 
but also situations analyzed from a more hydrodynamic perspective, in which a bifurcation such as the drift-pitchfork arise like in pattern forming systems~\cite{Coullet1985b,Coullet1989,Malomed1984,Douady1989,Brachet1987,Pan1994,Bensimon1989}, 
fluid dynamics~\cite{Xin1998,Thiele2004}, 
excitable media~\cite{Kness1992,Krischer1994}, and 
coupled lasers~\cite{Hassan2015,Clerkin2014,Soriano2013}.

Going beyond mean-field, \citet{Zelle2024} predicts using renormalization group calculations that mean-field critical exceptional lines that separate the static order phase from the rotating wave (chiral) phase are replaced, either by a sequence of transitions between static order/disorder/chiral phases, or by fluctuation-induced first-order phase transitions between the static and chiral phases, as shown in Fig.~\ref{figure_many_body_bifurcations}P.

\citet{Liu2025c} analyzed numerically and analytically a 1D non-reciprocal $O(2)$ model of the form \eqref{general_O2} with diffusive couplings of the form $D_{ab} \Delta \bm{\phi}_b$ and a Gaussian white noise (Fig.~\ref{figure_scalings}A). In 1D, no long-range order is expected nor observed, and the phase transition at zero noise becomes a crossover at finite noise strength. 
Nevertheless, different scaling regimes are observed in the roughness exponent $\alpha$ appearing in the Family-Vicsek scaling $w(L,t) = L^\alpha f(t/L^z)$ for the standard deviation of dephasings.
In addition to the Edwards-Wilkinson scaling where $\alpha_{\text{EW}}=1/2$, a regime associated with the critical exceptional point with $\alpha_{\text{CEP}}\simeq1.35$ is reported. 
These simulations also confirm the observation of \citet{Zelle2024} that there is no uniquely defined dynamical exponent $z$, because of the presence of both a ballistic and a diffusive components near the CEP, that scale differently with system size.
In addition, \citet{Liu2025c} report the existence of a regime of chiral disorder, roughly corresponding to the mean-field chiral phase, where the fact that phases are compact variables becomes crucial, leading to a very different logarithmic scaling $w^2 \propto 2 \gamma \log L$ with $\gamma=1/4$.

\subsubsection{The SNIC field theory}
\label{snic_field_theory}

We conclude this list with an example of a nonlocal bifurcation: the saddle-node on an invariant cycle (SNIC) bifurcation. 
The SNIC bifurcation describes the creation of a limit cycle by a saddle-node bifurcation that connects heteroclinic orbits together~\cite{Strogatz2018}.
As in the drift-pitchfork, the period of oscillation diverges at the bifurcation from the limit cycle side.
It arises, for instance, in the mean-field nonreciprocal Ising model (where multiple SNIC bifurcations (related by symmetry take place at the same time; see Fig.~\ref{figure_many_body_bifurcations}D and \citet{Guislain2024,Avni2025a,Avni2025b}) as well as in the effective dynamical system describing forced Kuramoto oscillators (Fig.~\ref{figure_many_body_bifurcations}G and \citet{Childs2008}). 

\citet{assis2011infinite} conjectured that the corresponding phase transition does not exist in spatially extended systems of arbitrary dimension based on the instability of the static ordered state of a three-state oscillator model introduced in \citet{wood2006critical,wood2006universality}. 
\citet{Avni2025a,Avni2025b} reported that a static ordered state can exist when the nonreciprocal couplings are not fully anti-symmetric thanks to a droplet-capture mechanism that stabilises the phase, but didn't observe a direct transition between static order and oscillating phases in simulations.
To the best of our knowledge, a field-theoretical analysis of this class of bifurcation has not yet been carried out.

\subsection{Nonreciprocal spin models}

Spin models are often used as paradigmatic in statistical physics~\cite{Baxter1982}. 
In this paragraph, we review a few nonreciprocal spin models that have been explored in the literature.

\subsubsection{Nonreciprocal Ising models}

\paragraph{How to make discrete models nonreciprocal?}

The equilibrium Ising model is one the simplest models of phase transitions. 
It consists of binary degrees of freedom (Ising spins) $\sigma_n = \pm 1$ (or $0,1$, etc.) that are coupled by interactions $J_{mn}$ that enter in a Hamiltonian $H(\bm{\sigma}) = \sum_{m,n} J_{mn} \sigma_m \sigma_n$ in which $\bm{\sigma}$ is a vector containing all the $\sigma_n$, and where the connectivity of the system is defined by the matrix $J_{mn}$. Note that only the symmetric part of the interaction matrix enters the Hamiltonian, so in equilibrium Ising models, we can always assume $J_{mn} = J_{nm}$ without loss of generality. 

This is not the case in nonreciprocal variants of the Ising model. 
Instead of starting with an energy function $H(\bm{\sigma})$, one considers a Markov chain giving the probability of going from a state $\bm{\sigma}$ to a state $\bm{\sigma'}$ (called a kinetic Ising model). 
In this formulation, it is possible to meaningfully consider an asymmetric $J_{mn}$ that encodes the fact that the way spin $m$ flips as a function of the value of spin $n$ is different from the way spin $n$ flips as a function of the value of spin $m$. 

For instance, this idea can be implemented in terms of a \enquote{selfish energy} \cite{Avni2025a} as follows.
We assume that each spin $n$ is endowed with a selfish energy function $E_n(\bm{\sigma})$ that it tends to minimize, which is implemented through transition rates $w(\vec{\sigma}|\vec{\sigma}')$ from configuration $\vec{\sigma}$ to $\vec{\sigma}'$ given by
\begin{equation}
\label{rates}
w(F_{n}\vec{\sigma}|\vec{\sigma})=\frac{1}{2\tau}\left[1-\tanh\left(\frac{\Delta E_n}{2k_B T}\right)\right]
\end{equation}
in which $T$ is the temperature, $k_B$ the Boltzmann constant, and $\tau$ is a characteristic time for single spin flips. Here, $\Delta E_n =  E_n(F_{n}\vec{\sigma}) - E_n(\vec{\sigma})$
is the difference in selfish energy between the states while $F_{n}$ is a single-spin-flipping operator that changes $\sigma_n$ to $-\sigma_n$ while keeping all the other spins unchanged. All other rates are taken to be zero.
This formulation encompasses several nonreciprocal spin dynamics~\cite{Lynn2021,Mello2003,Loos2023,Guislain2024,Lima2006,Sanchez2002,Lipowski2015}.
Note that we have used the generalized Glauber dynamics \cite{Glauber1963} to map energy functions to rates, but the same idea can be applied to other rules (such as the Metropolis dynamics).

\paragraph{Nonreciprocal Ising models.}
\label{nr_ising}
Several variants of nonreciprocal Ising models have been considered. For instance, \citet{Sanchez2002,Lima2006,Lipowski2015,Godreche2009,Godreche2011,Godreche2017} considered Ising models on directed networks (see \citet{Dorogovtsev2008} for a review of 
critical phenomena in complex networks). 
One of the simplest instantiation of this idea, discussed in \citet{Seara2023}, consists in having a regular lattice with translation invariant couplings that break mirror symmetry: for instance, in 1D, one can take $n \in \ZZ$ and $J_{m,n} = J_{L} \delta_{n,m+1} + J_{R} \delta_{n,m-1}$.
At the field theoretical level, this has been phenomenologically modeled as 
\begin{equation}
    \partial_t m = a m - b m^3+ (\bm{v} \cdot \nabla) m + D \Delta m + \text{noise}
\end{equation}
in which $m(t,x)$ is the magnetization, which follows a $\phi^4$ field theory with an advective term due to nonreciprocity in addition to the usual diffusive coupling (that may in general be anisotropic).
This situation, similar to models exhibiting a non-Hermitian skin effect like the Hatano-Nelson model \cite{Hatano1996}, leads to unidirectional propagation of perturbations. 
See \citet{Godreche2011,Weiderpass2025} for analytical studies of this model.

\citet{Avni2025a,Avni2025b} considered an Ising model with two species that interact nonreciprocally. 
More specifically, two species $a=A,B$ are put on each site $i \in \ZZ^d$ of a hypercubic lattice (so $m=(i,a)$ with the notations above). Intraspecies are reciprocal, but the interspecies couplings may have an arbitrary asymmetry, which is encoded in $J_{ij}^{ab} = \delta_{ij} (K_+ \delta^{ab} + K_{-} \epsilon^{ab}) + J_a \delta^{ab} [\sum_{k \text{n.n. of $i$}}\delta_{ik}]$ ($i$,$j$ label sites on $\ZZ^d$ while $a,b$ label species, $\delta$ is the Kronecker symbol while $\epsilon$ is the Levi-Civita symbol). 
\citet{Guislain2024,Guislain2024b} considered a similar but slightly different model with two species and a varying range of interactions.
In these models, the nonreciprocity can lead to oscillating phases through different kinds of bifurcations, as discussed in Secs.~\ref{hopf_field_theory}, \ref{snic_field_theory}, and \ref{time_crystals}.
See also \citet{Blom2025} for an analysis through moments with a Bethe-Guggenheim approximation and \citet{Garces2025} for other Ising models with nonreciprocal couplings.

\subsubsection{Nonreciprocal XY models}
\label{nonreciprocal_xy}

Nonreciprocal XY models are usually described by a Langevin equation of the form
\begin{equation}
    \partial_t \theta_m = \sum_{n} J_{mn} \sin(\theta_n - \theta_m) + \text{noise}
    \label{nrxy}
\end{equation}
in which the XY spins can be represented by angles $\theta_m(t) \in S^1 \simeq [-\pi,\pi]_{\text{per}}$, and where the couplings $J_{mn}$ have to be further specified.
In equilibrium systems, this equation can be derived from the relaxational dynamics $\partial_t \theta_m =-\partial H/\partial \theta_m$ of a Hamiltonian $H=-\sum_{m,n} J_{mn} \cos(\theta_n - \theta_m)$ in which the antisymmetric part of $J_{mn}$ drops out, so one can assume $J_{mn} = J_{n m}$ without loss of generality. 
As for other spin models, this assumption is lifted in nonreciprocal systems where we start from the dynamics \eqref{nrxy}. 
With two species (like in Sec.~\ref{nr_ising}), this dynamics may be seen as a particular case of the nonreciprocal Kuramoto dynamics considered in \citet{Fruchart2021}, with vanishing natural frequencies.

\citet{Loos2023,Bandini2025,Liu2025d,Popli2025,Dopierala2025,Shi2025} considered models with cones of vision in which the coupling strength $J_{mn}$ is not a constant, but couples the spin orientation with lattice positions as follows: one sets $J_{mn}(t) = J_{\text{CV}}(\psi_{mn}(t))$ in which $\psi_{mn} = \text{angle}(\bm{r}_n - \bm{r}_m) - \theta_m$, where $\bm{r}_n$ is the (fixed) position of the spin $n$. The function $J_{\text{CV}} : S^1 \to \RR$ can be interpreted as implementing a cone of vision (or field of view) that weights the interactions between spins.
This generalizes spin models with fore-aft asymmetry considered in \cite{Dadhichi2020}.
\citet{Loos2023} reported that such a XY model with vision cone interactions exhibits long-range order in space dimension $d=2$ (which would be prohibited at equilibrium by Hohenberg-Mermin-Wagner arguments), as well as directional propagation of defects.

\citet{Dopierala2025,Popli2025} showed that this model shares the same mean-field description as constant density flocks (see \citet{Toner2012,Besse2022}), rationalizing the presence of long-range order. They also showed that, like in constant density active clock models, this order is metastable to the nucleation of topological defects, eventually leading to an active foam of aster defects.
In addition, \citet{Dopierala2025,Popli2025} pointed out that the anisotropy of the underlying lattice is relevant \cite{Solon2022}: even though the dynamics is apparently isotropic, the anisotropy of the underlying lattice, which enters the dynamics through $\psi_{mn}$, is amplified by fluctuations and dominates the large-scale physics.

\subsubsection{Nonreciprocal Heisenberg model}

\citet{Das2002,Das2004,Bhatt2025} considered a nonreciprocal version of the classical
Heisenberg $O(3)$ spin chain where the exchange coupling is nonsymmetric. The dynamics is given by
\begin{equation}
    \dot{\bm{S}}_n = \bm{S}_n \times (\bm{S}_{n+1} + \alpha \bm{S}_{n-1})
    \label{heisenberg}
\end{equation}
in which $\bm{S}_n(t) \in S^2 \subset \RR^3$ with $\lVert \bm{S}_n(t) \rVert = 1$. The usual case is recovered when $\alpha = 1$ and has a canonical structure. It turns out that, with nearest neighbors couplings and open boundary conditions, Eq.~\eqref{heisenberg} has a symplectic structure in terms of transformed spin variables, although at the cost of an exponential position dependence in the change of
variables \cite{Bhatt2025}. This makes it applicable in systems with periodic boundary conditions only when $|\alpha| = 1$.
The model is reported to exhibit helical configurations without stochastic forcing, as well as a nonequilibrium critical point and a spatiotemporally chaotic phase in the mean-field.
We also refer to \citet{Galda2016,Galda2019} in which nonconservative classical spin dynamics are analyzed through the perspective of non-Hermitian Hamiltonians by taking the classical limit of a quantum Hamiltonian using $SU(2)$ spin-coherent states.

\subsection{Temporal crystals and beyond}
\label{time_crystals}

The notion of \enquote{time crystal} as a many-body system with spontaneously broken time-translation symmetry has emerged in the literature following the proposal of \citet{Wilczek2012,Shapere2012}, who borrowed the term from \citet{winfree1980geometry}.
Several definitions have been considered in the literature, and recent progresses are reviewed in \cite{Khemani2019,Yao2018,Zaletel2023,Sacha2017}.
In particular, continuous time crystals, also called boundary time crystals, arise from a continuous (time-independent) drive and therefore break continuous time translation symmetry, see \citet{Iemini2018,Buca2019,Kongkhambut2022} for the quantum versions.
As exemplified by the Hopf (Sec.~\ref{hopf_field_theory}) and exceptional field theories (Sec.~\ref{ept}), time-crystal phases are expected in nonreciprocal systems at the mean-field level.
Two intertwined questions arise in this context, about the \emph{spatial} coherence of oscillations and about their \emph{temporal} coherence.
Indeed, we can ask whether there is a macroscopic order parameter $z(t)$, defined as an average over the entire system, that oscillate forever and, if so, we can also ask whether the oscillation of this macroscopic order parameter is temporally coherent, i.e. if it keeps track of its initial phase. 
This is not the case for a noisy oscillator, where the noise eventually makes the oscillator loose temporal coherence \cite{Gaspard2002b}. The existence of perfectly coherent oscillations in the thermodynamic limit $L \to \infty$ can be seen as a breaking of ergodicity \cite{Chen2004,Liggett2012,Martinelli1999}.
It means that even at arbitrarily long times, the probability distribution function $P(t) \equiv U(t) P_0$ describing the systems does not converge to a constant stationary distribution $P^{\text{ss}}$ (such that $U(t) P^{\text{ss}} = P^{\text{ss}}$, where $U(t)$ is the evolution operator acting on probability distributions) but instead that there is a limit-cycle $P_{\text{LC}}$ satisfying $P_{\text{LC}}(t)=P_{\text{LC}}(t+T_{\rm osc})$ where $T_{\rm osc}$ is the oscillation period \cite{Carollo2022}. 
As a consequence, the distance between the distributions obtained from evolving two initial conditions $P_1$ and $P_2$ may not vanish at long times, but instead we can have
\begin{equation}
    \lim_{t \to \infty} \lVert U(t) P_1 - U(t) P_2 \rVert \neq 0.
    \label{pinfty_nonzero}
\end{equation}
In many-body systems, however, interactions can overcome noise, producing coherent oscillations over arbitrarily long periods.  These features can both arise in systems that follow their mean-field dynamics with an arbitrarily small noise. This is usually the case of all-to-all coupled systems as the number of individual units tends to infinity.
For instance, this arises in the Kuramoto model, as reviewed in \citet{Acebron2005} \footnote{One of the key takeaways of the Kuramoto model is that synchronization can occur despite the coupled oscillators all having different natural frequencies (following some continuous distribution, such as a Lorentzian). 
This form of disorder leads to several complication and is not present in the models discussed here.}.
One can also ask whether coherent oscillations can arise in locally coupled systems, or in other situations where it is not guaranteed that the mean-field dynamics is an appropriate description.

\begin{figure*}
    \centering
    \includegraphics[width=0.9\linewidth]{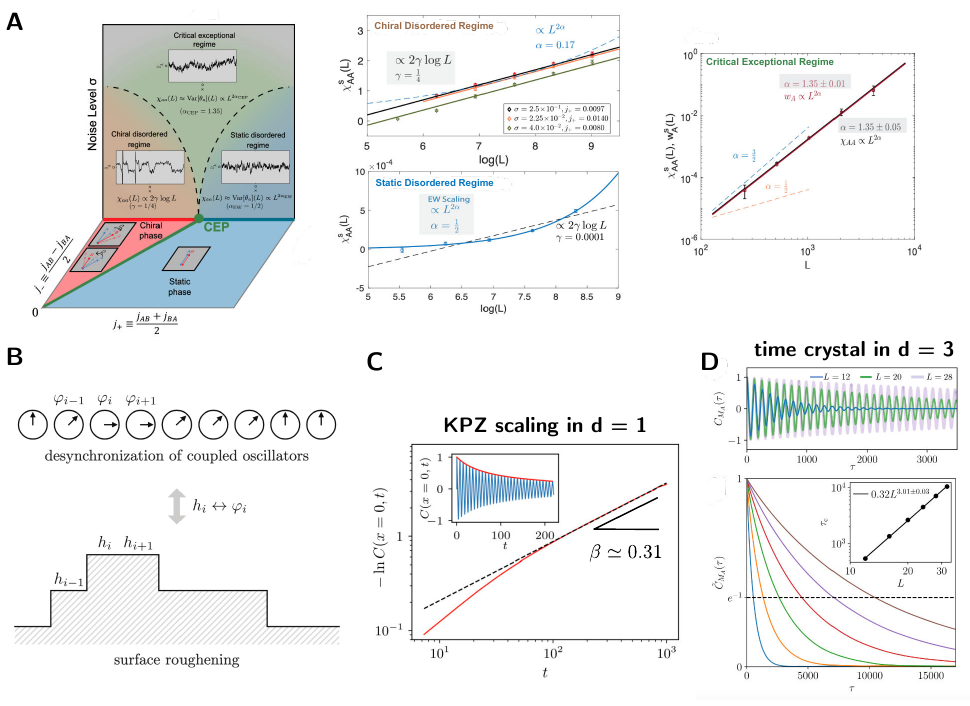}
    \caption{
    \label{figure_scalings}
    \textbf{Scaling behaviors of nonreciprocal critical systems.}
    (A) Scaling behavior of a stochastic 1D nonreciprocal $O(2)$ model. The model displays no long-range order, but it does display multiple scaling behaviors as shown in the phase diagram on the left and evidenced by the corresponding scaling behaviors of the logarithmic correlation function $\chi_{AA}(L) = - \log C_{AA}$ in which $C_{AA}$ is the temporal correlation function of the spatially-averaged order parameter for one of the two species. 
    Adapted from \citet{Liu2025c}.
    (B) Mapping between the desynchronization of coupled oscillators and surface roughening. 
    Adapted from \citet{Avni2025b}.
    (C) Autocorrelation function obtained from numerical simulations of a 1D van der Pol chain showing oscillations inside an envelope with KPZ scaling (the theoretical value is $\beta = 1/3$ in $d=1$).
    Adapted from \citet{Daviet2025}.
    (D) Behavior of the time correlation functions (top) and their envelopes (bottom) in a 3D nonreciprocal Ising model with two species. The inset displays the scaling $\tau(L) \sim L^d$ (with $d=3$) of the relaxation time $\tau$, showing that the system is a time crystal. 
    Adapted from \citet{Avni2025b}.
    }
\end{figure*}

\subsubsection{Stability of spatially coherent oscillations}
\label{stability_spatially_coherent}

The stability of spatially coherent oscillations against fluctuations depends on whether the time-translation symmetry that is spontaneously broken by the oscillations is discrete or continuous.
\citet{Bohr1987,Grinstein1988,bennett1990stability,Grinstein1994} argued that in spatially isotropic systems with short-range interactions, coherent oscillations are unstable in the thermodynamic limit if they arise from the spontaneous breaking of a discrete time-translation invariance. 
This is effectively the case when the period of oscillation is always commensurate with a discrete time unit characterizing the evolution.
In this case, the nucleation and growth of dephased droplets from fluctuations can destroy the time crystal in any dimension, provided that the system is isotropic and that the noise is unbounded (e.g. Gaussian), so that it can create arbitrarily large droplets~\cite{bennett1990stability,Grinstein1994}. 
Because the broken symmetry is discrete, the domain walls are robust (like Ising domain walls), and hence large enough droplets are always stable. Their successive nucleations eventually destroy the coherent oscillations.

In the case of spontaneously broken continuous time-translation invariance, domain walls between droplets progressively disappear, similar to XY domain walls \cite{Binder1997,Avni2025b}.
\citet{bennett1990stability,grinstein1993temporally,Grinstein1994} analyzed this case through a mapping between the desynchronization of local oscillators and the roughening of surfaces, described by the Kardar-Parisi-Zhang (KPZ) equation~\cite{Kardar1986}.
In the KPZ universality class, surfaces are always rough in $d \leq 2$, while two phases (rough or smooth) are possible for $d>2$ \cite{Livi2017,Kamenev2023}. This mapping suggests that temporal order can exist in $d>2$ but not in $d\leq 2$ in isotropic systems.
Similarly, \citet{chan2015limit,daviet2024nonequilibrium} suggested an analogy with the Mermin-Wagner theorem~\cite{Mermin1966,Hohenberg1967} in the time domain, according to which the fluctuations of the Goldstone modes associated with broken time-translation invariance destroy order in $d \leq 2$, leading to the same conclusion.

The idea can be understood from the normal form \eqref{hopf_normal_form} of a Hopf bifurcation: deep in the limit cycle regime, the radial direction is very stiff and quickly relaxes to its steady-state value when perturbed, so one can adopt a phase-only description in which the order parameter $z(t)$ is approximated as $z(t) \simeq R_0 \ee^{\ii \varphi(t)}$.
This method, known as phase reduction, works for general limit cycles \cite{kuramoto1984chemical,Teramae2009,Kuramoto2019,Nakao2015,
Pietras2019} except in very specific cases where there is no isochron foliation of the neighborhood of the limit cycle (see \citet{Weis2025} and references therein).
The phase variable that describes the position of the system on the limit cycle can, under certain approximations, be described by the equation
\begin{equation}
\label{kpz}
\frac{\partial\varphi}{\partial t}= \omega_0 + \nu\nabla^{2}\varphi+\lambda \left(\nabla\varphi\right)^{2} + \eta(t)
\end{equation}
where the local phase $\varphi(t,\bm{r})$ plays the role of the surface height in the KPZ equation \cite{kuramoto1984chemical,grinstein1993temporally,Chate1995,altman2015two,wachtel2016electrodynamic,Gutierrez2023}, as illustrated in Fig.~\ref{figure_scalings}B.
Here, $\nu$ and $\lambda$ are parameters, and $\eta$ is a noise.
As the variable $\varphi$ is an angle (defined on a compact space, the circle $S^1$), Eq.~\eqref{kpz} is known as the compact KPZ equation \cite{Chen2013,altman2015two,wachtel2016electrodynamic,Aranson1998,Sieberer2016,Zamora2017,Sieberer2018,Deligiannis2022,Daviet2025}.
Indeed, if the phase fluctuations follow the KPZ universality class, then 
$\langle
    [\varphi(\bm{r},t)-\varphi(0,t)]^2
    \rangle_{\text{c}}
    \sim
    \lVert \bm{r} \rVert^{2\chi}$ 
where $\chi$ is the roughness exponent~\cite{Kloss2012}.
The compactness of the phase allows for topological defects that, if they proliferate, eventually destroy long-range order; this is expected to occur in $d=1,2$.
In spite of this, KPZ scalings can be observed on intermediate time and length scales, as reported in numerical simulations in \citet{Daviet2025}, see Fig.~\ref{figure_scalings}C.
In 3D, defect rings are expected to be small in some regimes, and hence may not completely destroy long-range order \cite{grinstein1993temporally}, although to the best of our knowledge this has not yet been proven rigorously.

The theoretical arguments reviewed above are in agreement with results of numerical simulations of cellular automata and coupled map lattices in various space dimensions, in which periodic oscillations for $d \geq 3$ were reported \cite{Chate1991,Chate1992,Chate1995,Chate1997,Gallas1992,Binder1992,Hemmingsson1993,Losson1995,Pomeau1993,Chaté1996,Avni2025a,Avni2025b}.
Similar behaviors were reported in simulations of lattice predator-prey models~\cite{Lipowski1999,Lipowski2000,Mobilia2006,Antal2001b,Antal2001,Tauber2024,Odor2004,Tauber2014}.
Note that a similar question was considered for the synchronization of locally coupled oscillators with randomly distributed frequencies (like in the Kuramoto model), but the physics is different because of the random frequencies, as discussed in \citet[\S~IV.A]{Acebron2005}.

\subsubsection{Temporal coherence of oscillations}

In parallel of spatial coherence, one can also consider temporal coherence. To see what it means, consider a noisy phase oscillator
\begin{equation}
    \dot{\phi} = \omega + \sqrt{2 D_\phi} \eta(t)
\end{equation}
with $\phi(t) \in S^1$, where $\eta(t)$ is a white noise with $\langle\eta\rangle = 0$ and $\langle\eta(t) \eta(t')\rangle = \delta(t-t')$. 
The average value $\braket{\phi}$ always oscillates: averaging the equation of motion leads to $\braket{\phi(t)} = \phi(0) + \omega t$. 
However, this oscillator only keeps track of time (i.e. of its initial phase) for so long. 
Indeed, the temporal correlation function behaves as
\begin{equation}
    C(t) \equiv \langle e^{i [\phi(t) - \phi(0)]} \rangle \propto \cos(\omega t) e^{- t/\tau}
\end{equation}
in which $\tau = 1/D_\phi$, meaning that the oscillator loses memory of its original phase after $\mathcal{Q} = \omega \tau$ oscillations ($\mathcal{Q}$ is known as the quality factor of the noisy oscillator).
Any noisy oscillator behaves in this way at long times, as described in \citet{Gaspard2002b}.
Thermodynamic bounds on the number $\mathcal{Q}$ of coherent oscillations are described in \citet{Cao2015,Oberreiter2022,Shiraishi2023,Ohga2023,Santolin2025,Kolchinsky2025} (note that these do not directly apply when time-reversal-odd variables are present, like in underdamped mechanical systems, see \citet{Pietzonka2022}).
A bound conjectured in \citet{Oberreiter2022}, and proven under certain assumptions in \citet{Shiraishi2023}, provides a constraint on the average entropy per oscillation $\Delta S$ by the number of coherent oscillations $\mathcal{N} \equiv \tau_c/T_{\rm osc}$ through the inequality $\Delta S \geq 4\pi^2 \mathcal{N}$. 
As a consequence, the rate of entropy production per unit volume $\dot{s} \equiv {\Delta S}/(L^d T_{\rm osc})$
must satisfy $\dot{s} \geq ({4 \pi^2}/{T_{\rm osc}^2}) \, {\tau_c(L)}/{L^d}$.

In a time crystal, we expect an infinite quality factor in the thermodynamic limit, and so, provided that the frequency of oscillation does not depend on system size in the thermodynamic limit, this means that the coherence time $\tau(L)$ must diverge as $L \to \infty$. 
In classical systems with short-range interactions, this has for instance been reported in \citet{Oberreiter2021} in a periodically driven 2D system (a discrete time crystal) and in \citet{Avni2025a,Avni2025b} in a non-reciprocal version of the Ising model (a continuous time crystal), see Fig.~\ref{figure_scalings}D. 
In both cases, a scaling $\tau(L) \sim L^{d}$ was observed.

Finally, we note that ordered oscillations and infinite coherence time are not decoupled properties.
For instance, if the phases follow the KPZ universality class as discussed in Sec.~\ref{stability_spatially_coherent}, then the temporal phase correlations $\langle
    [\varphi(\bm{r}_0,t)-\varphi(\bm{r}_0,0)]^2
    \rangle_{\text{c}}
    \sim
    t^{2\beta}$ 
in which the growth exponent $\beta$ is related to the roughness exponent $\chi$ characterizing spatial correlations and the dynamical critical exponent $z$ through a dynamical scaling relation $z=\beta/\chi$.

\subsubsection{Time quasicrystals and other dynamical phases}

In addition to time crystals (many-body limit cycles), one can consider other many-body dynamical attractors such as limit toruses/tori or chaotic attractors. 
These indeed exist in mean-field: for instance, the modulated waves in Fig.~\ref{figure_symmetries} are limit tori.
Examples in many-body systems can be found in all-to-all models (see e.g. Fig.~\ref{figure_many_body_bifurcations}L,M,O), see e.g. \citet{Hong2011,Hong2011b,Hong2014,Hong2012,Fruchart2021} for Kuramoto-like oscillators and \citet{Nadolny2025} for quantum spins.
These have been predicted and observed in quantum systems such as spin gases and driven Bose-Einstein condensates \cite{Huang2025,Pizzi2019,Giergiel2019,Solanki2025,Cosme2025}.
Current evidences suggest that these can also survive beyond mean-field, at least in discrete-time lattice models \cite{Chate1992,Chate1995}, but the situation is much less clear.

\subsection{Nonreciprocity for engineering dynamic states}

Beyond time-crystals-like phases that are only well-defined as they are related to the thermodynamic limit, nonreciprocity can be used as a design tool to engineer dynamic states in finite-size systems observed during a finite time, for which the thermodynamic limit may be ultimately irrelevant.
From the conceptual perspective, methods are currently developed to address these situations from a topological perspective by making the dynamics singular enough to mathematically distinguish between different classes of behavior (Sec.~\ref{neqpt}). 
In this section, we take a complementary perspective and describe some of the dynamical phenomena that can be observed as a consequence of nonreciprocity in finite systems.

\begin{figure*}
    \centering
    \includegraphics[width=0.9\linewidth]{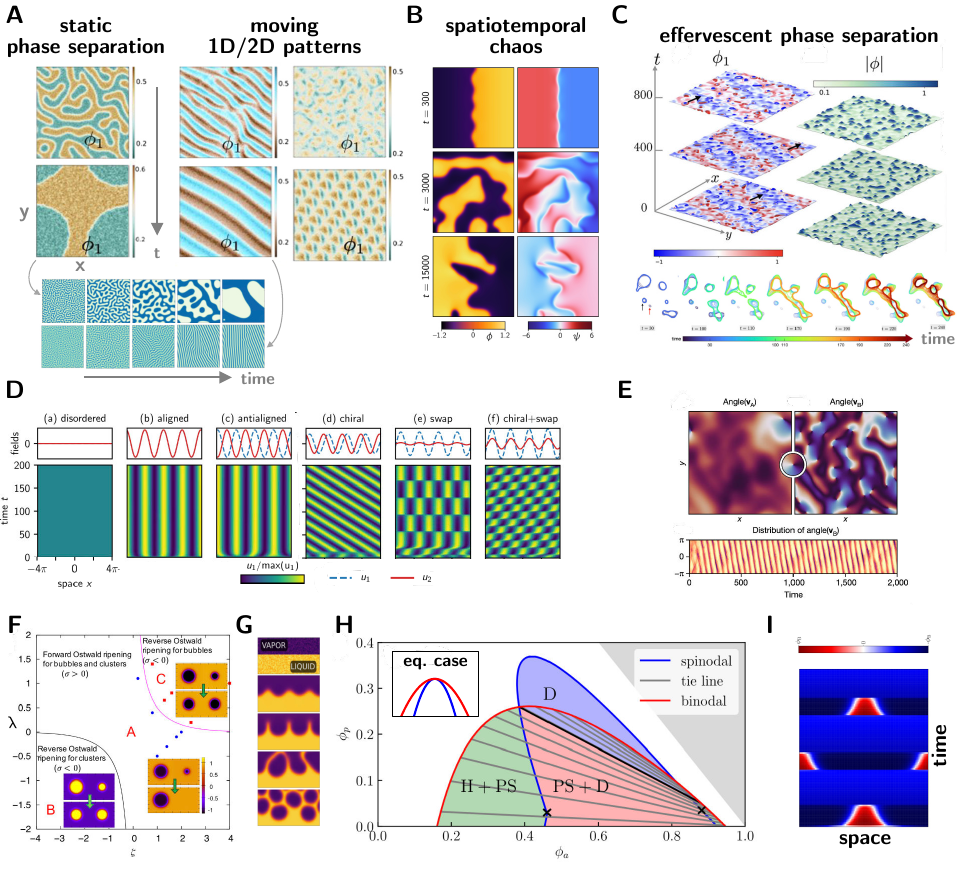}
    \caption{\textbf{Dynamic states.}
    \label{figure_dynamic_patterns}
    (A) Static phase separation (left) and moving patterns (lamellar or 2D; right) in a (conserved) nonreciprocal Cahn-Hilliard model. Adapted from \citet{Saha2020}.
    (B) Spatiotemporal chaos in a nonreciprocal Cahn-Hilliard model with no-flux boundary conditions. These arise from undulating wave front that would be stable in a system with periodic boundary conditions. Adapted from \citet{Brauns2024}.
    (C) Effervescent phase separation in a nonreciprocal Cahn-Hilliard model. Droplets are continuously formed and dissolved. Adapted from \citet{Saha2025}.
    (D) Phases of a (nonconserved) nonreciprocal Swift-Hohenberg model with $O(2)$ symmetry. The phases can be classified using Fig.~\ref{figure_symmetries}.
    Adapted from \citet{Fruchart2021}.
    (E) Deterministic spatiotemporal chaos in a nonreciprocal flocking model. Vortices continuously unbind and annihilate in the steady state, leading to noisy oscillations on average. Adapted from \citet{Fruchart2021}.
    (F) Inverse Ostwald ripening in an active model B, where both bubbles survive. Adapted from \citet{Tjhung2018}.
    (G) Interface instability in an active model B. Adapted from \citet{Fausti2021}.
    (H) In equilibrium systems (inset), the spinodal (limit of stability against phase separation) is entirely inside the binodal (coexistence curve) and meets it tangentially at the critical point.
    This is not the case out of equilibrium: the figure shows an active-passive mixture with nonreciprocal effective interactions in which the spinodal crosses the binodal, leading to a dynamical steady-state (D) in addition to the usual homogeneous (H) and phase-separated (PS) states. The white region is either H+D or H.
    Adapted from \citet{Mason2025}.
    (I) Standing-wave limit cycle in phase separation in a nonequilibrium model AB. Adapted from \citet{Li2020}.
    }
\end{figure*}

\begin{figure*}
    \centering
    \includegraphics[width=0.95\linewidth]{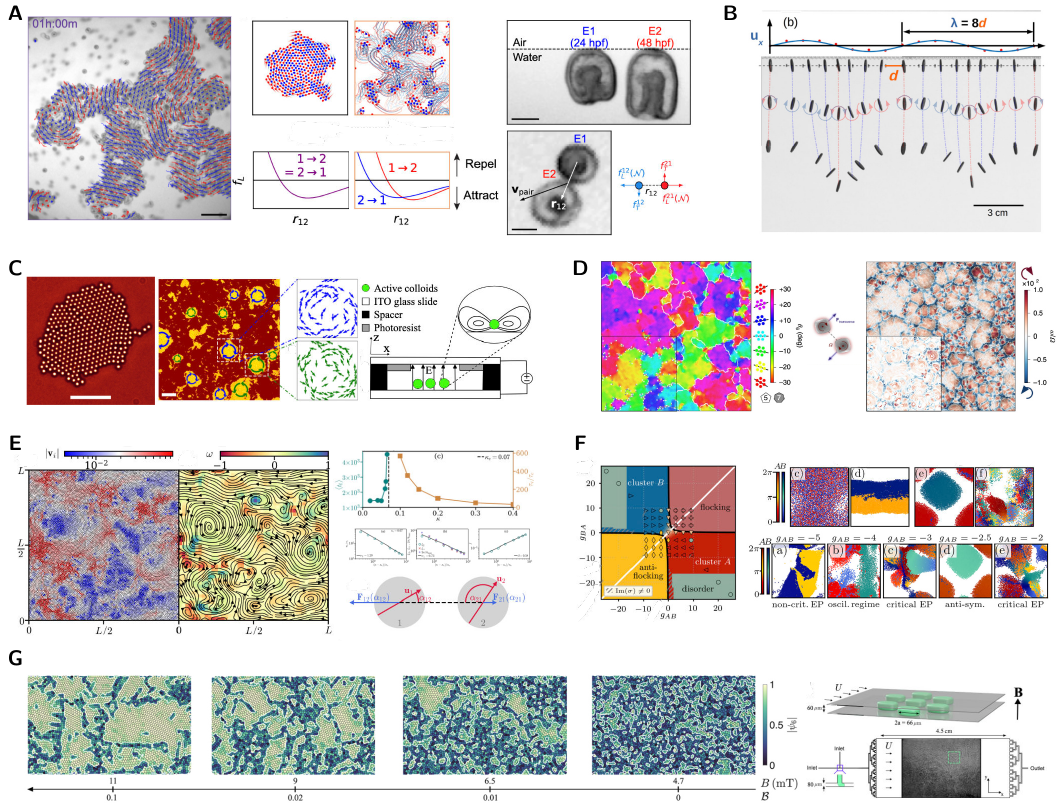}
    \caption{\textbf{Many-body nonreciprocal particle systems.}
    In contrast with Fig.~\ref{figure_dynamic_patterns}, which takes a field-theoretical perspective, here we review particle-based systems with nonreciprocal interactions.
    (A) Assemblies of different kinds of starfish embryo (with different ages) have been reported by \citet{Lee2025} to exhibit nonreciprocal interactions and to show different behaviors depending on the strength of nonreciprocity.  Adapted from \citet{Lee2025}.
    (B) Sedimentation of disks in a fluid. The fluids interact through hydrodynamic forces, which are usually nonreciprocal. A clumping instability takes place, which is discussed in \citet{Chajwa2020} in terms of nonnormal dynamics. Adapted from \citet{Chajwa2020}.
    (C) Spontaneous rotation of self-propelled achiral particles (Quincke rollers). The particles form dense hexatic clusters, which spontaneously rotate. This is discussed in \citet{Kole2025} in terms of non-reciprocally coupled broken symmetries.
    Adapted from \citet{Kole2025}.
    (D) Ensemble of spinning colloids that interact through both longitudinal and transverse pairwise interactions.
    The solid has been reported in \citet{Bililign2021} to self-knead into a dynamic polycrystal.
    Adapted from \citet{Bililign2021}.
    (E) Simulations of an ensemble of athermal overdamped particles interacting through nonreciprocal forces controlled by a frozen angle transported with the particles. \citet{Klamser2025} reported a directed percolation transition to active turbulence due to the nonreciprocal forces.
    Adapted from \citet{Klamser2025}.
    (F) Simulations of nonreciprocal active polar mixtures with self-propelled particles with mutual repulsion.
    \citet{Kreienkamp2024,Kreienkamp2025} reported asymmetric clustering in addition to (anti)flocking phases, and nonuniform nonreciprocity-induced spontaneous chirality.
    Adapted from \citet{Kreienkamp2024,Kreienkamp2025}.
    (G) Melting of a solid made of hexadecane droplets dispersed in an acqueous ferrofluid under a magnetic field. \citet{Guillet2025} report that nonreciprocal forces lead to the proliferation of self-propelled  dislocations in the hydrodynamically driven Wigner crystal, which gets progressively more disordered as the nonreciprocity increases (i.e. as the magnetic field $B$ leading to reciprocal interactions decreases at fixed nonreciprocal hydrodynamic interactions).
    Adapted from \citet{Guillet2025}.
    \label{figure_nr_particle_systems}
}
\end{figure*}

\subsubsection{From travelling waves to spatiotemporal chaos}
\label{tw_stc}

As we have already mentioned, one of the most visible consequences of nonreciprocity is to produce dynamic states such as travelling waves and oscillations. 
In many cases, we may ignore the rare events that potentially jeopardize these effects at very-late-time and focus on the deterministic dynamics. 
In particular, a spontaneous breaking of parity may happen, that typically leads to moving states, that can move in two different directions, e.g. to the left or to the right \cite{Coullet1989}. These usually stem from a drift-pitchfork bifurcation (Sec.~\ref{bifurcations_eps}) or from an imperfect bifurcation derived from it, like a drift-saddle node or drift-transcritical bifurcation. 
(Stationary parity-breaking states are also possible, but it has been argued that they require fine-tuning in nonvariational systems \cite{FrohoffHulsmann2023}.)
For instance, \citet{You2020,Saha2020} showed that nonreciprocal model B dynamics (Eq.~\eqref{nrch}) can lead to travelling wave states with periodic boundary conditions, which range from self-propelled lamellar patterns to more complex moving lattices (Fig.~\ref{figure_dynamic_patterns}A).
The same behaviors can happen in nonreciprocal model A dynamics \citet{Fruchart2021} (Fig.~\ref{figure_dynamic_patterns}D).
In situations where parity is explicitly broken by the dynamics (when the system is left-right asymmetric, or when it is chiral), then a Hopf-like bifurcation is likely to occur.
This may arise, for instance, in chiral active matter.

Nonreciprocity can also induce additional spatiotemporal instabilities (Sec.~\ref{instabilities_from_nonreciprocity}) that can lead to spatiotemporal chaos, even in purely deterministic dynamics \cite{Clerc2005,Clerc2013,Coullet1992,Bodenschatz1992}.  
This is illustrated in \citet{Fruchart2021} using the example of nonreciprocal flocking (Sec.~\ref{nonreciprocal_flocking}) in which a deterministic chaotic nonlinear dynamics is observed where vortex pairs continuously unbind and annihilate in the steady-state, without eventually coarsening as shown on Fig.~\ref{figure_dynamic_patterns}E (see also Sec.~\ref{defects} on defects).
This also arises in model B dynamics: \citet{Brauns2024} reported spatiotemporal Cahn-Hilliard with no-flux boundary conditions, emphasizing the role of boundary conditions, and \citet{Saha2025} observed a state with phase-separated droplets that are spontaneously created and annihilated on a background of oscillating densities or travelling waves, that was termed effervescence (Fig.~\ref{figure_dynamic_patterns}C).

\subsubsection{Static and dynamic phase separation}
\label{dynamic_phase_separation}

In conserved dynamics (in active phase separation and in nonreciprocal Cahn-Hilliard models), the spatiotemporal behaviors discussed in Sec.~\ref{tw_stc} are a manifestation of dynamic phase separation. 
A variety of phenomena arise in this class of systems, many of which are reviewed in \citet{Cates2025}, and we refer the reader to this for more details.
One of the key points is that the notion of surface tension (aka interfacial tension) becomes ill-defined, for the same reason than the dynamics is not described by a free energy.
This is for instance analyzed in detail in \citet{Ma2025} in a model of nonreciprocal ternary phase separation.
As a consequence, the behaviors of droplets is drastically modified: inverse Ostwald ripening can happen, where droplets split instead of merging (Fig.~\ref{figure_dynamic_patterns}F), and capillary waves can become unstable (Fig.~\ref{figure_dynamic_patterns}G). 
One of the consequences of these various processes is that they can interrupt coarsening, leading to a non-trivial distribution of droplet sizes (or more generally of length scales) in the steady-state (Sec.~\ref{scale_selection}).
As discussed in Sec.~\ref{tw_stc}, the different domains may also be moving or contain travelling waves \citet{Saha2025}, see also \citet{Dinelli2023} for a case where self-propulsion is also present.
These effects lead to a variety of classes of phase separation, which have been mostly mapped out in nonvariational models of active separation with a single order parameter \citet{Cates2025}, but not yet in models with multiple coupled fields.

Another effect of nonreciprocity is that the binodal line can cross the spinodal line (Fig.~\ref{figure_dynamic_patterns}H). This is forbidden in systems ruled by a free energy, in which the spinodal (limit of stability against phase separation) is entirely inside the binodal (coexistence curve) and meets it tangentially at the critical point (inset of Fig.~\ref{figure_dynamic_patterns}H), but as discussed in \cite{Bertin2024,Spinney2025,Mason2025}, it can happen in nonvariational systems, such as those with nonreciprocal interactions~\cite{Mason2025}.
This mechanism corresponds to a change in the nature of the instability (bifurcation) of the homogeneous state, which may for instance become oscillatory, leading to dynamical patterns instead of static phase separation.
Similar features are observed in a convective version of the Cahn-Hilliard equation \citet{Golovin2001}, where nonreciprocity arises at the level of a single species, in space.
Note also that in addition to travelling waves (in the meta-classification of Sec.~\ref{classes_dynamic_states}), it is in principle possible to have standing waves. 
This was indeed observed in \citet{Li2020} (Fig.~\ref{figure_dynamic_patterns}I).

\begin{figure*}
    \centering
    \includegraphics[width=0.99\linewidth]{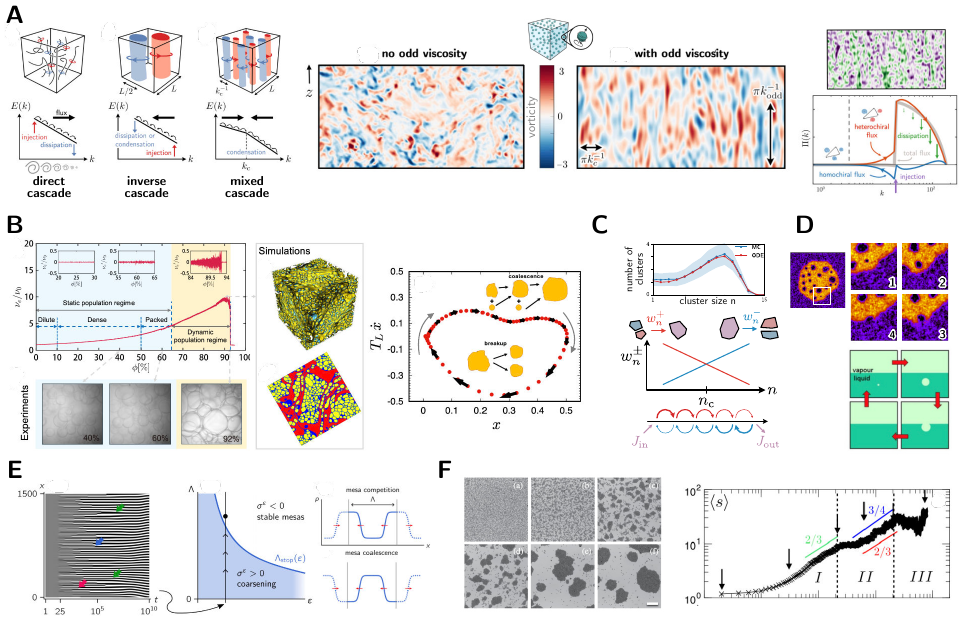}
    \caption{\textbf{Scale selection from nonreciprocity.}
    \label{figure_scale_selection}
    (A) Strongly nonlinear wavelength selection in an incompressible turbulent fluid. The presence of odd viscosity leads to a mixed cascade in which energy cascades towards an intermediate scale. It can also lead to a flux loop state (right). Adapted from \citet{deWit2024}.
    (B) Droplets in concentrated emulsions progressively coalesce into a large droplet that breaks up through an instability. A nonequilibrium phase space loop is shown in the right panel. Adapted from \citet{Girotto2025}.
    (C) A general mechanism for scale selection: the rate of merging/splitting of aggregates is such that large aggregates tend to split while small aggregates tend to merge. Adapted from \citet{deWit2024}.
    (D) Bubble phase separation in an active field theory. A nonequilibrium phase-space loop whereby vapor bubbles are created in the bulk liquid phase, and then expelled to the exterior vapor phase, which equilibrates with the rest. The process goes on indefinitely.
    Adapted from \citet{Tjhung2018,Cates2025}.
    (E) Coarsening processes of mesa and interrupting coarsening by weakly non-conserving processes. Adapted from \citet{Weyer2023}.
    (F) Experimental example of interrupted coarsening in a system of aligning active colloids. Adapted from \citet{vanderLinden2019}.
    }
\end{figure*}

\subsubsection{Interrupted coarsening and wavelength selection}
\label{scale_selection}

Nonvariational dynamics can play a role in various mechanisms of wavelength selection. 
The simplest mechanisms can be understood from the linear instability of an homogeneous state.
Nonreciprocal interactions can affect the type of instability, and for instance transform a large-scale (type II) instability into a finite-wavelength (type I) instability (see Sec.~\ref{classification_instabilities}).
This was demonstrated in \citet{FrohoffHulsmann2021} for the nonreciprocal Cahn-Hilliard model (Eq.~\eqref{nrch}).
Another example in a nonconserved nonreciprocal flocking model from \citet{Fruchart2021} is shown in Fig.~\ref{figure_dynamic_patterns}E.
A similar process is also at play in active turbulence \cite{Alert2022}.
In addition, more complex, nonlinear mechanisms may be at play. 
Examples range from sand dunes \cite{Andreotti2009}
to active colloids \cite{vanderLinden2019}.
This is illustrated in Fig.~\ref{figure_scale_selection}A in the example of turbulent fluids with odd viscosity, where no characteristic wavelength is present in the growth rate of the linearized theory, but where patterns may emerge from a strongly nonlinear turbulent cascade mechanism \citet{deWit2024}.
One of the key distinct mechanisms unique to nonequilibrium systems is the possibility of loops on phase space that break detailed balance, which are naturally triggered by nonreciprocity. 
Turbulent cascades provide a measurable example of this through so-called flux loop states, see \citet{Alexakis2018} for an introduction and the right panel of Fig.~\ref{figure_scale_selection}A for an example.
Another instance of phase-space loop, in concentrated emulsions, is shown in Fig.~\ref{figure_scale_selection}B: droplets progressively coalesce into a large droplet that breaks up through an instability \cite{Girotto2025}.
The general mechanism is illustrated in Fig.~\ref{figure_scale_selection}C: large objects tend to split while small objects tend to merge (here, objects may represent vortices, clusters, droplets, and so on). 
\citet{deWit2024} constructed a minimal model inspired from shell models in turbulence that summarizes these trends.

In equilibrium diffusive mass-conserving systems, the overall tendency is for the system to coarsen, either by volume transfer between droplets (Ostwald ripening) or by droplet coalescence. Additional processes such as inverse Ostwald ripening or droplet destabilization can take place in nonvariational systems (an example in \enquote{active model B+} is shown in Fig.~\ref{figure_scale_selection}E-F; see \citet{Cates2025} for consequences in active matter).
These additional processes can lead to phase-space loops, as shown in Fig.~\ref{figure_scale_selection}D in the example of an active mixture exhibiting bubbly phase separation \cite{Tjhung2018}.
These effects can strongly affect the coarsening dynamics, potentially leading to systems exhibiting interrupted coarsening (also called arrested coarsening\footnote{\citet{Singh2019} differentiate between arrested and interrupted coarsening, reserving the later for dynamic deterministic steady-states. It is not obvious to us that this nomenclature is broadly followed in the literature.
}).
In this case, the system is neither fully phase-separated nor fully homogeneous in the steady-state. 
Instead, the average droplet size is finite, typically independent of system size. 
We refer to \citet{FrohoffHulsmann2021} for a thorough discussion of these points in the context of the nonreciprocal Cahn-Hilliard model.

Criteria for when coarsening is expected to occur have been put forward in \citet{Politi2004,Politi2006,Politi2015,Misbah2009,Misbah2010,Nicoli2013}. 
In its simplest instantiation, this criterion based on the existence of periodic steady-state solutions states that coarsening takes place when $d\lambda/dA >0 $, in which $\lambda(A)$ gives the wavelength $\lambda$ of the periodic steady-state as a function of the amplitude $A$ of the pattern. 
Extensions relate the fate of coarsening to phase instabilities of more general patterns. 
In the context of reaction-diffusion systems, \citet{
Halatek2018,Brauns2020,Brauns2021,Weyer2023,Brauns2024} developed a theory for the arrest of coarsening by terms that weakly break mass conservation (Fig.~\ref{figure_scale_selection}E).
\citet{Nepomnyashchy2015,Misbah2010} review some of these aspects.
An experimental example of interrupted coarsening in mass-conserving systems is shown in Fig.~\ref{figure_scale_selection}F, where the size of the droplets reaches a steady-state value in a system of aligning active colloids \cite{vanderLinden2019}.

\begin{figure}
    \centering
    \includegraphics[width=\linewidth]{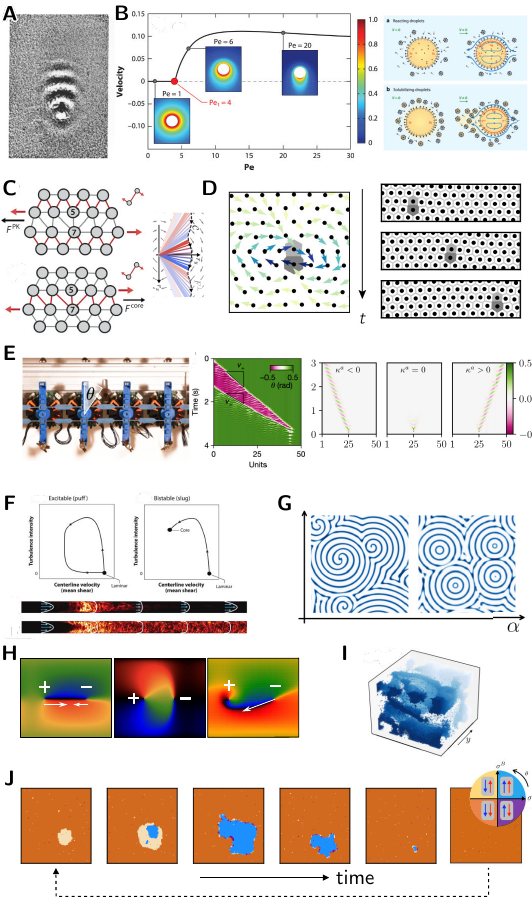}
    \caption{
    \textbf{Localized structures.}
    (A) Pulses of travelling-wave convection in ethanol-water mixtures.
    These are stabilized and tend to move due to nonvariational effects.
    Adapted from \citet{Bodenschatz1992}.
    (B) Self-propulsion of chemically active droplets. Propulsion speed is related to droplet asymmetry.
    Adapted from \citet{Michelin2023}.
    (C) Mechanisms of dislocation motion in odd elastic solids. A core force due to nonreciprocal interactions, not captured from continuum theories, is present. 
    Adapted from \citet{Braverman2021,Fruchart2023}.
    (D) Glide of a dislocation in a flowing crystal. Left panel shows forces on particles. Adapted from \citet{Poncet2022}.
    (E) Nonreciprocal solitons in mechanical metamaterials. Solitons and antisolitons can go in the same direction (middle panel).
    Solitons are stabilized by nonreciprocity (right panel). 
    Adapted from \citet{Veenstra2024,Veenstra2025b}.
    (F) Puffs and slugs in pipe turbulence correspond to homoclinic and heteroclinic orbits.
    Adapted from \citet{Avila2023}.
    (G) Spiral and target defects in a nonreciprocal Cahn-Hilliard model. 
    Adapted from \citet{Rana2024}.
    (H) Interaction between positive and negative XY defects in a model with cones of vision, for different defect shapes. Arrows indicate velocity.
    Adapted from \citet{Rouzaire2025}.
    (I) Scroll wave in a nonreciprocal Ising model with two species. Adapted from \citet{Avni2025b}.
    (J) Droplet-capture mechanism in a nonreciprocal Ising model with two species.
    This process forms a full cycle, and can repeat (stochastically).
    Adapted from \citet{Avni2025b}.
    \label{figure_localized_structures}
    }
\end{figure}

\subsubsection{Localized structures: defects, domain walls, droplets}
\label{defects}

Nonreciprocity has several notable consequences for defects and other localized structures.
As emphasized by \citet{Bodenschatz1992}, the nonvariational nature of the dynamics can lead to effects that would not be present in purely relaxational systems.

\paragraph{Self-propulsion of localized structures}

In particular, defects and domain walls tend to become self-propelled and can move at constant speed~\cite{Pomeau1983,Siggia1981,vanSaarloos1990,Coullet1990b,Colinet2002,Tsimring1996,Kozyreff2007,Clerc2005,Houghton2011,Pismen2013}.
The onset of self-propulsion is typically associated to the spontaneous breaking of parity, which provides a direction of motion and tends to determine the self-propulsion velocity \cite{Coullet1990b,Knobloch1995}.
This has been observed, for instance, for travelling wave pulses in binary-fluid convection \cite{Kolodner1991,Mercader2013,Niemela1990,Heinrichs1987,Bodenschatz1992}, although the effect has been reported to be very sensitive to experimental imperfections (Fig.~\ref{figure_localized_structures}A).
The determination of the defect self-propulsion speed and of defect interactions from continuum theories has been discussed \cite{Pocheau1984,Siggia1981b,Siggia1981c,Pismen1990,Pomeau1983} and is reviewed in \citet{Pismen2006,Pismen1999}.
In the case of defects in roll patterns, for instance, nonvariational effects can lead to gliding motion (perpendicular to the rolls) that cannot be captured by a Peach-Köhler-like force \cite{Pomeau1983}.

A similar process underlies common mechanisms of self-propulsion. 
This is reviewed in \citet{Michelin2023} in the case of chemically active droplets: a pitchfork-like bifurcation\footnote{See \citet{Farutin2024} for subtleties about the bifurcation normal form.} controls both the self-propulsion field of the droplets and the asymmetry of the chemical concentration field around them (Fig.~\ref{figure_localized_structures}B).
Indeed, one usually expects parity-breaking structures to move in nonequilibrium systems \cite{FrohoffHulsmann2023}, irrespective of whether the parity breaking is spontaneous, like in these droplets, or explicit, like in Janus particles. 

Beyond pattern-forming systems, the mechanisms of defect self-propulsion have been explored in particle models in \citet{Braverman2021,Poncet2022} (Fig.~\ref{figure_localized_structures}C-D). 
Like in the case of pattern formation \cite{Siggia1981b,Siggia1981c,Pomeau1983}, some but not all aspects can be rationalized through an effective Peach-Köhler force. 
In particular, forces due to small length-scale effect at the lattice scale evade such a description \cite{Braverman2021}, see Fig.~\ref{figure_localized_structures}C.

Note that domain walls between domains with different symmetries can also acquire a velocity, as discussed in \citet{Coullet1990b}. 
This effect is distinct from the propagation of fronts into unstable states \cite{vanSaarloos2003}, and instead arises between domains with equal stability.

\citet{Veenstra2024,Veenstra2025b,Sun2025} also considered solitons in nonreciprocal systems, leading for instance to unidirectional breathers.
As the medium includes gain, long-lived unidirectional solitons are observed. Notably, \citet{Veenstra2024} found that solitons and antisolitons can travel in the same direction thanks to the nonreciprocal drive (Fig.~\ref{figure_localized_structures}E, middle panel; here they annihilate but it is also possible to have identical velocities). 
In the right panel of Fig.~\ref{figure_localized_structures}E, solitons are stabilized by nonreciprocity (compare the center case, reciprocal, to the left and right ones, which are nonreciprocal).

\paragraph{Proliferation of defects and destruction of order}

In addition, defects may proliferate thanks to various mechanisms.
For instance, in the complex Ginzburg-Landau equation, stable localized patterns can turn into expanding turbulent domains depending on the parameters \citet{Hakim1990,vanSaarloos1992}.
One situation where this transition can be understood are simplified models of turbulent flow in pipes, where a transition between localized turbulent excitable \enquote{puffs} (homoclinic orbit) to bistable \enquote{slugs} (heteroclinic orbit), as reviewed in \citet{Barkley2016,Avila2023}, see Fig.~\ref{figure_localized_structures}F. 
\citet{Tsimring1996,Colinet2002,Young2003} discuss the case of penta-hepta defects in hexagonal patterns.
The underlying pattern can also undergo an oscillatory instability which tends to promote defects \cite{Young2003c,Madruga2006,Madruga2007}.

The interplay between defect self-propulsion in crystals \cite{Braverman2021,Poncet2022} and the proliferation of defects has been recently observed in flowing crystals \cite{Guillet2025} and in chiral active matter including bacteria crystals \cite{Petroff2015}, starfish embryo \cite{Tan2022} and artificial colloids \cite{Bililign2021}. 
Similar dynamical states are also observed in nonchiral active colloids \cite{Ginot2018,vanderLinden2019}.
Indeed, the self-propulsion mechanisms present at the scale of a single defect also take place at grain boundaries
A few examples are shown in Fig.~\ref{figure_nr_particle_systems}A,C,D,G: one common theme is that nonreciprocal interactions tend to induce disorder, through rotation (panels A,C,D) or other means (panel G).
Despite substantial progress \cite{Holl2025,Caprini2025b,Choi2025,Huang2025b}, there is currently no overarching theoretical understanding of these order-disorder transitions.

\paragraph{Nature of the localized structures and their interactions}

The nature of defects (topological or not) or localized structures can be affected by the presence of nonvariational couplings or nonreciprocal forces.
In addition, the effective interactions between these can be substantially modified.
In the complex Ginzburg-Landau equation, for instance, rotating spiral states appear \cite{Aranson2002} (these are indeed observed experimentally, see \citet{Bodenschatz1992}). 
These defects can have different interactions compared to equilibrium counterparts: \citet{Aranson1998b} show that, in contrast with equilibrium XY vortices, vortices in the CGLE have a nonzero mobility and a modified effective interaction that is screened so that they respond only to perturbations at a distance smaller that a screening length from the spiral core, with potential implications for the behavior of many vortices (like in the BKT transition).
As another example, \citet{Rana2024,Rana2024b} analyzed the defects in a nonreciprocal Cahn-Hilliard model (see Eq.~\eqref{nrch}), and showed that different kinds of defects occur (spirals and targets, with respective topological charges $\pm 1$ and $0$, see Fig.~\ref{figure_localized_structures}G), with defect interactions controlled by their topological charges and by the strength of nonreciprocity in the underlying model.
\cite{Loos2023,Rouzaire2025,Dopierala2025,Popli2025} also analyzed the anisotropic propagation of defects in models with cones of vision (see Sec.~\ref{nonreciprocal_xy}).
In these systems, the interaction between defects is also substantially affected \cite{Rouzaire2025}, see Fig.~\ref{figure_localized_structures}H.
In 3D, scroll waves appear instead of spirals \cite{Winfree1973,Keener1988,Gabbay1997} (these are illustrated in the nonreciprocal Ising model in Fig.~\ref{figure_localized_structures}I).
Beyond topological defects, the behavior of localized structures in nonvariational systems is reviewed in \citet{Knobloch2015}, see also \citet{vanSaarloos1992} for the case of the complex Ginzburg-Landau equation. 
These include travelling and oscillatory localized states \cite{Houghton2011} that can be stabilized by nonvariational effects \citet{Thual1988,Malomed1987,Bodenschatz1992}.
\citet{FrohoffHulsmann2021b} discuss these in a nonreciprocal Cahn-Hilliard model (Eq.~\eqref{nrch}).

\paragraph{Droplet nucleation and destruction}

Localized structures also include droplets, which can play a crucial role in destroying or preserving order in nonreciprocal systems (Sec.~\ref{snic_field_theory}).
The nucleation of droplets in variational systems can be understood from the perspective of large-deviation theory: it can be phrased in terms of an escape problem in the quasipotential (Secs.~\ref{transverse_decomposition} and \ref{escape_times_and_paths}).
In nonvariational systems, the quasipotential is still well-defined, but various subtleties arise (Sec.~\ref{escape_times_and_paths}), and so it is not obvious that classical nucleation theory can be readily extended. 
This issue has been discussed in \citet{Ohta1989} in the context of excitable media, where it was found that the nucleation rate is not affected by the probability current.
\citet{Redner2016,Richard2016,Cates2023} found similar results in context of active fluid phase separation. \citet{Cates2023} argue that a key element is that classical nucleation theory only involves the droplet radius, which allows detailed balance to be restored along the instanton. 
These results suggest that, under certain conditions, the qualitative description of droplet nucleation is unaffected by nonvariational contributions (see \citet{Cates2023} for potential limitations), even though substantial quantitative changes do occur. 
Indeed, \citet{Cho2023} found that nucleation kinetics can be accelerated out of equilibrium (see also Sec.~\ref{accelerated_convergence}).
See \citet{Heller2024} for a numerical algorithm to evaluate the transition rates.

After nucleation, a variety of behavior can take place (see e.g. Sec.~\ref{scale_selection}).
For instance, Fig.~\ref{figure_localized_structures}J shows a droplet-capture mechanism that arises in a nonreciprocal Ising model with two species (see Sec.~\ref{nr_ising}), and originates from the different front velocities of different droplets \cite{Avni2025a,Avni2025b}.
The blue-yellow droplet is faster to grow than the yellow-orange one, and so it captures it by full invasion. 
However, the blue-orange droplet is unstable, so it shrinks and disappears.
This can then repeat (stochastically), as shown by the cycle in the figure, manifestly violating detailed balance. 
This mechanism can salvage long-range static order in the thermodynamic limit (Sec.~\ref{snic_field_theory}).

\subsubsection{Nonreciprocal flocking}
\label{nonreciprocal_flocking}

Flocking is an iconic manifestation of active matter in which self-propelled particles align their velocities into a coherent state of collective motion \citet{Marchetti2013,Bechinger2016}.
Flocking is an out-of-equilibrium phenomenon, and several approaches have been developed to characterize its irreversibility \cite{Bialek2012,Ferretti2022,Cavagna2014} which, in some cases, can be cast in terms of nonreciprocal interactions \cite{Cavagna2017,Dadhichi2020}.  
Indeed, equilibrium models of flocking such as proposed in \citet{LolandBore2016,Casiulis2020} can produce collective motion, but \citet{Tasaki2020} showed that Hohenberg-Mermin-Wagner type theorems apply and prevent long-range order.

In situations where more than one species of self-propelled particles is present, another class of nonreciprocity can arise, between these species. This has been considered in \citet{Yllanes2017} where the effect of random {intruders} on a flock, that do not align with the other agents, but with which the other agents align: it turns out that a small fraction of these dissenters are enough to disrupt the flocking state, way smaller than the fraction of static obstacles that would be required for the same effect.
\citet{Fruchart2021} analyzed a similar situation, where two species of self-propelled particles may either try to align or antialign with the other species. In mean-field, this leads to flocking and antiflocking steady-states, as well as a other dynamic phases such as a chiral state where the agents move in circle, either clockwise or counterclockwise, spontaneously breaking parity (Fig.~\ref{figure_many_body_bifurcations}L). 
Because of the underlying symmetry, these phases turn out to correspond to the ones in the classification of Fig.~\ref{figure_symmetries}.

Going beyond the mean-field picture, \citet{Kreienkamp2022,Kreienkamp2024,Kreienkamp2024b,Kreienkamp2025,Kreienkamp2025b} performed extensive two-dimensional microscopic simulations of nonreciprocally aligning self-propelled particles that also include repulsive interactions (that may lead to motility-induced phase separation). As a consequence, in addition to homogeneous phases, phase separation with asymmetric clusters of different species can emerge and chase each other (Fig.~\ref{figure_nr_particle_systems}F). 
The demixing of both species and appearance of structures is also analyzed in \citet{Myin2025}. 
Using an exact hydrodynamic coarse-graining procedure in the spirit of macroscopic fluctuation theory, \citet{Martin2025} derived the fluctuating hydrodynamic theory describing a nonreciprocal version of an active spin model and analyzed the corresponding phase diagram.
A similar model is analyzed in \citet{Mangeat2025}.

On the experimental side, \citet{Maity2023} observed the spontaneous demixing of a bidispersed mixture of colloidal rollers, that was attributed to a combination of active speed heterogeneity and nonreciprocal repulsive interactions. \citet{Chen2024} also observed hyperuniform states with spontaneous emergence of chirality in binary mixtures of programmable robots.

\begin{figure*}
    \centering
    \includegraphics[width=0.95\linewidth]{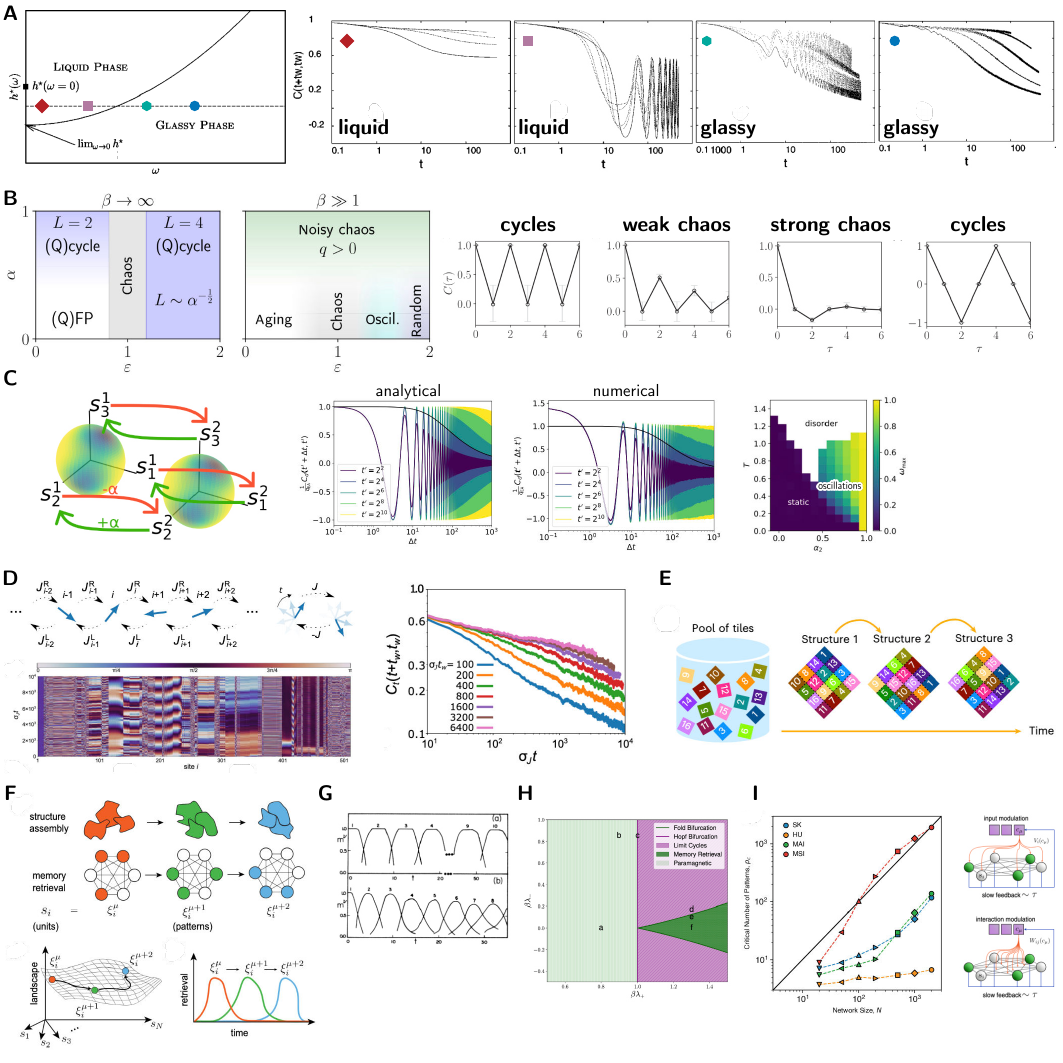}
    \caption{
    \label{figure_disorder}
    \textbf{Nonreciprocity in disordered systems.}
    (A) Transition between glassy and non-glassy (liquid) phases in a system submitted to time-dependent driving forces. The relaxation dynamics (glassy or not) may be accompanied by oscillations. Adapted from \citet{Berthier2001}.
    (B) Phase diagram of a Sherrington-Kirkpatrick game without noise (left) and with weak noise (right), where agents attempt to learn a strategy in the game. In addition to fixed-points (FP) where learning converges, oscillations due to impatient learning and chaotic behavior are observed. Whe the noise is finite, these are replaced with quasifixed points and quasicycles (Q), where aging is observed. Adapted from \citet{GarnierBrun2024}.
(C) Behavior of a nonreciprocal spherical Sherrington-Kirkpatrick model with two species, realizing a nonreciprocal spin glass. Here, the model is not submitted to a time-periodic drive, but it is still out of equilibrium. A phase with oscillating aging is observed (labelled \enquote{oscillations} in the phase diagram), in addition to the more familiar static spin-glass phase and the disorder phase. Adapted from \citet{Garcia2025,Garcia2025b}.
(D) Dynamics of a disordered glassy system with nonreciprocal frustration. The 1D spin chain with random asymmetric nearest-neighbor coupling exhibits complex trajectories with properties reminiscent of a spin glass. Adapted from \citet{Hanai2024}.
(E) Nonreciprocal self-assembly can lead to a sequence of assembly and disassembly. Here, a predefined sequence of transitions between three different target configurations is considered. Adapted from \citet{Osat2022}.
(F) Nonreciprocal associative memories can store and retrieve dynamic sequences (limit cycles). In the context of self-assembly, this can be interpreted as a sequence of structures, as in panel E. Adapted from \citet{Herron2023}.
(G) An example of sequence retrieval in an asymmetric Hopfield network. Adapted from \citet{Sompolinsky1986}.
(H) Phase diagram of a two-memory cyclic memory retrieval in a non-reciprocal Hopfield network. Adapted from \citet{Xue2025}.
(I) The retrieval of dynamic sequences can be controlled by feedback, through input or interaction modulation, with different consequences on the capacity of the memory. Adapted from \citet{Herron2023}.
	}
\end{figure*}

\subsection{Non-reciprocity in disordered systems}

\subsubsection{Nonreciprocal glasses and beyond}

There are at least two broad ways that a system can be out of equilibrium: (i) it is externally driven out of equilibrium or (ii) it fails to equilibrate when left to its own device.
The first case corresponds to systems connected to multiple reservoirs and that is maintained in a non-equilibrium steady-state~\cite{Bertini2015}. 
The second case corresponds to phenomena such as glassy dynamics and aging \cite{Binder1986,Henkel2010,Berthier2011,Keim2019,Arceri2022,Cugliandolo2024}, many-body localization~\cite{Nandkishore2015,Abanin2019}, or non-thermal fixed points \cite{Schmied2019,Berges2015}, in which a system thermalizes very slowly or even fails to thermalize whatsoever. 
These tend to arise in disordered systems, and are the object of this paragraph.
Note that both cases are, to some extent, conceptually related, a bit like fluctuations and response; both can coexist in the same physical system \cite{Kurchan1997,Tapias2024}. While a large body of work on glasses focused on equilibrium systems \cite{Mezard1986,Charbonneau2023}, there are many systems which are both disordered and nonreciprocal, ranging from ecological systems \cite{Altieri2021,Hatton2024,Hu2022,Biroli2018,Bunin2025,Bunin2017} to active matter \cite{Ghimenti2023,Ghimenti2024,Ghimenti2024b,Berthier2019,Berthier2013}.

One of the main features of glasses is aging, where the local relaxation time scale of a system increases with time, meaning that the system evolves slower as it gets older \cite{Houches2002,Keim2019,Arceri2021}.
The possible coexistence of oscillations with aging dynamics has been demonstrated in \citet{Berthier2001} in the case of periodically-driven systems (Fig.~\ref{figure_disorder}A).
Nevertheless, \citet{Crisanti1987,Parisi1986} showed that in a spherical Sherrington-Kirkpatrick model, random all-to-all non-reciprocal interactions between spins suppresses the finite temperature spin-glass transition as well as the aging dynamics, which is replaced with a chaotic dynamics.
\citet{Cugliandolo1997,Berthier2000,Berthier2001,Horner1996,Fyodorov2025} extended these results to more general glassy models. 
In marginal systems, it is believe that any amount of non-reciprocal interactions destroy glassiness, while a finite amount is needed in discrete models \cite{Iori1997,GarnierBrun2024}.
This is illustrated on the case discussed by \citet{GarnierBrun2024} of the process of learning a Sherrington-Kirpatrick game, shown in Fig.~\ref{figure_disorder}B.

However, it turns out that aging can survive when the nonreciprocity is structured, and that it can indeed coexist with oscillations \cite{Garcia2025}.
This is illustrated in Fig.~\ref{figure_disorder}C.
\citet{Guislain2024b} showed that oscillatory dynamics persists in mean-field Mattis-like models in which disorder can be gauged away by a change of variables \cite{Mattis1976,Toulouse1987}.
A glassy state arising from nonreciprocal frustration in a random system was also reported in \citet{Hanai2024}, in which a dynamic version of order-by-disorder is analyzed in detail (Fig.~\ref{figure_disorder}D). This phenomenon stems from a dynamical frustration \cite{Fruchart2021,Hanai2024} that shares similarities but is distinct from the geometrical frustration found in equilibrium disordered systems.
\citet{Garcia2025,Garcia2025b} considered a bipartite spherical Sherrington-Kirpatrick model, in which the nonreciprocity arises between two identical spin glasses that each have symmetric internal couplings. Using dynamical
mean field theory calculations, it was showed that a non-reciprocal glassy phase heralded by an exceptional-point mediated transition can exist, in which the dynamics exhibits both aging and oscillations.

Similar questions were explored in \citet{Daido1992,Daido1987} who developed the idea of \enquote{oscillator glasses}, see 
\citet{OttinoLoffler2018,Pruser2024,Pruser2024b,Leon2025} for recent progresses and \citet{Kaluza2010,Ionita2013,Ionita2014} for discussions of frustration in oscillating and excitable systems.

\subsubsection{Neural networks and learning}
\label{NN}

In neuroscience, neurons can be classified into two categories, called excitatory and inhibitory, depending on whether they promote or reduce the propensity of others neurons to fire \cite{Wilson1972,Buzsaki2012}, and the balance of excitation and inhibition is believed to play an important role in the function of assemblies of neurons \cite{Okun2009}.
Asymmetric interactions also play a role in artificial neural networks~\cite{Amit1989,Amit1988,Derrida1987,Lecun1985,Fasoli2018,Sompolinsky1986,Rabinovich2006,Khona2022}. 
For instance, backpropagation can be directly applied to feedforward neural networks, one of the most simple instances of neural networks, because the flow of information is unidirectional \cite{LeCun2015}.

On the opposite side, \citet{Hopfield1982} introduced a model of associative memory that can store and retrieve patterns, see \citet{Krotov2016,Krotov2023,Ramsauer2020} for more recent developments. In this Hopfield model, couplings are symmetric: this simplification makes the dynamics effectively variational and guarantees that it converges \cite{Amit1989}.
\citet{Sompolinsky1986} explored the effect of asymmetric couplings in Hopfield networks to encode and retrieve dynamic states, such as limit cycles. This strategy has been developed in several works \cite{Kanter1987,Kleinfeld1986,Herron2023,Dehaene1987,Buhmann1987,Kleinfeld1988,Xue2025} and has also been applied to self-assembly \cite{Osat2022,Metson2025,Liu2024,OuazanReboul2023,Kim2024} and physical learning \cite{Mandal2024b} (Fig.~\ref{figure_disorder}E-I).
We expect extensions of these ideas to apply to other paradigms for behavior, for instance in learning of strategies for games \cite{Sandholm1996,Hein2020} or for pursuit and escape \cite{Borra2022}.

\subsection{Non-normal amplification and noise}
\label{non_normal_amplification_and_noise}

The dynamics generated by non-normal operators leads to two effects with wide applications: (i) they can lead to a transient amplification of perturbations, which is not predicted by their eigenvalues, and (ii) they tend to increase the variance in response to noise. These effects and others are reviewed in detail in \citet{Trefethen2005}, and in \citet{Farrell1996a,Farrell1996b,Neubert1997}.
Consequences range from fluid mechanics \cite{Trefethen1993,Schmid2007,Chomaz2005} 
and pattern formation \cite{Neubert2002,Ridolfi2011,Biancalani2017,Klika2017,Muolo2019} to network theory \cite{Asllani2018,Asllani2018a,Hennequin2012}, and ecology \cite{Neubert1997,Neubert2004,Tang2014,Yang2023}.
See also \citet{Charan2020,Chattoraj2019} for giant amplification of perturbations and instabilities due to non-normality in frictional matter and \citet{Sornette2023,Wang2025,Troude2025} for applications to socio-economic instabilities.

\subsubsection{Non-normal operators and transient amplification}
\label{transient_amplification}

\begin{figure}
    \centering
    \includegraphics[width=\linewidth]{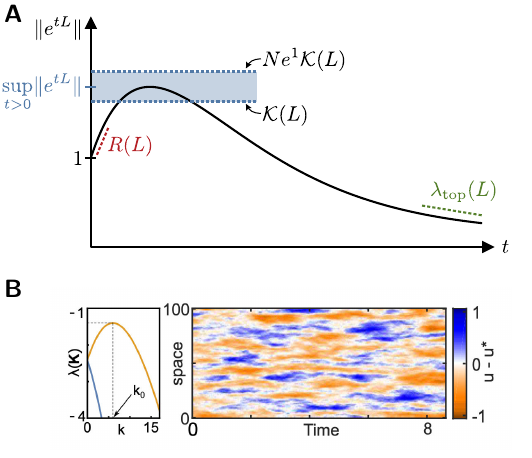}
    \caption{
    \textbf{Non-normal amplification.}
    (A) The behavior of $\lVert \ee^{t L} \rVert$ summarized non-normal amplification. The initial slope (in red) is given by the reactivity $R(L)$ in Eq.~\eqref{reactivity}, while the long-time behavior is given by the top Lyapunov exponent $\lambda_{\text{top}}$ in Eq.~\eqref{lambda_top}.
    In addition, the Kreiss theorem \eqref{kreiss_theorem} bounds the maximum normal amplification from above and below through the Kreiss constant $\mathcal{K}(L)$ in Eq.~\eqref{kreiss_constant}.
    Adapted from \citet{Trefethen2005}. 
    (B) Non-normal amplification may lead to a strong enhancement of the amplitude of noise-induced patterns compared to what is expected from naive dimensional analysis (Sec.~\ref{noise_induced_patterns}).
    The picture shows a noise-induced Turing-like pattern in stochastic simulations of a two-species model, that has substantial amplitude despite the fact that the growth rates $\lambda(k)$ are negative and that the large system size makes the noise small.
    Adapted from \citet{Biancalani2017}.
    \label{figure_non_normal}
    }
\end{figure}

Consider a linear system of the form
\begin{equation}
    \frac{d \bm{x}}{dt} = L \bm{x}
    \label{ls_nna}
\end{equation}
in which $L$ is a finite-dimensional matrix (\citet{Trefethen2005} discusses the case of infinite-dimensional operators).
We are interested in the behavior of $\bm{x}(t) = \ee^{t L} \bm{x}(0)$ and in particular of its norm $\lVert\bm{x}(t)\rVert$.
When $L$ is a normal matrix, it is determined by the eigenvalues of $L$. 
This is not the case when $L$ is not a normal matrix or operator: in particular, $\lVert\bm{x}(t)\rVert$ may increase, for a time, despite all eigenvalues being negative. 
This phenomenon is known as non-normal amplification. 
It is summarized in Fig.~\ref{figure_non_normal}A.

In order to characterize non-normal amplification, it is convenient to consider the matrix norm
\begin{equation}
\lVert \ee^{t L} \rVert = \sup_{\bm{x}_0 \neq \bm{0}} \frac{\lVert \ee^{t L} \bm{x}_0 \rVert}{\lVert \bm{x}_0 \rVert}
\end{equation}
in which we maximize over initial conditions $\bm{x}_0$ to get rid of them.
The quantity 
\begin{subequations}
    \label{reactivity}
\begin{equation}
    R(L) = \left. \frac{d \lVert \ee^{t L} \rVert}{dt} \right|_{t=0}
    = \sup_{\bm{x}_0 \neq \bm{0}} \frac{1}{\lVert \bm{x}_0 \rVert} \left. \frac{d \lVert \ee^{t L} \bm{x}_0 \rVert}{dt} \right|_{t=0}
\end{equation}
is called the \enquote{reactivity} or \enquote{numerical abscissa} of $L$,
which can be expressed using Eq.~\eqref{ls_nna} as the largest eigenvalue of the symmetric or Hermitian part $[L + L^\dagger]/2$ of $L$, namely
\begin{equation}
    R(L) = \max \sigma([L + L^\dagger]/2).
\end{equation}
\end{subequations}
in which $\sigma(A)$ is the spectrum (set of eigenvalues) of the matrix $A$.
The reactivity gives the initial slope of $t \mapsto \lVert \ee^{t L} \rVert$ (red line in Fig.~\ref{figure_non_normal}A).
Hence, we know that there must be transient amplification whenever $R(L) > 0$.

For a $N \times N$ matrix $L$, bounds on how much transient amplification may occur is given by the Kreiss matrix theorem (blue region in Fig.~\ref{figure_non_normal}A)
\begin{equation}
    \label{kreiss_theorem}
    \mathcal{K}(L) \leq \sup_{t \geq 0}\,\lVert\mathrm{e}^{t L}\rVert \leq N \exp(1) \, \mathcal{K}(L)
\end{equation}
in which $\mathcal{K}(L)$ is the Kreiss constant of $L$ (with respect to the left-half plane), defined as
\begin{equation}
    \label{kreiss_constant}
    \mathcal{K}(L) \equiv \sup_{z \text{ s.t. }\text{Re}(z) > 0}[ \text{Re}(z) \lVert (z-L)^{-1} \rVert].
\end{equation}
\citet{Mitchell2020} describes globally convergent algorithms for computing the Kreiss constant $\mathcal{K}(L)$.

The eigenvalues of $L$ still rule the long-time dynamics even for non-normal operators (green line in Fig.~\ref{figure_non_normal}A). 
In particular, the asymptotic growth rate of $\lVert \ee^{t L} \rVert$ is given by
\begin{equation}
    \label{lambda_top}
    \lambda_{\text{top}}(L) \equiv\lim_{t \to \infty} \frac{1}{t} \log \lVert \ee^{t L} \bm{x}_0 \rVert = \max \text{Re}\, \text{Sp}(L)
\end{equation}
in which $\text{Sp}(L)$ is the spectrum of $L$. 
This quantity is known as the spectral abscissa of $L$, or as its top Lyapunov exponent.

\subsubsection{Enhanced fluctuations}
\label{enhanced_fluctuations}

Consider now a version of Eq.~\eqref{ls_nna} with noise added, namely the SDE
\begin{equation}
    \frac{d \bm{x}}{dt} = L \bm{x} + M \bm{\eta}(t)
    \label{ls_nna_noise}
\end{equation}
in which $M$ is a matrix and $\bm{\eta}(t)$ is a vector of uncorrelated standard Gaussian white noises, namely $\langle \eta_i(t) \eta_j^*(t')\rangle = \delta_{ij} \delta(t-t')$.
Then, the equal-time covariance matrix with components $C_{ij}(t) = \langle x_i(t)x_j^*(t) \rangle$ satisfies
\begin{equation}
    \frac{d C}{dt} = L C + C L^\dagger + M M^\dagger
\end{equation}
so in the steady-state, the covariance matrix $C_{\text{ss}}$ satisfies the Lyapunov equation
\begin{equation}
    L C_{\text{ss}} + C_{\text{ss}} L^\dagger + M M^\dagger = 0.
\end{equation}
\citet{Ioannou1995} discusses the case where $M$ is the identity matrix (or unitary), so $M M^\dagger = \Id$. 
In this case, the steady-state modal energy (called maintained variance by \citet{Ioannou1995})
\begin{equation}
    E_\infty \equiv 
    \lim_{t \to \infty} \langle x^\dagger(t) x(t)\rangle
    = \lim_{t \to \infty} \text{tr} \, C(t)
\end{equation}
is bounded by
\begin{equation}
    E_\infty \geq \sum_{i} \frac{1}{-(\lambda_i + \lambda_i^*)}
\end{equation}
in which $\lambda_i$ are the eigenvalues of $L$, with equality if and only if $L$ is a normal operator.

The general case where $M$ is arbitrary is discussed in \citet{Weiss2003}, where the noise amplification is defined in a coordinate-invariant way and related to broken detail balance. 
\citet{Farrell1996a,Ioannou1995} also discuss the response to an arbitrary forcing as a function of frequency.
In short, the forced system $(d/dt)\bm{x} = L \bm{x} + \bm{f}(t)$ in which $\bm{f}(t)$ is an arbitrary forcing is analyzed in terms of the resolvent operator $R(\omega) = [\ii \omega - L]^{-1}$ which satisfies $\text{tr}\, [R^\dagger(\omega) R(\omega)] \geq \sum_i [\ii \omega - \lambda_i]^{-2}$, with equality only when $L$ is normal.
\citet{Farrell1996b} extends these results to time-dependent linear systems, where $L = L(t)$ depends on time.

\subsubsection{Non-normal enhancement of noise-induced patterns}
\label{noise_induced_patterns}

A strong enhancement of noise-induced patterns can be observed in non-normal dynamics \cite{Ridolfi2011,McKane2013,Biancalani2017,Klika2017,Muolo2019}. 
The main mechanism is analyzed by \citet{Biancalani2017}, see Fig.~\ref{figure_non_normal}B for an example.
By dimensional analysis, the (mean square) amplitude of noise-induced patterns can be estimated to be of order $\tau \sigma^2$ in which $\tau$ is a characteristic decay time associated to the linearized dynamics and $\sigma$ the strength of the noise (so the linearized dynamics looks like $\dot{x} = - 1/\tau x + \sigma \xi$ where $\xi$ is a standard Gaussian white noise).
This suggests that noise-induced patterns may be unobservable in situations where the noise amplitude is small.
For instance, this argument suggests that the demographic noise due to a finite population size $N$ will be of order $1/\sqrt{N}$.
As it turns out, the dimensional argument is only valid for normal dynamics, both because of the linear enhanced fluctuations (Sec.~\ref{enhanced_fluctuations}) and because of the non-normal transient amplification (Sec.~\ref{transient_amplification}) that may trigger nonlinearities.

\subsubsection{Overdamped oscillations}
\label{overdamped_oscillations}

We have seen in paragraph \ref{cep} that the presence of a critical exceptional point may lead to the conversion from diffusive to propagative dynamics. 
More generally, exceptional points typically mark the boundary between \enquote{PT-broken} and \enquote{PT-unbroken} regions of a family of operators (Sec.~\ref{PT_symmetry}), a mechanism that allows for the sudden appearance of imaginary parts in a complex growth rate. 
This means that oscillations can arise in an overdamped system as a consequence of non-reciprocity.
You can for instance observe this in the elasticity of chiral active solids, which involves so-called odd elastic moduli in addition to the standard ones \cite{Scheibner2020}, see \citet{Fruchart2023} for a review. 
These odd moduli, that violate Maxwell-Betti reciprocity (Sec.~\ref{nonreciprocal_responses}), can only occur in solids with internal sources of energy, and may lead to overdamped elastic oscillations \cite{Scheibner2020}.

\subsubsection{Instabilities induced by non-reciprocity}
\label{instabilities_from_nonreciprocity}

In this section, we consider deterministic instabilities induced by non-reciprocity (contrary to section \ref{noise_induced_patterns} discussing noise-induced phenomena).
First, the same phenomenon leading to overdamped oscillations (Sec.~\ref{overdamped_oscillations}) in first-order dynamics ($\partial_t X = \cdots)$ can produce instabilities in second-order dynamics ($\partial_t^2 X = \cdots)$. Examples range from pattern formation \cite{Coullet1985b,Coullet1985c} to elasticity \cite{Scheibner2020} and nonequilibrium spinor condensates \cite{Bernier2014}.
Similar mechanisms associated to the non-normal dynamics of non-reciprocal systems take place in a variety of systems, and we review a few of them.
As we shall see, the common mechanism is that a perturbation mixes the only eigenvector at the exceptional point with the missing direction(s), i.e. the other generalized eigenvectors in the Jordan chain.
We also refer to \citet{Marsden1994,Krechetnikov2007} for a discussion of dissipation-induced instabilities, from the perspective of perturbed Hamiltonian dynamics.

\paragraph{Cross-diffusion}

\citet{Shraiman1986} analyzed the Kuramoto-Sivashinsky (KS) equation
\begin{equation}
    \partial_t u + \partial_x^2 u + \partial_x^4 + u \partial_x u + \kappa u = 0
\end{equation}
on a system of size $L$ with periodic boundary conditions, in which a damping $\kappa$ has been artificially added to analyze the effect of an explicit symmetry breaking.
This equation has been introduced to describe the instabilities of a laminar flame front, and has since become a model equation for chaotic patterns \cite{Kevrekidis1990,Hyman1986} that arises in many contexts~\cite{Craster2009}.
The soft modes of a perturbation about a stationary periodic solutions of the KS equation with a wavelength $\lambda_0$ follow the equation
\begin{equation}
    \partial_t \begin{pmatrix}
        \phi \\
        \xi
    \end{pmatrix}
    =
    -
    \left[
    \begin{pmatrix}
        0 & 1 \\
        0 & \kappa
    \end{pmatrix}
    +
    \begin{pmatrix}
        \alpha & \gamma \\
        \sigma & \nu
    \end{pmatrix}
    \partial_x^2
    \right]
    \begin{pmatrix}
        \phi \\
        \xi
    \end{pmatrix}
\end{equation}
in which $\phi$ is a phase variable associated to the translation of the original pattern (Goldstone mode) and $\xi$ is an additive perturbation to the field $u$, and where the elements of the matrix depend on $\lambda_0$.
Note that there is an exceptional point at momentum $k=0$ when the artificial symmetry-breaking term $\kappa$ vanishes.
In the general case, the soft modes are
diffusive when $k \ll \kappa$, but at finite wavelengths $k \gg \kappa$, this leads to a dispersion relation $s = (\alpha + \nu) k^2 \pm \ii \sqrt{\sigma} |k|$.
Even though these viscoelastic modes (see \citet{Frisch1986}) are irrelevant for certain wavelengths of the original pattern (when $\sigma(\lambda_0) > 0$ and $\alpha + \nu < 0$), \citet{Shraiman1986} argue that they are unstable to finite-amplitude dilations which effectively change $\lambda_0$ and change the sign of $\sigma(\lambda_0)$, making the system unstable and leading to the formation of phase shocks.

\paragraph{Cross-advection}

Consider the equation \cite{Fruchart2021}
\begin{equation}
    \partial_t \begin{pmatrix}
        \phi_1 \\
        \phi_2
    \end{pmatrix}
    =
    \left[
    \begin{pmatrix}
        0 & 1 \\
        0 & 0
    \end{pmatrix}
    + 
    M (\bm{v}_0 \cdot \nabla) 
    + 
    D \nabla^2 
   \right]
    \label{ep_plus_flow}
\end{equation}
in which we have considered an advective term that mixes the degrees of freedom through the matrix $M$, and a scalar diffusive term with diffusion constant $D$.
As the eigenvalues of a perturbed exceptional point typically behave as the square root of the perturbation~\cite{Kato1984},  we expect the momentum space complex growth rate to behave as $s_{\pm}(k) \simeq \pm \ii\sqrt{\ii v_{\text{ss}} \, k}$ at small wavevector $k$, and as $- D \, k^2$ at large~$k$, leading to a maximum in the growth rate at finite momentum.
An explicit calculation shows that it is indeed the case, except in particular cases (for instance when $M$ is a scalar matrix).
In models of multiple-species nonreciprocal flocking, this typically leads to deterministic pattern-forming instabilities near the homogeneous-state critical lines \cite{Fruchart2021}.

\begin{figure}
    \centering
    \includegraphics[width=85mm]{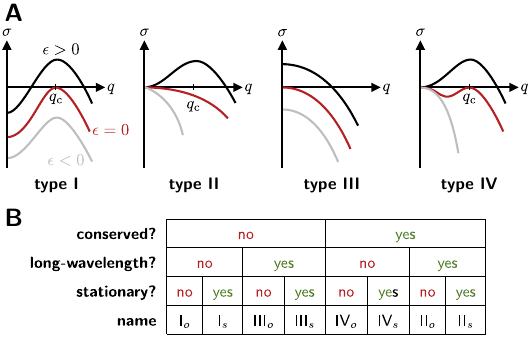}
    \caption{\textbf{Classification of single-mode linear instabilities of isotropic homogeneous states.}
    The classification is implemented by asking three questions: is there a conserved quantity? is the critical mode long-wavelength or finite-wavelength? is the instability stationary?
    Classes I-III are defined in \citet{Cross1993}, and class IV in \citet{FrohoffHulsmann2023b}.
    The following names are sometimes used:
    ($\text{I}_{\text{o}}$) finite-wavelength Hopf,
    ($\text{I}_{\text{s}}$) Turing,
    ($\text{II}_{\text{o}}$) conserved-Hopf,
    ($\text{II}_{\text{s}}$) Cahn-Hilliard,
    ($\text{III}_{\text{o}}$) Hopf,
    ($\text{III}_{\text{s}}$) Allen-Cahn,
    ($\text{IV}_{\text{o}}$) conserved finite-wavelength Hopf
    ($\text{IV}_{\text{s}}$) conserved-Turing.
\label{figure_classification_instabilities}
    }
\end{figure}

\paragraph{Classification of instabilities}
\label{classification_instabilities}

In this paragraph, we review a classification of the linear instabilities of uniform steady-states due to \citet{Cross1993} and recently extended in \citet{FrohoffHulsmann2023b} in the context of nonreciprocal Cahn-Hilliard models (Eq.~\eqref{nrch}).
This classification relies on the following hypotheses: (i) before the instability, the system is uniform, so we can study its linear stability in reciprocal space, (ii) there is a single branch of dispersion relation $\bm{k} \mapsto s(\bm{k})$ giving the complex growth rate $s = \sigma + \ii \omega$ which is relevant to the instability, meaning that all other modes are sufficiently quickly damped to be ignored (ii) the dispersion is isotropic, so we can consider only the norm $k=\lVert\bm{k}\rVert$ of the momentum in the dispersion $s(k)$.
One can expect these hypotheses to typically occur in homogeneous isotropic systems described by scalar fields, although accidental degeneracies are possible. 
We then ask three binary questions:
\begin{itemize}[nosep,left=0pt,label=--]
    \item does the dynamics involve a conservation law or not?
    \item is the instability short-wavelength or long-wavelength?
    \item is the instability stationary or oscillatory
\end{itemize}
leading to eight distinct classes.
These questions are made precise as follows.
Let $\epsilon$ be the control parameter of the instability, such that the system is stable when $\epsilon < 0$ and unstable when $\epsilon > 0$.
We define $q_{\text{c}}(\epsilon)$ as the position of the maximum of the growth rate curve $\sigma(q)$. 
Then,
\begin{itemize}[nosep,left=0pt,label=--]
    \item if $\sigma(q \to 0) = 0$ for all $\epsilon$, the instability is said to be {conserved} (else, {nonconserved})
    \item if $q_{\text{c}} = 0$ as $\epsilon \to 0$, the instability is said to be {long-wavelength} (else, {short-wavelength})
    \item if $\omega(q_{\text{c}}) = 0$ when $\epsilon \to 0$, the instability is said to be {stationary} (else, {oscillatory}).
\end{itemize}
A summary is given in Fig.~\ref{figure_classification_instabilities}, in which we give the standard names from \citet{Cross1993} except for the classes IV discussed in \citet{FrohoffHulsmann2023b}.
Further discussions on linear instabilities especially in the context of pattern formation are given in \cite{Cross1993} and in the textbooks \citet{Cross2009} and \citet{Hoyle2006}, as well as \citet{Shklyaev2017} especially for long-wavelength instabilities.

\subsubsection{Generation of correlated noise}

Consider the stochastic dynamics described by
\begin{equation}
\label{eom_cn}
    \frac{d \bm{x}}{dt} = \bm{f}(\bm{x}) + \bm{\zeta}(t)
\end{equation}
in which $\bm{\zeta}(t)$ is a Gaussian correlated noise, namely, satisfying $\langle \bm{\zeta}(t) \bm{\zeta}(t') \rangle = \bm{G}(t-t')$ with some matrix-valued correlation function $\bm{G}(\tau)$. 
Such a noise is sometimes called \enquote{colored}, to distinguish it from a white noise for which $\bm{G}$ is the Dirac distribution \cite{Hanggi1994}.
The simplest instance of this is the Ornstein-Uhlenbeck noise, which can be constructed from the Langevin equation
\begin{align}
    \label{ou_langevin}
    \frac{d \bm{\zeta}}{dt} = - \frac{1}{\tau} \bm{\zeta} + \bm{\eta}(t)
\end{align}
in which $\tau$ is a time scale characterizing the Ornstein-Uhlenbeck noise and $\bm{\eta}(t)$ is a Gaussian white (i.e., uncorrelated) noise.
The Langevin dynamics describing the \emph{joint} evolution of $\bm{x}$ and $\bm{\zeta}$ is not variational, even if $\bm{f}$ is variational. Like for delayed interaction (Sec.~\ref{time_delayed}), the presence of a correlated noise makes the description of the dynamics non-Markovian, and adding Eq.~\eqref{ou_langevin} to Eq.~\eqref{eom_cn} can be seen as performing a Markovian embedding of the dynamics. 
As described in \citet{Mori1965,Mori1965b,Grigolini1982,Hanggi1985,Hanggi1993,Hanggi1994}, this can be generalized to more general correlated noises.
Here, the nonreciprocity stems from the hypothesis that (by construction) the noise in Eq.~\eqref{eom_cn} is not affected by the variable $\bm{x}$ (but $\bm{x}$ is affected by the noise).

\begin{figure}
    \centering
    \includegraphics[width=9cm]{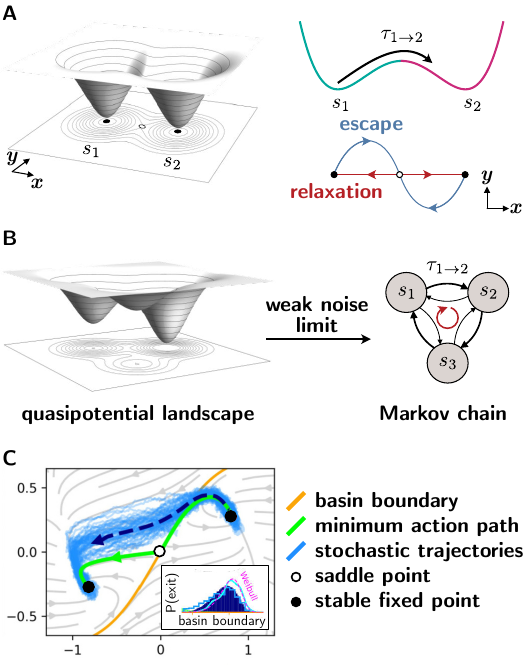}
    \caption{
    \label{figure_weak_noise_limit}
    \textbf{Weak noise limit in nonvariational systems.}
    (A) Quasipotential and Freidlin–Wentzell instantons in a nonvariational double-well potential.
    In nonvariational systems, the forward (relaxation, in red) and backward (escape, in blue) instantons are the mapped to each other by time-reversal (i.e. reversing the trajectory).
    See \citet{Kikuchi2020} for an explicit example.
    (B) In nonvariational systems, the Markov chain between discrete states corresponding to the basins of attraction of the deterministic attractors, obtained in the weak-noise limit from the escape rates $\tau_{i \to j}$, usually does not satisfy detailed balance, leading to fluxes in the steady-state (red circle).
    (C) In nonvariational systems (here a FitzHugh-Nagumo model), the stochastic trajectories (in blue) do not closely follow the Freidlin–Wentzell instanton (here the right-to-left instanton, in green), even at weak noise.
    This phenomenon has been called saddle avoidance: as illustrated here, the stochastic trajectories avoid the saddle point (white circle).
    In the inset, the measured distribution $P(\text{exit})$ of exit points along the orange basin boundary, shown as a blue line, is not gaussian but closer to a Weibull distribution.
    Adapted from \citet{Borner2024}.
    }
\end{figure}

\subsubsection{Escape times and paths in the weak-noise limit}
\label{escape_times_and_paths}

In equilibrium systems, the mean escape time from a potential $V$ is given by the Arrhenius-Kramers-Eyring law
\begin{equation}
    \tau^* \equiv \mathbb{E}[\tau] = \frac{2\pi}{|V''(x^*)| \, V''(x_{0})} \ee^{[V(x^*) - V(x_{0})] / T}
\end{equation}
in the limit where $T \to 0$, in which $x^*$ is the position of the barrier (maximum of $V$) and $x_{0}$ is the position of the minimum of $V$. 
In the case of irreversible systems, the weak-noise limit is described by the Freidlin–Wentzell large deviation theory (Sec.~\ref{transverse_decomposition}), see Fig.~\ref{figure_weak_noise_limit}A-B for pictures.
The consequences of having a non-gradient drift field have been analyzed in \citet{Maier1992,Maier1993,Maier1993b}. 
In particular, these references showed that the most probable exit path from the basin of attraction of a metastable point (defined as the exiting classical path of least action, aka instanton) can end at an unstable fixed point on the boundary (rather than a saddle point). As a consequent, the distribution of exit locations may spread over the boundary as the actual exit trajectory avoid the unstable fixed point, and may move sideways along the boundary until they read saddle points, making it non-Gaussian (inset in Fig.~\ref{figure_weak_noise_limit}C). 
In fact, it turns out that with small but finite noise, observed trajectories can sometimes avoid saddles and strongly deviate from the Freidlin-Wentzell instanton \cite{Maier1997,Borner2024}, see Fig.~\ref{figure_weak_noise_limit}C. 
In addition, the mean exit time (given in equilibrium by the Kramers-Eyring formula) is also modified, and can acquire a pre-exponential factor that depends on the noise strength (temperature) as a power law (rather than being constant) with an exponent that varies continuously as a function of the parameters of the model.
In contrast with systems with time-reversal invariance (detailed balance), the most probable path to leave a state is not necessarily the time-reversed of the most probable path to enter it (Fig.~\ref{figure_weak_noise_limit}A, compare blue and red paths). 
This manifestation of time-reversal invariance breaking has recently been analyzed in detail in \citet{Schuttler2025} in a dynamical system representing the mean-field dynamics of a nonreciprocal Ising model with two species (see Sec.~\ref{nr_ising}), where recurrent but aperiodic \enquote{Escher cycles} between metastable attractors are observed.
This illustrates a general feature, that the effective Markov chain whose states corresponds to the deterministic attractors (and their basins of attraction) does in general not satisfy detailed balance when the original stochastic dynamics does not satisfy detailed balance (Fig.~\ref{figure_weak_noise_limit}B).

\subsubsection{Accelerated convergence towards steady-states}
\label{accelerated_convergence}

In a stochastic differential equation, one of the effects of nonvariational drift fields is to accelerate the convergence of the stochastic process to the stationary distribution \cite{Pavliotis2014}. 
The rate of convergence to equilibrium is usually characterized by the \enquote{spectral gap} $\Lambda(\mathscr{L})$ of the generator $\mathscr{L} \equiv \hat{\mathbb{W}}^\dagger$ of the Fokker-Planck equation \eqref{fokker_planck}, namely the quantity
\begin{equation}
    \Lambda(\mathscr{L}) \equiv \sup \big\{  \text{Re}(\mu) \,|\;\mu \in \text{Sp}(\mathscr{L}) \backslash \{0\} \big\}
\end{equation}
where $\text{Sp}(\mathscr{L})$ is the eigenvalue spectrum of $\mathscr{L}$; this reduces to the first nonzero eigenvalue in simple cases. 
When $\Lambda(\mathscr{L}) < 0$, the corresponding stochastic process is said to \enquote{have a spectral gap}.
In this case, it has an exponential convergence rate  towards the steady-state distribution $p_{\text{ss}}$ in the sense that 
$\lVert p(t) - p_{\text{ss}} \rVert \leq e^{-t/\tau(\mathscr{L})} \lVert p(t) - p_{\text{ss}} \rVert$
in which we have defined the characteristic time
\begin{equation}
    \label{sgap_ctime}
    \tau(\mathscr{L})\equiv|\Lambda(\mathscr{L})|^{-1}
\end{equation}

\citet{Hwang2005,Hwang1993} considered a family of stochastic differential equations on $\RR^D$ of the form
\begin{equation}
    \label{acc_sde}
    dX = [- \nabla U(X) + C(X)] dt + \sqrt{2 \epsilon} \, dW.
\end{equation}
where $W(t)$ is the Brownian motion on $\RR^d$.
The corresponding stochastic process is denoted by $X_{C,\epsilon}$. 
When $C$ vanishes identically, the drift field is variational and the system has a stationary distribution $p_{\text{ss}}(X) = 1/Z_\epsilon \ee^{-U(X)/\epsilon}$. 
The nonvariational perturbation $C$ is assumed to satisfy $\nabla\cdot(C \, p_{\text{ss}}) = 0$ which ensures that $p_{\text{ss}}$ is also a stationary distribution of Eq.~\eqref{acc_sde}.
The corresponding contribution to the infinitesimal generator is antisymmetric under a suitable inner product.
Under these conditions, and assuming that these diffusive processes are well-behaved (and in particular that $\mathscr{L}_0$ has a spectral gap), \citet{Hwang2005} showed that
\begin{equation}
    \Lambda(\mathscr{L}_C) \leq \Lambda(\mathscr{L}_0) < 0
\end{equation}
in which $\mathscr{L}_C$ is the infinitesimal generator associated to Eq.~\eqref{acc_sde}.
In terms of characteristic times $\tau(C)\equiv\tau(\mathscr{L}_C)$ in Eq.~\eqref{sgap_ctime}, $\tau(C) \leq \tau(0)$, meaning that adding an irreversible drift accelerates the convergence towards the steady-state.

This acceleration can also be observed in the weak noise limit.
\citet{Lee2022} considered a similar setup, starting from the SDE \eqref{acc_sde} under the assumption that $\nabla U \cdot C = 0$ and $\nabla \cdot C = 0$ everywhere.
This implies that $\nabla\cdot(C \, p_{\text{ss}}) = 0$, ensuring that $p_{\text{ss}}$ is a stationary distribution, and the first condition (corresponding to the transverse decomposition of Sec.~\ref{transverse_decomposition}) guarantees that $U$ coincides with the quasipotential of \eqref{acc_sde} when both are defined (see Sec.~\ref{transverse_decomposition}).
Under these hypotheses, the stable points of the stochastic process $X_{C,\epsilon}$ coincide with the stable points of $X_{0,\epsilon}$, which are the local minima of $U$, which allows one to compare the mean transition times between the states.
\citet{Lee2022} proved a Kramers-Eyring formula for the process $X_{C,\epsilon}$ and showed that resulting mean transition time between two minima of $U$ under the process with a nonvariational drift $X_{C,\epsilon}$ is always faster than that of the overdamped (variational) dynamics $X_{0,\epsilon}$. 

This acceleration can be used as a practical tool in order to efficiently sample an equilibrium distribution, for instance through Markov chain Monte Carlo algorithms, see \cite{Suwa2010,Ichiki2013,Ohzeki2015,Duncan2016,Lelievre2013,ReyBellet2015,Kaiser2017,Coghi2021} for further discussions and \cite{Ghimenti2023,Ghimenti2024,Ghimenti2024b,Ghimenti2024c} for applications to dense and glassy liquids using transverse forces.

From a more fundamental perspective, it is related to effects such as the enhancement of the diffusion coefficient by advection (see Sec.~\ref{modes_same_field}) in a shear flow (known as Taylor-Aris dispersion; see \citet{Aris1956,Taylor1953}), or to similar enhancements in odd diffusive systems (see Sec.~\ref{hall_odd_effects}; see \citet{Kalz2022}).
See also \citet{Lam2014} for a similar situation in noisy integrable systems.
Enhanced diffusion has also been reported in mixtures of particles with non-reciprocal interactions \cite{Benois2023,Damman2024}.

\subsection{Universality and renormalization group}
\label{nonvariational_renormalization_group}

Broadly defined, \enquote{universality} refers to the idea that natural phenomena can be captured and understood through simplified descriptions in which not all the microscopic details are encoded, so that different physical systems may have \enquote{the same} important properties, because they only differ by unimportant microscopic details. 
In statistical physics, and in particular in the study of critical phenomena, the notion of universality can be made more precise through the renormalization group \citet{Lesne2012,Lesne1998,Cardy1996,Goldenfeld2018,Kardar2007,Fisher1998,Tauber2014,Kamenev2023}.
Nonequilibrium systems challenge some of the assumptions underlying universality. For instance, that stationary nonequilibrium states generically exhibit long range correlations \cite{Bertini2015}, boundary conditions including disorder at the boundary can strongly affect bulk phase behaviour \cite{Granek2024}, 
and is also possible to observe behaviors usually peculiar to critical systems, such as scale invariance and power laws, in systems that are not tuned to a critical point \cite{Grinstein1991,Bonachela2009,Munoz2018,Henkel2008}, an idea sometimes known as self-organized criticality~\cite{Watkins2015}

Nevertheless, it turns out the idea of universality has been extended to many classes of nonequilibrium systems across domains \cite{Vicsek1992,Odor2004,Odor2008,Kardar2007,Henkel2008,Henkel2010,Munoz2018,Dupuis2021,Sieberer2025}. 
In this paragraph, we discuss nonvariational renormalization group flows, a particular feature that tends to occur in nonequilibrium systems, in particular nonreciprocal field theories, but that turns out to be even possible at equilibrium.

\paragraph{The renormalization group.}

The renormalization group (RG) describes an evolution in the space of mathematical models (theory space) as \enquote{fast} or \enquote{irrelevant} degrees of freedom are systematically integrated out while \enquote{slow} or \enquote{relevant} degrees of freedom are kept. 
This coarse-graining process is encoded as a flow on theory space through a dynamical system
\begin{equation}
    \frac{d \bm{g}}{dt} = \bm{\beta}(\bm{g})
    \label{rg_ds}
\end{equation}
where the vector field $\bm{\beta}$ is called the \enquote{beta function}, $\bm{g}$ labels points in theory, and $t=- \log \mu$ is the RG coarse-graining scale.

The intuitive picture we have of a well-behaved RG flow is as follows:
\begin{itemize}[nosep,left=0pt,label=--]
  \item Even though theory space is high-dimensional, we quickly end up on a low-dimensional invariant submanifold after a bit of coarse-graining, after which we can assume that $\bm{g}$ only has a few components.
  \item There are only a few isolated fixed points on this attracting invariant manifold: stable fixed points correspond to thermodynamic phases, while saddle points correspond to phase transitions. These RG fixed points are, by definition, scale invariant.
  \item Critical phenomena (phenomena that arise near a phase transition) can be studied by analysing the neighbourhood of the corresponding fixed points: the eigenvalues of the Jacobian of $\bm{\beta}$ at the fixed points determine the critical exponents that summarize the scaling behaviour of the phase transition.
  \item A universality class is defined as the set of all points in theory space that flow from or to such a RG fixed point, and all share the same critical exponents.
\end{itemize}
This picture implicitly assumes that $\bm{\beta}$ defines a Morse-Smale gradient/relaxational flow, where fixed points are hyperbolic (and in particular isolated). Universality classes are therefore discrete, and no trouble arise. Mild troubles arise when the fixed points are not hyperbolic (for instance, not isolated); this can for instance lead to continuously varying critical exponents~\cite{Baxter1982,Mukherjee2023}, but the general picture of universality is not entirely lost.

\begin{figure}
    \centering
    \includegraphics[width=9cm]{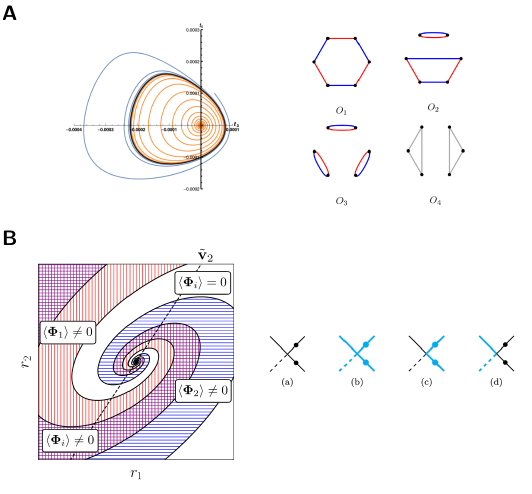}
    \caption{
    \label{figure_nonvariational_rg}
    \textbf{Nonvariational renormalization group flows.}
    (A) RG flow exhibiting a limit cycle in the space of couplings (corresponding to the diagrams on the right) in a $O(N)$ model with sextic interactions. Here, the theory is continued to noninteger values of $N$ where it is nonunitary.
    Adapted from \citet{Jepsen2021}.
    (B) Phase diagram of a nonreciprocal $O(n_1) \times O(n_2)$ model in a situation where the critical exponent $\nu$ is complex, leading to logarithmic spirals with discrete scale invariance in the phase diagram. White indicates disorder, red and blue describes phases where only one of the order parameters is nonzero, and purple a fully ordered phase.
    Adapted from \citet{Young2024}.
    }
\end{figure}

\paragraph{Nonvariational renormalization group flows.}

More serious troubles can however arise when the dynamical system \eqref{rg_ds} is not variational (Sec.~\ref{nonvariational_dynamics}): in this case, the RG flow can exhibit limit cycles, limit tori, or chaotic sets~\cite{Wilson1974,Damgaard1992,Morozov2003,Gukov2017} (Fig.~\ref{figure_nonvariational_rg}A). 
This has been theoretically demonstrated in a variety of systems~\cite{Bernard2001,Glazek2002,Jepsen2021b,Bosschaert2022,Jepsen2021} such as certain models of spin glasses and disordered systems \cite{Fisher1986,Banavar1987,McKay1982,Derrida1983,LeDoussal2006}.
The presence of chaotic repellers (attractors for the reverse dynamics) leads to fractal basin boundaries \cite{Grebogi1987} which can be interpreted as a strong sensitivity to microscopic details.
Similarly, the eigenstructure of fixed points can be modified, including complex eigenvalues (leading to complex critical exponents and discrete scale invariance \cite{Sornette1998}) and non-diagonalizable Jacobians.
\citet{Young2024,Young2020} discusses these phenomena in coupled $O(n_1) \times O(n_2)$ models out-of-equilibrium; they show in particular that nonvariational RG flows can arise from a non-reciprocal coupling between the fields (Fig.~\ref{figure_nonvariational_rg}B).

In spite of all these works, the physical consequences of nonvariational RG flows are not fully elucidated.
In particular, it is not obvious that even chaotic RG flows completely rule out the possibility of some form of universality (for instance if the dynamics is still attracted to a low-dimensional manifold), although how to precisely define it is still unclear.
To the best of our knowledge, there is no general proof that, under a given set of assumptions, the RG flow is a gradient flow or admits a Lyapunov function.
The closest results, known as \enquote{c-theorems} (or $a$ or $F$ theorems depending on the space dimension), show that under certain conditions, there is a function $c$ that monotonically decreases under the RG flow of unitary quantum field theories.
However, this function is constructed from a section of a line bundle over theory space which is not necessarily trivializable, and as a consequence the function $c$ may be multivalued (and hence not a proper Lyapunov function) \cite{Morozov2003,Curtright2012}.
It is possible that information-theoretic perspectives on RG will permit progress in this direction \cite{Gaite1996,Beny2015,Apenko2012,Machta2013,KochJanusz2018,Gordon2021}.

\section{Conclusions}
Nonreciprocity, as a perspective on nonequilibrium systems, is currently providing fruitful intuitions across the physical sciences.
These invite us to reinvestigate time-honored topics from a different point of view, consolidating existing insights but also raising new questions.
These range from the effect of nonreciprocal responses on turbulent flows, of possible relevance in tokamak plasma, to the influence of interactions that do not satisfy Newton's third law on the glass transition. Similarly, recent generalizations of key many-body phenomena such as aging and replica-symmetry broken phases to systems where non-reciprocity is structured have potential applications to fields as diverse as neuromorphic computing, quantum transport, machine learning and ecology.

Even though nonreciprocity is often directly accessible to observations, sometimes even with the unaided eye, improvements in the design of dedicated experimental platforms are needed to tell apart the precise underlying mechanisms at play, especially in biology where the complexity of the experimental data may require new tools to be adequately handled.
From a more abstract perspective, nonreciprocity sheds a new light on non-unitary field theories, which play a key role in high-energy theory and mathematical physics, with applications to particle physics and cosmology, and which are at the root of some of the key conceptual tools such as PT-symmetry that are underlying nonreciprocal field theories.

In many-body nonreciprocal systems, the statistical physics notion of universality appears to extend to dynamical phases (many-body limit cycles or time crystals and generalizations), going from quantum limit cycles to biological oscillators.
Perhaps even more importantly, we are starting to recognize distinct classes of dynamic behavior that are observable in real, finite and noisy experimental systems, and the underlying mechanisms can be, to some extent, rationalized.
Yet, a robust and usable classification scheme is not presently available, that would allow us to infer, predict, and control -- with guarantees -- the behavior of a given system.

We have reviewed some of the recent progress that go towards such a scheme. In our view, a many-body theory of non-reciprocal systems capable of capturing the rich experimental landscape we have explored would blend together a Conley-Morse description of dynamical systems and bifurcations, a large deviation picture of noisy systems, a Halperin-Hohenberg picture of critical phenomena, a Cross-Hohenberg classification of instabilities, and probably new mathematical tools to be invented.

\clearpage

\onecolumngrid
\appendix

\clearpage
\global\bibsep=0.05cm

\end{document}